\documentclass[twocolumn, twocolappendix, resetfootnote, trackchanges]{aastex702}%longauthor

\usepackage{soul}

\usepackage{enumitem}
\usepackage{lipsum}

\usepackage{CJK}
\usepackage[whole]{bxcjkjatype} 
\usepackage{CJKutf8}
\usepackage[T1]{fontenc}

\usepackage{amsmath}

\usepackage{booktabs}

\usepackage{xspace}

\definecolor{fed_blue}{HTML}{07004D}
\definecolor{steel_blue}{HTML}{2D82B7}
\definecolor{steel_blue_dark}{HTML}{1C71A6}
\definecolor{aqua_marine}{HTML}{42E2B8}
\definecolor{dutch_white}{HTML}{F3DFBF}
\definecolor{light_coral}{HTML}{EB8A90}
\definecolor{light_coral_dark}{HTML}{BA5A60}

\hypersetup{
    colorlinks  = true,
    linkcolor   = purple,
    urlcolor    = steel_blue_dark,
    citecolor   = steel_blue_dark,
    anchorcolor = black
}

\newcommand{\gaia}{\textit{Gaia}\xspace}
\newcommand{\galex}{\textit{GALEX}\xspace}

\newcommand{\jwst}{\textit{JWST}\xspace}
\newcommand{\swift}{\textit{Swift}\xspace}

\newcommand{\wise}{\textit{WISE}\xspace}
\newcommand{\nustar}{\textit{NuSTAR}\xspace}

\newcommand{\kms}{~\rm km\,s^{-1}}
\newcommand{\program}[1]{\textsc{#1}}

\newcommand{\sn}{SN\,2023iop\xspace}

\shorttitle{\sn \& the Death of Very Massive Stars}
\shortauthors{Schulze et al.}
\submitjournal{ApJ}
\begin{document}

\title{
SN\,2023iop: An Extreme 1987A-like Supernova and the Death of Very Massive Stars
}

\author[0000-0001-6797-1889, gname=Steve, sname=Schulze]{Steve Schulze}
\affiliation{Department of Particle Physics and Astrophysics, Weizmann Institute of Science, 234 Herzl St, 76100 Rehovot, Israel}
\affiliation{Center for Interdisciplinary Exploration and Research in Astrophysics (CIERA), Northwestern University, 1800 Sherman Ave., Evanston, IL 60201, USA}
\email[show]{steve.schulze@weizmann.ac.il}

\author[0000-0003-2091-622X, gname=Avinash, sname=Singh]{Avinash Singh}
\affiliation{The Oskar Klein Centre, Department of Astronomy, Stockholm University, Albanova University Center, 106 91 Stockholm, Sweden}
\email{avinash.singh@astro.su.se}

\author[0000-0003-2700-1030, gname=Nikhil, sname=Sarin]{Nikhil Sarin}
\affiliation{Kavli Institute for Cosmology, University of Cambridge, Madingley Road, CB3 0HA, UK}
\affiliation{Institute of Astronomy, University of Cambridge, Madingley Road, CB3 0HA, UK}
\email{ns2043@cam.ac.uk}

\author[0000-0002-0301-8017, gname=Ofer, sname=Yaron]{Ofer Yaron}
\affiliation{Department of Particle Physics and Astrophysics, Weizmann Institute of Science, 234 Herzl St, 76100 Rehovot, Israel}
\email{ofer.yaron@weizmann.ac.il}

\author[0000-0003-1546-6615, gname=Jesper, sname=Sollerman]{Jesper Sollerman}
\affiliation{The Oskar Klein Centre, Department of Astronomy, Stockholm University, Albanova University Center, 106 91 Stockholm, Sweden}
\email{jesper@astro.su.se}

\author[0000-0003-1169-1954, gname=Takashi J., sname=Moriya]{Takashi~J. Moriya}
\affiliation{National Astronomical Observatory of Japan, National Institutes of Natural Sciences, 2-21-1 Osawa, Mitaka, Tokyo 181-8588, Japan}
\affiliation{Graduate Institute for Advanced Studies, SOKENDAI, 2-21-1 Osawa, Mitaka, Tokyo 181-8588, Japan}
\affiliation{School of Physics and Astronomy, Monash University, Clayton, VIC 3800, Australia}
\email{takashi.moriya@nao.ac.jp}

\author[0000-0002-3653-5598, gname=Avishay, sname=Gal-Yam]{Avishay Gal-Yam}
\affiliation{Department of Particle Physics and Astrophysics, Weizmann Institute of Science, 234 Herzl St, 76100 Rehovot, Israel}
\email{avishay.gal-yam@weizmann.ac.il}

\author[0009-0001-3767-942X, gname=Bart, sname=van Baal]{Bart van Baal}
\affiliation{Finnish Centre for Astronomy with ESO (FINCA), FI-20014 University of Turku, Finland}
\affiliation{Department of Physics and Astronomy, FI-20014 University of Turku, Finland}
\affiliation{The Oskar Klein Centre, Department of Astronomy, Stockholm University, Albanova University Center, 106 91 Stockholm, Sweden}
\email{barteld.vbaal@astro.su.se}

\author[0000-0003-4800-2737, gname=Stan, sname=Barmentloo]{Stan Barmentloo}
\affiliation{The Oskar Klein Centre, Department of Astronomy, Stockholm University, Albanova University Center, 106 91 Stockholm, Sweden}
\email{stan.barmentloo@astro.su.se}

\author[0000-0002-9646-8710, gname=Conor M. B., sname= Omand]{Conor~M.~B. Omand}
\affiliation{Astrophysics Research Institute, Liverpool John Moores University, Liverpool Science Park, 146 Brownlow Hill, Liverpool L3 5RF, UK}
\email{c.m.omand@ljmu.ac.uk}
    
\author[0009-0004-1442-619X, gname=Hannah, sname=Wichern]{Hannah Wichern}
\affiliation{DTU Space, Technical University of Denmark, Elektrovej 327/238, Kongens Lyngby, DK-2800, Denmark}
\email{hawic@space.dtu.dk}

\author[0000-0002-5477-0217, gname=Tuomas, sname=Kangas]{Tuomas Kangas}
\affiliation{Finnish Centre for Astronomy with ESO (FINCA), FI-20014 University of Turku, Finland}
\affiliation{Department of Physics and Astronomy, FI-20014 University of Turku, Finland}
\email{tjakan@utu.fi}

\author[0000-0002-8041-8559, gname=Priscila J., sname=Pessi]{Priscila~J. Pessi}
\affiliation{Astrophysics Division, National Centre for Nuclear Research, Pasteura 7, 02-093 Warsaw, Poland}
\email{Priscila.Pessi@ncbj.gov.pl}

\author[0000-0001-9515-478X, gname=Adam A., sname=Miller]{Adam~A. Miller}
\affiliation{Department of Physics and Astronomy, Northwestern University, 2145 Sheridan Road, Evanston, IL 60208, USA}
\affiliation{Center for Interdisciplinary Exploration and Research in Astrophysics (CIERA), Northwestern University, 1800 Sherman Ave., Evanston, IL 60201, USA}
\affiliation{NSF—Simons AI Institute for the Sky (SkAI), 172 E. Chestnut St., Chicago, IL 60611, USA}
\email{amiller@northwestern.edu}

\author[0000-0002-4410-5387, gname=Armin, sname=Rest]{Armin Rest}
\affiliation{Space Telescope Science Institute, 3700 San Martin Drive, Baltimore, MD 21218, USA}
\email{arest@Space Telescope Science Institute, 3700 San Martin Drive, Baltimore, MD 21218, USA.edu}

\author[0000-0003-1710-9339, gname=Lin, sname=Yan]{Lin Yan \begin{CJK}{UTF8}{gbsn}(严琳)\end{CJK}}
\affiliation{Division of Physics, Mathematics and Astronomy, California Institute of Technology, Pasadena, CA 91125, USA}
\affiliation{The Caltech Optical Observatories, California Institute of Technology, Pasadena, CA 91125, USA}
\email{lyan@caltech.edu}

\author[0000-0003-3460-0103, gname=Alexei V., sname=Filippenko]{Alexei~V. Filippenko}
\affiliation{Department of Astronomy, University of California, Berkeley, CA 94720-3411, USA}
\email{afilippenko@berkeley.edu}

\author[0000-0002-7866-4531, gname=Chang, sname=Liu]{Chang Liu \begin{CJK}{UTF8}{gbsn}(刘畅)\end{CJK}}
\affiliation{Department of Physics and Astronomy, Northwestern University, 2145 Sheridan Road, Evanston, IL 60208, USA}
\affiliation{Center for Interdisciplinary Exploration and Research in Astrophysics (CIERA), Northwestern University, 1800 Sherman Ave., Evanston, IL 60201, USA}
\affiliation{NSF—Simons AI Institute for the Sky (SkAI), 172 E. Chestnut St., Chicago, IL 60611, USA}
\email{ptg.cliu@u.northwestern.edu}

\author[0000-0003-4663-4300, gname=Miika, sname=Pursiainen]{Miika Pursiainen}
\affiliation{Department of Physics, University of Warwick, Gibbet Hill Road, Coventry, CV4 7AL, UK}
\email{miika.pursiainen@warwick.ac.uk}

\author[0000-0002-6535-8500, gname=Yi, sname=Yang]{Yi Yang \begin{CJK}{UTF8}{gbsn}(杨轶)\end{CJK}}
\affiliation{Department of Physics, Tsinghua University, Qinghua Yuan, Beijing 100084, China}
\email{yi_yang@mail.tsinghua.edu.cn}

\author[0000-0001-9494-179X, gname=Georgios, sname=Dimitriadis]{Georgios Dimitriadis}
\affiliation{School of Physics and Astronomy, Lancaster University, Lancaster, LA1 4YB, UK}
\email{g.dimitriadis@lancaster.ac.uk}

\author[0000-0002-8989-0542, gname=Kishalay, sname=De]{Kishalay De}
\affiliation{Department of Astronomy and Columbia Astrophysics Laboratory, Columbia University, 550 W 120th St. MC 5246, New York, NY 10027, USA}
\affiliation{Center for Computational Astrophysics, Flatiron Institute, 162 5th Ave., New York, NY 10010, USA}
\email{kd3038@columbia.edu}

\author[0000-0003-1325-6235, gname=Seán J., sname=Brennan]{Seán~J. Brennan}
\affiliation{Max-Planck-Institut f{\"u}r Extraterrestrische Physik, Giessenbachstra{\ss}e 1, 85748, Garching, Germany}
\email{sbrennan@mpe.mpg.de}

\author[0000-0001-5955-2502, gname=Thomas G., sname=Brink]{Thomas~G. Brink}
\affiliation{Department of Astronomy, University of California, Berkeley, CA 94720-3411, USA}
\email{tgbrink@berkeley.edu}

\author[0000-0003-0853-6427, gname=Ping, sname=Chen]{Ping Chen}
\affiliation{Institute for Advanced Study in Physics, Zhejiang University, Hangzhou 310058, China}
\affiliation{Institute for Astronomy, School of Physics, Zhejiang University, Hangzhou 310058, China}
\email{chenp1220@gmail.com}

\author[0000-0001-7101-9831, gname=Aleksandar, sname=Cikota]{Aleksandar Cikota}
\affiliation{Gemini Observatory / NSF's NOIRLab, Casilla 603, La Serena, Chile}
\email{aleksandar.cikota@noirlab.edu}

\author[0000-0003-2910-6565, gname=Felice, sname=Cusano]{Felice Cusano}
\affiliation{INAF - Osservatorio di Astrofisica e Scienza dello Spazio, via Piero Gobetti 93/3, 40129 Bologna, Italy}
\email{felice.cusano@inaf.it}

\author[0000-0001-6627-9903, gname=Nicholas, sname=Earley]{Nicholas Earley}
\affiliation{Division of Physics, Mathematics and Astronomy, California Institute of Technology, Pasadena, CA 91125, USA}
\email{nearley@caltech.edu}

\author[0000-0001-8532-3594, gname=Claes, sname=Fransson]{Claes Fransson}
\affiliation{The Oskar Klein Centre, Department of Astronomy, Stockholm University, Albanova University Center, 106 91 Stockholm, Sweden}
\email{claes@astro.su.se}

\author[0000-0002-1296-6887, gname=Lluís, sname=Galbany]{Lluís Galbany}
\affiliation{Institute of Space Sciences (ICE, CSIC), Campus UAB, Carrer de Can Magrans, s/n, 08193 Barcelona, Spain}
\affiliation{Institut d'Estudis Espacials de Catalunya (IEEC), 08860 Castelldefels (Barcelona), Spain}
\email{lluisgalbany@gmail.com}

\author[0009-0000-9383-2305, gname=Anamaria, sname=Gkini]{Anamaria Gkini}
\affiliation{The Oskar Klein Centre, Department of Astronomy, Stockholm University, Albanova University Center, 106 91 Stockholm, Sweden}
\email{anamaria.gkini@astro.su.se}

\author[0000-0001-8005-4030, gname=Anders, sname=Jerkstrand]{Anders Jerkstrand}
\affiliation{The Oskar Klein Centre, Department of Astronomy, Stockholm University, Albanova University Center, 106 91 Stockholm, Sweden}
\email{anders.jerkstrand@astro.su.se}

\author[0000-0001-9454-4639, gname=Ragnhild, sname=Lunnan]{Ragnhild Lunnan}
\affiliation{The Oskar Klein Centre, Department of Astronomy, Stockholm University, Albanova University Center, 106 91 Stockholm, Sweden}
\email{ragnhild.lunnan@astro.su.se}

\author[0000-0003-2611-7269, gname=Keiichi, sname=Maeda]{Keiichi Maeda}
\affiliation{Department of Astronomy, Kyoto University, Kitashirakawa-Oiwake-cho, Sakyo-ku, Kyoto 606-8502, Japan}
\email{keiichi.maeda@kusastro.kyoto-u.ac.jp}

\author[0000-0002-7409-8114, gname=Diego, sname=Paris]{Diego Paris}
\affiliation{INAF - Osservatorio Astronomico di Roma, via Frascati 33, I-00040 Monteporzio Catone, Italy}
\email{diego.paris@inaf.it}

\author[0000-0002-8860-6538, gname=Andrea, sname=Rossi]{Andrea Rossi}
\affiliation{INAF - Osservatorio di Astrofisica e Scienza dello Spazio, via Piero Gobetti 93/3, 40129 Bologna, Italy}
\email{andrea.rossi@inaf.it}

\author[0000-0002-5683-2389, gname=Nabeel, sname=Rehemtulla]{Nabeel Rehemtulla}
\affiliation{Department of Physics and Astronomy, Northwestern University, 2145 Sheridan Road, Evanston, IL 60208, USA}
\affiliation{Center for Interdisciplinary Exploration and Research in Astrophysics (CIERA), Northwestern University, 1800 Sherman Ave., Evanston, IL 60201, USA}
\affiliation{NSF—Simons AI Institute for the Sky (SkAI), 172 E. Chestnut St., Chicago, IL 60611, USA}
\email{nabeelrehemtulla2027@u.northwestern.edu}

\author[0009-0006-7528-0168, gname=Claudia, sname=Skoglund]{Claudia Skoglund}
\affiliation{The Oskar Klein Centre, Department of Astronomy, Stockholm University, Albanova University Center, 106 91 Stockholm, Sweden}
\email{claudia.skoglund@astro.su.se}

\author[0000-0002-2636-6508, gname=WeiKang, sname=Zheng]{WeiKang Zheng}
\affiliation{Department of Astronomy, University of California, Berkeley, CA 94720-3411, USA}
\email{weikang@berkeley.edu}

\author[0009-0007-2731-5562, gname=Alaa, sname=Alburai]{Alaa Alburai}
\affiliation{Institute of Space Sciences (ICE, CSIC), Campus UAB, Carrer de Can Magrans, s/n, 08193 Barcelona, Spain}
\affiliation{Institut d'Estudis Espacials de Catalunya (IEEC), 08860 Castelldefels (Barcelona), Spain}
\email{alburai@ice.csic.es}

\author[0000-0002-4848-5508]{Ke-Jung Chen}
\affiliation{Academia Sinica, Institute of Astronomy and Astrophysics, Taipei 106319, Taiwan, R.O.C.}
\affiliation{Heidelberger Institut f\"ur Theoretische Studien, Schloss-Wolfsbrunnenweg 35, 69118 Heidelberg, Germany}
\email{chenken1229@gmail.com}

\author[0000-0002-0129-806X, gname=K.-Ryan, sname=Hinds]{K.-Ryan Hinds}
\affiliation{Division of Physics, Mathematics and Astronomy, California Institute of Technology, Pasadena, CA 91125, USA}
\email{khinds@caltech.edu}

\author[0000-0002-3884-5637, gname=Anjasha, sname=Gangopadhyay]{Anjasha Gangopadhyay}
\affiliation{The Oskar Klein Centre, Department of Astronomy, Stockholm University, Albanova University Center, 106 91 Stockholm, Sweden}
\email{anjashagangopadhyay@gmail.com}

\author[0000-0002-0264-7356, gname=Jonay Isaí, sname=González Hernández]{Jonay Isaí González Hernández}
\affiliation{Instituto de Astrof\'{\i}sica de Canarias, V\'{\i}a L\'actea, 38205 La Laguna, Tenerife, Spain}
\affiliation{Universidad de La Laguna, Departamento de Astrof\'{\i}sica, 38206 La Laguna, Tenerife, Spain}
\email{jonay@iac.es}

\author[0000-0002-5936-1156, gname=Assaf, sname=Horesh]{Assaf Horesh}
\affiliation{Racah Institute of Physics, The Hebrew University of Jerusalem, Jerusalem, Israel}
\email{assafh@mail.huji.ac.il}

\author[0000-0002-8597-0756, gname=Giorgos, sname=Leloudas]{Giorgos Leloudas}
\affiliation{DTU Space, Technical University of Denmark, Elektrovej 327/238, Kongens Lyngby, DK-2800, Denmark}
\email{giorgos@space.dtu.dk}

\author[0000-0002-8691-7666, gname=Eliana, sname=Palazzi]{Eliana Palazzi}
\affiliation{INAF - Osservatorio di Astrofisica e Scienza dello Spazio, via Piero Gobetti 93/3, 40129 Bologna, Italy}
\email{eliana.palazzi@inaf.it}

\author[0000-0001-8472-1996, gname=Daniel A., sname=Perley]{Daniel~A. Perley}
\affiliation{Astrophysics Research Institute, Liverpool John Moores University, Liverpool Science Park, 146 Brownlow Hill, Liverpool L3 5RF, UK}
\email{D.A.Perley@ljmu.ac.uk}

\author[0000-0002-3181-8652, gname=Frédérick, sname=Poidevin]{Frédérick Poidevin}
\affiliation{Instituto de Astrof\'{\i}sica de Canarias, V\'{\i}a L\'actea, 38205 La Laguna, Tenerife, Spain}
\affiliation{Universidad de La Laguna, Departamento de Astrof\'{\i}sica, 38206 La Laguna, Tenerife, Spain}
\email{fpoidevin@iac.es}

\author[0009-0007-0178-2875, gname=Ramon, sname=Sanfeliu]{Ramon Sanfeliu}
\affiliation{Institut d'Estudis Espacials de Catalunya (IEEC), 08860 Castelldefels (Barcelona), Spain}
\affiliation{Institute of Space Sciences (ICE, CSIC), Campus UAB, Carrer de Can Magrans, s/n, 08193 Barcelona, Spain}
\email{rsanfeliu@ice.csic.es}

\author[0000-0003-0733-2916, gname=Jacob L., sname=Wise]{Jacob~L. Wise}
\affiliation{Astrophysics Research Institute, Liverpool John Moores University, Liverpool Science Park, 146 Brownlow Hill, Liverpool L3 5RF, UK}
\email{j.l.wise@2022.ljmu.ac.uk}

\author[0000-0001-8985-2493, gname=Erez A., sname=Zimmerman]{Erez~A. Zimmerman}
\affiliation{Department of Particle Physics and Astrophysics, Weizmann Institute of Science, 234 Herzl St, 76100 Rehovot, Israel}
\email{erez.zimmerman@weizmann.ac.il}

\author[0009-0001-0983-623X, gname=Alice, sname=Cai]{Alice Cai}
\affiliation{Center for Interdisciplinary Exploration and Research in Astrophysics (CIERA), Northwestern University, 1800 Sherman Ave., Evanston, IL 60201, USA}
\affiliation{Department of Physics and Astronomy, Northwestern University, 2145 Sheridan Road, Evanston, IL 60208, USA}
\email{acai@u.northwestern.edu}

\author[0000-0002-0987-3372, gname=Joahan, sname=Castaneda Jaimes]{Joahan Castaneda Jaimes}
\affiliation{IPAC, California Institute of Technology, 1200 E. California Blvd, Pasadena, CA 91125, USA}
\email{jocastan@ipac.caltech.edu}

\author[0000-0002-1066-6098, gname=Ting-Wan, sname=Chen]{Ting-Wan Chen}
\affiliation{Graduate Institute of Astronomy, National Central University, 300 Jhongda Road, 32001 Jhongli, Taiwan, R. O. C.}
\email{twchen@astro.ncu.edu.tw}

\author[orcid=0000-0002-8262-2924, gname=Michael W., sname=Coughlin]{Michael~W. Coughlin}
\affiliation{School of Physics and Astronomy, University of Minnesota, Minneapolis, MN 55414, USA}
\email{cough052@umn.edu}

\author[0000-0002-5884-7867, gname=Richard, sname=Dekany]{Richard Dekany}
\affiliation{The Caltech Optical Observatories, California Institute of Technology, Pasadena, CA 91125, USA}
\email{rgd@astro.caltech.edu}

\author[0000-0003-2238-1572, gname=Ori~D., sname=Fox]{Ori~D. Fox}
\affiliation{Space Telescope Science Institute, 3700 San Martin Drive, Baltimore, MD 21218, USA}
\email{ofox@Space Telescope Science Institute, 3700 San Martin Drive, Baltimore, MD 21218, USA.edu}

\author[0000-0002-4223-103X]{Christoffer Fremling}
\affiliation{The Caltech Optical Observatories, California Institute of Technology, Pasadena, CA 91125, USA}
\affiliation{Division of Physics, Mathematics and Astronomy, California Institute of Technology, Pasadena, CA 91125, USA}
\email{fremling@caltech.edu}

\author[0000-0002-3168-0139, gname=Matthew~J., sname=Graham]{Matthew~J. Graham}
\affiliation{Division of Physics, Mathematics and Astronomy, California Institute of Technology, Pasadena, CA 91125, USA}
\email{mjg@caltech.edu}

\author[0000-0001-5668-3507, gname=Steven L., sname=Groom]{Steven~L. Groom}
\affiliation{IPAC, California Institute of Technology, 1200 E. California Blvd, Pasadena, CA 91125, USA}
\email{sgroom@ipac.caltech.edu}

\author[0000-0002-5619-4938, gname=Mansi M., sname=Kasliwal]{Mansi~M. Kasliwal}
\affiliation{Division of Physics, Mathematics and Astronomy, California Institute of Technology, Pasadena, CA 91125, USA}
\email{mansi@astro.caltech.edu}

\author[0000-0001-8594-8666, gname=Marek, sname=Kowalski]{Marek Kowalski}
\affiliation{Deutsches Elektronen-Synchrotron DESY, Platanenallee 6, D-15738 Zeuthen, Germany}
\affiliation{Institut f\"ur Physik, Humboldt-Universit\"at zu Berlin, Newtonstr. 15, 12489 Berlin, Germany}
\email{marek.kowalski@desy.de}

\author[0000-0003-2451-5482, gname=Russ E., sname=Laher]{Russ~R. Laher}
\affiliation{IPAC, California Institute of Technology, 1200 E. California Blvd, Pasadena, CA 91125, USA}
\email{laher@ipac.caltech.edu}

\author[0009-0002-9885-3388, gname=Paraskevas, sname=Lampropoulos]{Paraskevas Lampropoulos}
\affiliation{Gravitation and Astroparticle Physics Amsterdam (GRAPPA), University of Amsterdam, 1098 XH Amsterdam, The Netherlands}
\email{paris.lampropoulos@student.uva.nl}

\author[0000-0003-2593-4355, gname=Elisabetta, sname=Maiorano]{Elisabetta Maiorano}
\affiliation{INAF - Osservatorio di Astrofisica e Scienza dello Spazio, via Piero Gobetti 93/3, 40129 Bologna, Italy}
\email{elisabetta.maiorano@inaf.it}

\author[0000-0002-6894-1267, gname=Ester, sname=Marini]{Ester Marini}
\affiliation{INAF - Osservatorio Astronomico di Roma, via Frascati 33, I-00040 Monteporzio Catone, Italy}
\email{ester.marini@inaf.it}

\author[0000-0001-8342-6274, gname=Jakob, sname=Nordin]{Jakob Nordin}
\affiliation{Institut f\"ur Physik, Humboldt-Universit\"at zu Berlin, Newtonstr. 15, 12489 Berlin, Germany}
\email{jnordin@physik.hu-berlin.de}

\author[0000-0003-1227-3738, gname=Josiah, sname=Purdum]{Josiah Purdum}
\affiliation{The Caltech Optical Observatories, California Institute of Technology, Pasadena, CA 91125, USA}
\email{jpurdum@caltech.edu}

\author[0000-0002-0387-370X, gname=Reed, sname=Riddle]{Reed Riddle}
\affiliation{The Caltech Optical Observatories, California Institute of Technology, Pasadena, CA 91125, USA}
\email{riddle@caltech.edu}

\author[]{Niharika Sravan}
\affiliation{Department of Physics, Drexel University, Philadelphia, PA 19104, USA}
\email{niharika.sravan@gmail.com}

\author[0009-0005-8230-030X, gname=Aswin, sname=Suresh]{Aswin Suresh}
\affiliation{Center for Interdisciplinary Exploration and Research in Astrophysics (CIERA), Northwestern University, 1800 Sherman Ave., Evanston, IL 60201, USA}
\affiliation{Department of Physics and Astronomy, Northwestern University, 2145 Sheridan Road, Evanston, IL 60208, USA}
\email{aswinsuresh2029@u.northwestern.edu}

%% Mark off the abstract in the ``abstract'' environment. 
\begin{abstract}
Supernova (SN) 1987A showed that some hydrogen-rich core-collapse supernovae (SNe) rise slowly to peak, defining a distinct subclass. Wide-field surveys have uncovered more extreme members, such as OGLE-2014-SN-73, SN\,2020faa, and SN\,2023vbw, that challenge neutrino-driven explosion models. We present ultraviolet-to-infrared, radio, X-ray, and spectropolarimetric observations of \sn at $z = 0.1020$, the most extreme 87A-like event to date. It rose for $\gtrsim88$ rest-frame days to $M_r \approx -20.1$~mag, also placing it among the hydrogen-rich superluminous SNe, and radiated $\gtrsim4\times10^{50}$~erg. Its spectra resemble those of typical Type II SNe and remain photospheric at $\sim330$ days post-maximum. Continuum polarization of 0.8--1.0\% indicates an aspherical system. Hydrogen-rich pair-instability supernova (PISN) models of a $190~M_\odot$ star reproduce the light-curve up to 150 days after peak, but not the late-time emission. A late-time spectrum shows a feature near 7300~\AA, tentatively identified as [\ion{Ni}{2}]\,$\lambda$7378. The implied $\approx0.02~M_\odot$ of stable $^{58}$Ni is reached only in the most massive PISN models. The low-mass host galaxy ($\sim10^{8.2}~M_\odot$) is consistent with a metal-poor progenitor as required by the PISN channel. Fallback accretion reproduces the light-curve morphology, but not the colors. We consider a pair-instability explosion the most plausible scenario, with the late-time light-curve excess powered by circumstellar interaction, which shows no signatures in the spectra. \sn shows that the most luminous 87A-like events occupy an underexplored region of massive-star evolution.
\end{abstract}

%% Keywords should appear after the \end{abstract} command. 
%% The AAS Journals now uses Unified Astronomy Thesaurus (UAT) concepts:
%% https://astrothesaurus.org
%% You will be asked to selected these concepts during the submission process
%% but this old "keyword" functionality is maintained in case authors want
%% to include these concepts in their preprints.
%%
%% You can use the \uat command to link your UAT concepts back its source.
\keywords{
\uat{Type~II supernovae}{1731},
\uat{Core-collapse supernovae}{304}
}

%% From the front matter, we move on to the body of the paper.
%% Sections~are demarcated by \Section~and \subsection, respectively.
%% Observe the use of the LaTeX \label
%% command after the \subSection~to give a symbolic KEY to the
%% subSection~for cross-referencing in a \ref command.
%% You can use LaTeX's \ref and \label commands to keep track of
%% cross-references to sections, equations, tables, and figures.
%% That way, if you change the order of any elements, LaTeX will
%% automatically renumber them.

\section{Introduction} 
The observational landscape of hydrogen-rich core-collapse Type II supernovae \citep[SNe II;][]{Minkowski1941a, Arcavi2017a} was fundamentally altered by the discovery of SN\,1987A. At the time, stellar-evolution calculations and models of SN~II light curves generally favored extended red-supergiant progenitors for H-rich explosions \citep[e.g.,][]{Falk1977a,Arnett1980a}. However, SN\,1987A was traced to the compact blue supergiant Sanduleak $-69^{\circ}202$ \citep{Arnett1989a,Podsiadlowski1999a}, demonstrating that core collapse can also occur in a markedly different evolutionary state.

This event defined a distinct subclass of 87A-like SNe, characterized by a long, slow rise to a typically faint photometric peak (at $\sim 100$ days) and a broad light-curve morphology that differs markedly from that of more common SNe~II \citep{Arcavi2017a,Zampieri2017a,Pumo2023a,Sit2023a}, but spectroscopically they are similar to normal SNe~II \citep{Filippenko1997a, Gal-Yam2017a}. These properties are commonly associated with compact progenitors ($R\lesssim100\,R_\odot$), although the observed light-curve evolution also depends on the ejecta mass, explosion energy, and source of luminosity \citep{Nakar2016a,Dessart2019b}.

In the classical interpretation, a substantial fraction of the shock-deposited energy is lost to adiabatic expansion within the dense envelope, causing the early-time luminosity to remain low until energy diffusing from deeper layers becomes visible \citep{Nakar2016a,Zampieri2017a,Pumo2023a,Pumo2025a}. While 87A-like SNe remain relatively rare \citep[$\sim 1.4\times10^{-6}~{\rm Mpc^{-3}\,yr^{-1}}$, corresponding to $\sim 1.4\%$ of the total core-collapse SN rate;][]{Sit2023a}, $\sim 30$ such events have now been identified through systematic surveys \citep{Taddia2013a,Sit2023a}. This growing population provides a valuable laboratory for understanding the evolutionary pathways of massive stars, progenitor structure and mixing, and the diverse outcomes of H-rich explosions \citep{Zampieri2017a}.

Recent years have seen the discovery of extreme transients that share the defining photometric characteristics of the 87A-like subclass while extending its properties to much higher luminosities and longer characteristic timescales. Three well-studied examples of this extreme population are OGLE-2014-SN-073 \citep[hereafter called OGLE-73;][]{Terreran2017a}, SN\,2020faa \citep{Yang2021a, Salmaso2023a}, and SN\,2023vbw \citep{Hiramatsu2026a}. OGLE-73 displayed an unusually broad light curve and a high peak luminosity, suggesting a remarkably large ejecta mass and explosion energy \citep{Terreran2017a}. Similarly, SNe 2020faa and 2023vbw exhibited sustained high luminosity that challenged conventional radioactive-decay models, implying that an additional, more potent energy source was at play \citep{Yang2021a, Salmaso2023a, Hiramatsu2026a}. Whether the three extreme events represent the high-luminosity tail of the classical 87A-like population or a physically distinct class of explosions remains unclear.

These extraordinary objects have necessitated the exploration of powering mechanisms beyond standard explosion mechanisms and radioactive decay \citep{Terreran2017a, Kozyreva2018a, Orellana2018a, Moriya2018c, Dessart2019b, Salmaso2023a}. One candidate is the pair-instability supernova (PISN) mechanism, thought to occur in stars with helium core masses between $\sim 65$ and $130 \, M_{\odot}$, where the production of electron-positron pairs leads to thermonuclear disruption and $^{56}\mathrm{Ni}$ yields of up to several $10~M_\odot$ \citep{Fowler1964a, Barkat1967a, Rakavy1967a, Kozyreva2018a}. Alternatively, a magnetar central engine can inject rotational energy into the ejecta via magnetic multipole radiation \citep{Kasen2010a, Dessart2018a, Dessart2019b, Terreran2017a, Orellana2018a, Omand2024a}. Fallback accretion provides another potential engine, where gravitational potential energy is released as material accretes onto the central compact remnant following a failed or partial shock revival \citep{Moriya2018c, Terreran2017a}. Finally, ``hidden'' circumstellar-matter (CSM) interaction might also boost luminosity if the interaction occurs within an asymmetric disk or torus embedded beneath the optically thick photosphere. In this scenario, the CSM interaction zone is within the optically thick photosphere and is not directly observed, masking the characteristic narrow emission lines of Type~IIn SNe \citep{Smith2017a, Nyholm2020a, Salmaso2023a}.

It is currently unclear which combinations of progenitor radii, ejecta masses, and explosion energies are required to produce the high luminosities observed in the extreme 87A-like events. Furthermore, distinguishing observationally between PISN, magnetar-powered, fallback-powered, and hidden-CSM-interaction scenarios remains a primary challenge owing to the lack of unique spectral signatures in the photospheric phase and the degeneracy of light-curve models. Resolving these questions is vital for mapping the upper end of the core-collapse mass distribution and the final fates of the most massive stars.

Here, we present \sn, the most extreme 87A-like event observed to date. It exhibits a light-curve morphology that mirrors the SN\,1987A prototype, but on an unprecedented scale. Its light curve is broader and $\sim 0.9$ mag more luminous at peak than that of the previous record holder, OGLE-73. When modeled using radioactive decay, the required radioactive $^{56}\mathrm{Ni}$ mass exceeds what is achievable in neutrino-driven explosion models, necessitating either a PISN explosion or the presence of an additional energy source (e.g., energy injection from a compact stellar object, and CSM interaction). \sn therefore provides a useful test of the models that have been proposed for the most luminous 87A-like events.

This paper is structured as follows. We report the discovery of \sn in Section~\ref{sec:discovery} and describe the observations in Section~\ref{sec:obs}. In Section~\ref{sec:results}, we derive the photometric, spectroscopic, spectropolarimetric, and host-galaxy properties of \sn. Section~\ref{sec:discussion} constrains the ejecta composition, evaluates different powering mechanisms, compares \sn with other 87A-like SNe, and outlines a multiwavelength strategy for identifying the power source of such extreme events. We summarize our conclusions in Section~\ref{sec:conclusion}.

Throughout the paper, we provide all uncertainties at the $1\sigma$ level. The photometry is reported in the AB system \citep{Oke1983a}. Phase information is reported in the rest frame with respect to the time of $r$-band maximum brightness at $\rm MJD=60169.63$. UTC dates are used throughout.

\section{Discovery}\label{sec:discovery}

\begin{figure}
    \centering
    \includegraphics[width=1\columnwidth]{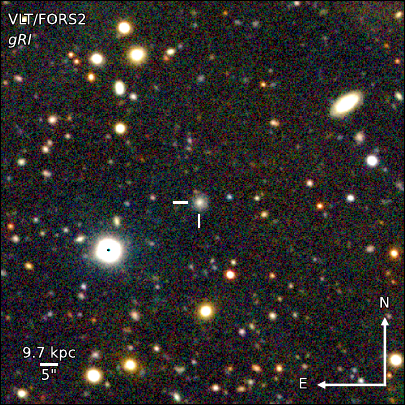}
    \caption{False-color image built from VLT FORS2 data acquired in June 2025 using the software package \program{STIFF} \citep{Bertin2012a}. The crosshair marks the host galaxy ($r\approx21.7$~mag; Table~\ref{tab:host_phot}). The SN location is 1.3 kpc ($\sim 0\farcs7$) northwest of the galaxy center. Fitting the observed galaxy spectral energy distribution (SED) from the $g$ to the $K$ bands with \program{Prospector} reveals that the host is a regular low-luminosity ($M_r\approx-16.9$ mag), low-mass ($\sim 10^{8.2}~M_\odot$) galaxy with a star-formation rate of $\sim 0.01~M_\odot\rm\,yr^{-1}$ (Section~\ref{sec:host_galaxy}).
    }
    \label{fig:host:rgb}
\end{figure}

\sn, located at $\alpha$, $\delta$ (J2000) = $00^{\rm hr}14^{\rm m}42\fs699$, $+21^\circ51'33\farcs77$, was detected in the transient alert stream of the Zwicky Transient Facility \citep[ZTF;][]{Bellm2019a, Graham2019a} on 11 May 2023 by \citet[][]{Fremling2023a} (internal name: ZTF23aajucmw). Other surveys also reported detections: the Asteroid Terrestrial-impact Last Alert System \citep[ATLAS;][]{Tonry2018b} survey on 29 May 2023 (internal name: ATLAS23lqe) and the Panoramic Survey Telescope and Rapid Response System (Pan-STARRS) Survey for Transients \citep{Huber2015a} on 23 August 2023 (internal name: PS23hhw). A false-color image of the field, built from our observations with the ESO Very Large Telescope, is shown in Figure~\ref{fig:host:rgb}.

\citet{Sollerman2023a} classified \sn as a Type~II SN at redshift $z=0.095$ on 26 July 2023 using the SED Machine \citep{Blagorodnova2018a}, based on a comparison with spectra of SNe II. Late-time spectra that are presented in this paper reveal galaxy emission lines at $z=0.1020$, which we adopt as the SN redshift (Section~\ref{sec:redshift_host_attenuation}). This converts the peak apparent magnitude of $r=18.33$ to a K-corrected absolute magnitude of $M_r=-20.14$ (Table~\ref{tab:lc:properties}), placing \sn in the regime of H-rich superluminous supernovae \citep[SLSNe\footnote{\citet{Pessi2025a} defined an H-rich SLSN as an object that is brighter than $-19.9$~mag in any optical filter.};][]{GalYam2012a,Inserra2018b,Gal-Yam2019a,Kangas2022a, Pessi2025a}. 

\section{Observations and data reduction}\label{sec:obs}

Our follow-up campaign of \sn combines optical, near-infrared (NIR), ultraviolet (UV), radio, and X-ray observations obtained between discovery and more than 600 rest-frame days after maximum light. A summary of all observations and their data reduction is provided in Section~\ref{appendix:observations} of the Appendix. Table \ref{tab:SN2023iop:photometry} in the Appendix summarizes the SN photometry.

\section{Results}\label{sec:results}

\begin{figure*}
    \centering
    \includegraphics[width=1\linewidth]{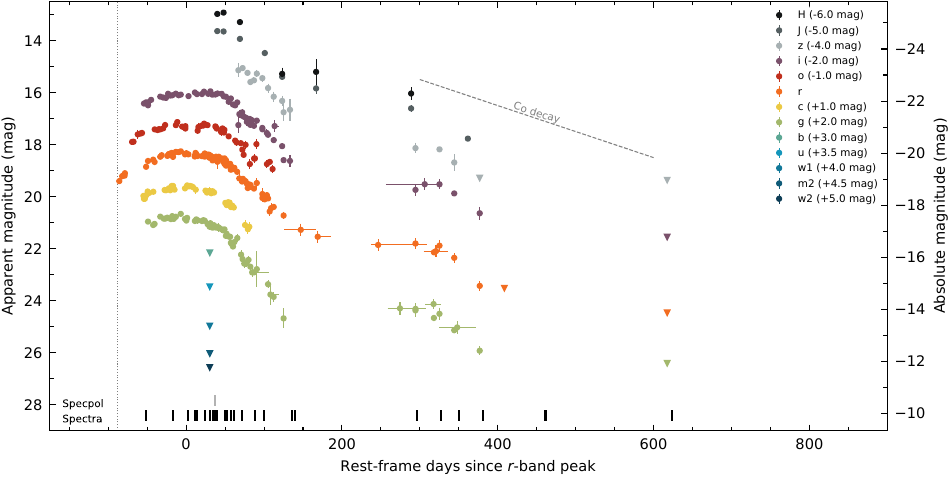}
    \caption{\sn's light curve from the UV to the NIR, corrected for Milky Way extinction. At the bottom, vertical bars indicate the epochs of spectroscopy and spectropolarimetry. To guide the eye, the $^{56}$Co decay slope is displayed. The absolute magnitude is computed with $M=m - {\rm DM}(z) + 2.5\,\log \left(1+z\right)$, where DM is the distance modulus and $z$ the redshift. Upper limits are indicated by downward-pointing triangles.}
    \label{fig:2023iop:lc}
\end{figure*}

\subsection{Redshift and Host-galaxy Attenuation}\label{sec:redshift_host_attenuation}

The late-time spectra obtained between days +380.6 and +623.8 show narrow emission lines from \ion{H}{2} regions in the host galaxy. The day~+623.8 spectrum has the highest signal-to-noise ratio (S/N) and no longer contains any contribution from \sn. From the H$\alpha$, H$\beta$, and [\ion{O}{3}]\,$\lambda$5007 lines in this spectrum, we infer $z = 0.1020 \pm 0.0001$. The redshifts measured from the other epochs agree with this value within $1\sigma$. Assuming a $\Lambda$CDM cosmological model with H$_0=67.8~{\rm km\,s}^{-1}\,{\rm Mpc}^{-1}$, $\Omega_{\rm M}=0.308$, and $\Omega_\Lambda=0.692$ \citep{Planck2016a}, this redshift corresponds to a luminosity distance of 485~Mpc and a distance modulus of 38.4~mag.

The X-shooter spectra obtained on days~+71.7 and +99.9 (Appendix, Table~\ref{tab:spectroscopy}) have a nominal resolution of $R \equiv \lambda/\Delta\lambda \approx 8900$ in the optical (VIS arm, observer frame; \citealt{Vernet2011a}). Narrow \ion{Na}{1} doublet absorption lines from the host-galaxy interstellar medium, a commonly used proxy for line-of-sight extinction \citep{Poznanski2012a}, are not detected in either spectrum. We normalize the spectra around the doublet and measure rest-frame equivalent widths (EWs) for each component: $0.08\pm0.10$~\AA\ and $0.00\pm0.11$~\AA\ in the first epoch, and $-0.15\pm0.09$~\AA\ and $0.03\pm0.09$~\AA\ in the second. Summing the two lines and averaging over both epochs gives a total \ion{Na}{1} EW of $-0.03\pm0.10$~\AA, consistent with a nondetection. The \citet{Poznanski2012a} relation converts this into a $3\sigma$ upper limit of $E(B-V)_{\rm host} < 0.03$~mag. Host attenuation along the line of sight to \sn is therefore negligible.

\subsection{Light-curve Parameters}\label{sec:lc:properties}

Figure~\ref{fig:2023iop:lc} shows \sn's multiband light curve from the \textit{w2} to the \textit{H} band, covering $\sim 700$ rest-frame days of its evolution. \sn was detected in the first image acquired by the private ZTF twilight survey after the field emerged from behind the Sun. \sn took 87.6 rest-frame days\footnote{Measured from the ZTF $r$-band data before applying dynamic rebinning.} to reach its maximum brightness of $r=18.33$~mag. This corresponds to an absolute magnitude of $-20.14$~mag after applying a K-correction \citep{Hogg2002a} derived from the NOT/ALFOSC spectrum obtained 13.6 days after peak brightness (Appendix, Table~\ref{tab:spectroscopy}). After peak, \sn's brightness declined gradually for $\sim 50$ days, then faded rapidly at $\sim 2.7$~mag~(100~day)$^{-1}$ over the following 100 days. This rapid decline is visible in all bands, but it is more pronounced in the $g$ band than in the $i$ band. The rapid decline ended around day~$+150$, after which the $r$-band brightness decreased by only $0.2$~mag~(100~day)$^{-1}$ until it began fading again at $\sim3$~mag~(100~day)$^{-1}$ from day~+325.

\begin{table*}
\caption{Light curve properties}
\label{tab:lc:properties}
\hspace{-0.75cm}
\begin{tabular}{cccccc}
\hline\hline
Property                      & ZTF/$g$             & ATLAS/$c$& ZTF/$r$ & ATLAS/$o$ & ZTF/$i$ \\
\hline
Peak time (MJD)               &$ 60155.70 \pm 0.86 $&$60163.85\pm1.23 $&$60169.63\pm0.54 $&$ 60172.85\pm1.47$&$60180.46\pm0.93 $\\
$m_{\rm peak}$ (mag)          & $18.81\pm0.02$& $18.62\pm0.02$& $18.33\pm0.01$& $18.24\pm0.03$& $18.14\pm0.02$\\
$M_{\rm peak}$ (mag)          & $-19.79$& $-19.86$& $-20.14$& $-20.10$& $-19.98$\\
$t_{1/2,\rm rise}$ (day)      & $>36†.2      $&$>49.4     $&$75.1^{+1.3}_{-1.1} $&$ >72.1 $& $>63.1$\\
$t_{1/e,\rm rise}$ (day)      & $>36.2      $&$>49.4     $&$>85.8              $&$ >72.1 $& $>63.1$\\
$t_{1/10,\rm rise}$ (day)   & $>36.2      $&$>49.4     $&$>85.8              $&$ >72.1 $& $>63.1$\\
$t_{1/2,\rm decline}$ (day)   & $66.1\pm0.5 $&$63.5\pm1.0$&$66.4\pm0.5         $&$ 65.4\pm1.3  $& $65.6\pm0.8$\\
$t_{1/e,\rm decline}$ (day)   & $75.6\pm0.6 $&$74.5\pm1.5$&$77.2\pm0.7         $&$ 78.4\pm2.0 $& $78.7\pm1.4$\\
$t_{1/10,\rm decline}$ (day)  & $110.1\pm1.5$&$>86.5     $&$131.7^{+3.8}_{-3.1}$&$ >108.2  $& $123.2^{+3.0}_{-2.1}$\\
$X-r~({\rm rest})$ (mag)      & 0.44 & 0.28 & --- & $-0.14$ & 0.18 \\
\hline
\end{tabular}
\tablecomments{All magnitudes are corrected for Galactic extinction. The absolute magnitudes include a $k$-correction measured from the spectrum at day~$+13.6$. All time scales are reported in the rest-frame. In the final row, $X$ refers to the filter of the corresponding column; the reported value is the rest-frame colour $X-r$ measured from the spectrum at day~$+13.6$. All uncertainties reflect the $1\sigma$ statistical errors.
}
\end{table*}

To put \sn in the context of other SNe~II and H-rich SLSNe, we measure light-curve parameters from the multiband $cgroi$ photometry. The apparent magnitude in each band was first interpolated using Gaussian Processes \citep[GP;][]{Rasmussen2006a} with a Mat\'{e}rn-5/2 kernel. The peak epochs and magnitudes were then determined by fitting a weighted second-degree polynomial within a $\pm75$\,day window centered on the GP-derived maximum, with uncertainties obtained from 5000 Monte Carlo draws from the curve-fitting covariance matrix. The peak time depends strongly on wavelength: the light curve peaks first in the $g$ band, at MJD~$60155.70\pm0.86$, and 22.5 rest-frame days later in the $i$ band (Table~\ref{tab:lc:properties}). Throughout the paper, we use the peak time in the $r$ band as a reference point ($\rm MJD = 60169.63$; Table~\ref{tab:lc:properties}). Rest-frame decline timescales, defined as the elapsed time for the flux to fall to a given fraction of its peak value and measured from the GP light curves, are listed in Table~\ref{tab:lc:properties}.

During the decline of the early-time light curve, \sn's colors trace the ejecta's thermal evolution. Up to 150 days after peak brightness, the color reddens steadily. The $g-r$ color increases from $\sim 0.5$~mag near maximum to $\sim 1.5$--2.0~mag roughly by day~+120, tracking the cooling and recombination of the hydrogen envelope. On day~+30, \sn's $u-g\gtrsim1.4$~mag and $u-r\gtrsim2.0$~mag are already far redder than any object in the broad-H$\alpha$ SLSN~II sample observed by ZTF at comparable phase \citep[$u-g$ spans only $0.3$--$1.2$~mag and $u-r$ spans $0.3$--$1.2$~mag;][]{Kangas2022a}. \sn's day~+30 $u-r$ color is instead comparable to the $u-r>2$~mag reached by ordinary SNe~II at a similar phase \citep{deJaeger2018a, Zimmerman2024a, Singh2024a}. Once \sn reaches the plateau, the color settles to $g-r\approx0.6$~mag and remains constant to within $0.1$~mag between days~+250 and +400. The optical-to-near-IR colors ($r-J$, $i-J$, and $z-J$) likewise show no significant evolution over the same interval, including across the onset of the rapid decline after day~+325, though the large uncertainties and sparse NIR sampling preclude a more nuanced conclusion.

The $r$-band light curve spans the longest time interval, allowing accurate measurements of the rise and decline timescales. The $e$-folding rise and decline timescales are $>85.8$ and $77.2\pm0.7$ rest-frame days, respectively (additional timescales, including band-dependent values, are listed in Table~\ref{tab:lc:properties}). Two properties stand out in the context of H-rich SLSNe. The lower limit on the $r$-band rest-frame rise time firmly places \sn among the slowest-rising H-rich SLSNe with broad H$\alpha$ emission lines \citep{Inserra2018a, Kangas2022a, Pessi2025a}. However, \sn's combination of a very slow rise and a comparatively fast decline breaks the positive rise-decline correlation seen in H-rich SLSNe \citep{Pessi2025a}. This light-curve shape is also markedly different from that of a regular SN~II. Instead, \sn's light-curve morphology resembles that of SN\,1987A, though it is significantly broader and more luminous than even the most extreme 87A-like objects, OGLE-73 and SN\,2020faa. In Sections~\ref{section:discussion:SLSN-II} and \ref{section:discussion:87A}, we compare in detail \sn's light-curve properties with those of SLSNe-II and 87A-like objects.

\subsection{Bolometric Light Curve}\label{sec:lc:bolometric}

\begin{figure}
    \centering
    \includegraphics[width=1\columnwidth]{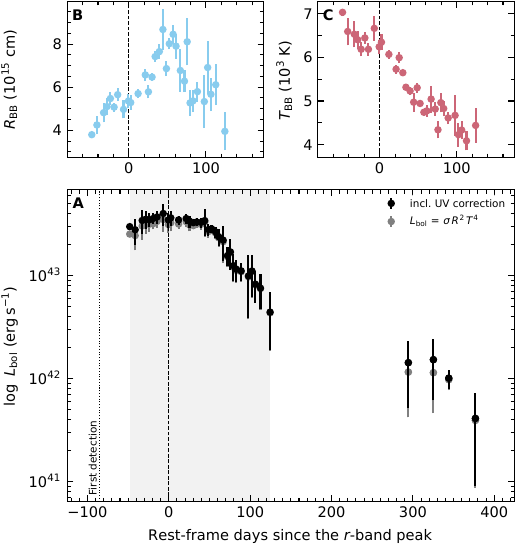}
    \caption{Bolometric light curve (panel A), and evolution of the blackbody photosphere's radius (panel B) and temperature (panel C). The gray-shaded region indicates the time interval where the blackbody radius and temperature could be measured.
    }
    \label{fig:2023iop:bolometrics}
\end{figure}

We constructed the bolometric light curve of SN\,2023iop (shown in Panel A of Figure \ref{fig:2023iop:bolometrics}) with the \textsc{estimate\_bolometric\_luminosity} routine in \textsc{redback} \citep{Sarin2024a} version 1.15.1, fitting the extinction-corrected photometry in 5-day observer-frame bins (each with $\geq3$ measurements) to a single-temperature blackbody parameterized by $\log~T_{\rm BB}$ and $\log~R_{\rm BB}$, with model magnitudes integrated over the filter transmission curves rather than effective wavelengths. The luminosity of each epoch then follows from the Stefan-Boltzmann law, $L_{\rm bol} = 4\pi \sigma R_{\rm BB}^2 T_{\rm BB}^4$.

This procedure relies on several simplifying assumptions. First, the spectral energy distribution (SED) is a pure blackbody at all wavelengths. Second, $L_{\rm bol}$ includes flux at wavelengths that our data do not cover. Our $griz$ coverage spans only 4700--8960~\AA, with the bluest constraint at the $g$ band in 34 of 37 epochs, so both the UV and the IR contributions are pure extrapolation, and the majority of the reported $L_{\rm bol}$ therefore rests on extrapolation rather than on measured flux. As the ejecta cool, the unobserved fraction migrates from the UV to the IR, so the infrared dominates the correction at late times. Although we have detections in the IR, they are too sparse to constrain this extrapolation directly. Third, we apply no line-blanketing correction; because line blanketing suppresses the emergent blue flux relative to a blackbody, the extrapolation likely overestimates the true UV contribution. Finally, the routine discards epochs with a fractional uncertainty on $T_{\rm BB}$, $R_{\rm BB}$, or $L_{\rm bol}$ exceeding unity, and six of the 37 surviving epochs rely on fewer than three filters, including one based on a single band, so their temperatures are weakly constrained.

At peak brightness, \sn reached a luminosity of $(4.0\pm1.0)\times10^{43}\,\rm erg\,s^{-1}$, and integrating over the entire light curve yields a radiated energy of $(4.3\pm0.3)\times10^{50}~\rm erg$. The radiated energy is a strict lower limit. If \sn's luminosity was (2--$3)\times10^{43}\,\rm erg\,s^{-1}$ between the first detection (day~$-87.6$) and the first epoch of the bolometric light curve (day~$-48$), the radiated energy increases by $\sim20\%$. The uncertain explosion date may further increase the radiated energy, probably bringing it closer to $\rm 10^{51}~erg$. While \sn has a low peak absolute magnitude for an H-rich SLSN \citep{Inserra2018a, Kangas2022a, Pessi2025a}, the lower limit on its radiated energy places it in the upper half of the observed distribution of H-rich SLSNe with broad Balmer emission lines \citep{Kangas2022a}.

Such an energy release is difficult to reconcile with a neutrino-driven explosion. The energy budget of a neutrino-driven explosion is set by the gravitational binding energy released when the neutron star forms, $\sim 3\times10^{53}~\rm erg$. Because the neutron-star mass distribution is narrow, this budget is fixed to within a factor of $\sim2$ \citep{Lattimer2001a, Oezel2016a, Janka2025a}. Only a small fraction of it reaches the ejecta. Current simulations of neutrino-driven explosions convert $\sim1\%$ of this budget into ejecta kinetic energy, or $\lesssim2\times10^{51}~\rm erg$ \citep{Bollig2021a, Burrows2021a, Janka2025a}. A similarly small fraction of that kinetic energy is radiated. Ordinary SNe~II emit only $\sim10^{49}~\rm erg$ \citep{Pejcha2015a, Martinez2022a}, about $1\%$ of their kinetic energy \citep{Kasen2009a, Martinez2022b}. 

Reproducing the $\gtrsim4\times10^{50}~\rm erg$ radiated by \sn would therefore require the combined neutrino-to-ejecta coupling and kinetic-to-radiation conversion efficiency to be $\sim50$ times larger than in normal SNe~II. Such an increase could not occur silently. A substantially larger explosion energy would manifest itself in higher ejecta velocities, and a more efficient conversion of kinetic energy into radiation would typically require strong ejecta-CSM interaction, which would generally imprint clear signatures on the spectra. Neither is observed: the photospheric velocities of \sn are typical of normal SNe~II (Section~\ref{sec:spectroscopy:line_evolution}), and its spectra show no evidence of strong interaction (Section~\ref{sec:spectroscopy:sequence}). A purely neutrino-driven origin therefore appears highly contrived. Instead, the energetics point to an additional energy source: weak or geometrically confined ejecta-CSM interaction, a contribution from a compact object (a rapidly spinning neutron star or an accreting black hole), or a rare explosion channel (e.g., a PISN). We discuss these scenarios in Section~\ref{section:discussion:lightcurve_modeling}.

Ongoing interaction between the ejecta and dusty CSM would be expected to produce a mid-infrared (MIR) excess from thermal dust emission. \sn is not detected in the MIR (Appendix, Table \ref{tab:SN2023iop:photometry:wise}). \textit{WISE} $W1$ and $W2$ imaging obtained at day~$\sim295$ yields $3\sigma$ upper limits of 41 and $128~\mu$Jy, respectively. Neither band shows an excess over the optical and NIR SED, arguing against strong ongoing CSM interaction at this epoch. The limits are shallow, however, and exclude only a luminous warm-dust component.

Panels 3.B and 3.C in Figure~\ref{fig:2023iop:bolometrics} show the evolution of the blackbody radius and temperature computed with \program{Redback}. The temperature evolution is well described by a linear decline of $1800 \pm 100~\rm K\,(100~day)^{-1}$, whereas the extent of the photosphere follows a piecewise linear function: (i) a linear expansion of $4300\pm200\kms$ up to $57\pm3$ days after peak, reaching almost $9\times10^{15}~\rm cm$, and (ii) a linear decline of $6500\pm1200\kms$ later. The expansion velocity is similar to that extracted from the absorption minima of \ion{Fe}{2}\,$\lambda$5018 (Section~\ref{sec:spectroscopy:line_evolution}). Although \sn's light curve is similar to that of 87A-like objects, the maximum blackbody radius of $9\times10^{15}~\rm cm$ is 15 times larger than that of SN\,2018hna (regular 87A-like; \citealt{Xiang2023a}) and 2--4 times larger than that of the extreme 87A-like SNe OGLE-73 and SN\,2020faa \citep{Yang2021a, Salmaso2023a}. We will discuss the light-curve properties in more detail in Section~\ref{section:discussion:87A}.

\subsection{Spectroscopy}\label{sec:spectroscopy}

\subsubsection{Spectral Template Matching}\label{sec:spectroscopy:template_matching}

\begin{figure}
\includegraphics[width=\columnwidth]{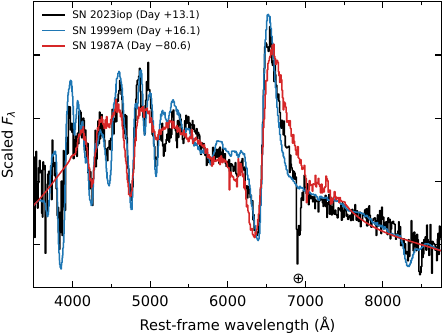}
\caption{Spectral template matching of the day~+13.6 spectrum of \sn\
with \program{SNID-SAGE}. The highest-ranked matches are
the prototypical 87A-like SN\,1987A and the normal Type~IIP SN\,1999em. Both templates reproduce the overall shape and the principal spectral features remarkably well,
despite their different photometric evolution. The location of the telluric A band is marked by ``$\earth$''.}
\label{fig:2023iop:spec:template_matching}
\end{figure}

Template matching of the NOT/ALFOSC spectrum obtained on day~$+13.6$ with SN~II templates using \program{NGSF} \citep{Goldwasser2022a} and \program{SNID-SAGE} \citep{Stoppa2026a} yields excellent matches to normal SNe~IIP, most notably SN\,1999em, but also to the prototypical 87A-like event SN\,1987A (Figure~\ref{fig:2023iop:spec:template_matching}). Repeating this analysis for all subsequent spectra yields similarly good matches with SNe~II throughout the entire spectroscopic sequence \citep[e.g., SNe 2004et, 2005cs, 2015bs, 2016hmq, 2018emt, 2018kpo;][]{Sahu2006a, Pastorello2006a, Anderson2018a}, including the final spectrum obtained on day~$+326$ (Figure~\ref{fig:2023iop:spec:template_matching}). Good matches with SN\,1987A are also recovered at most epochs (maximum light epoch shown in Figure~\ref{fig:2023iop:spec:template_matching}), although no satisfactory SN\,1987A match is found for the day~$+326$ spectrum. This close correspondence demonstrates that, despite its extreme luminosity and unusually broad light curve, \sn occupies a familiar spectroscopic regime and closely resembles common SNe~II during the photospheric phase.

\begin{figure*}[ht!]
    \centering  
    \includegraphics[width=1\textwidth]{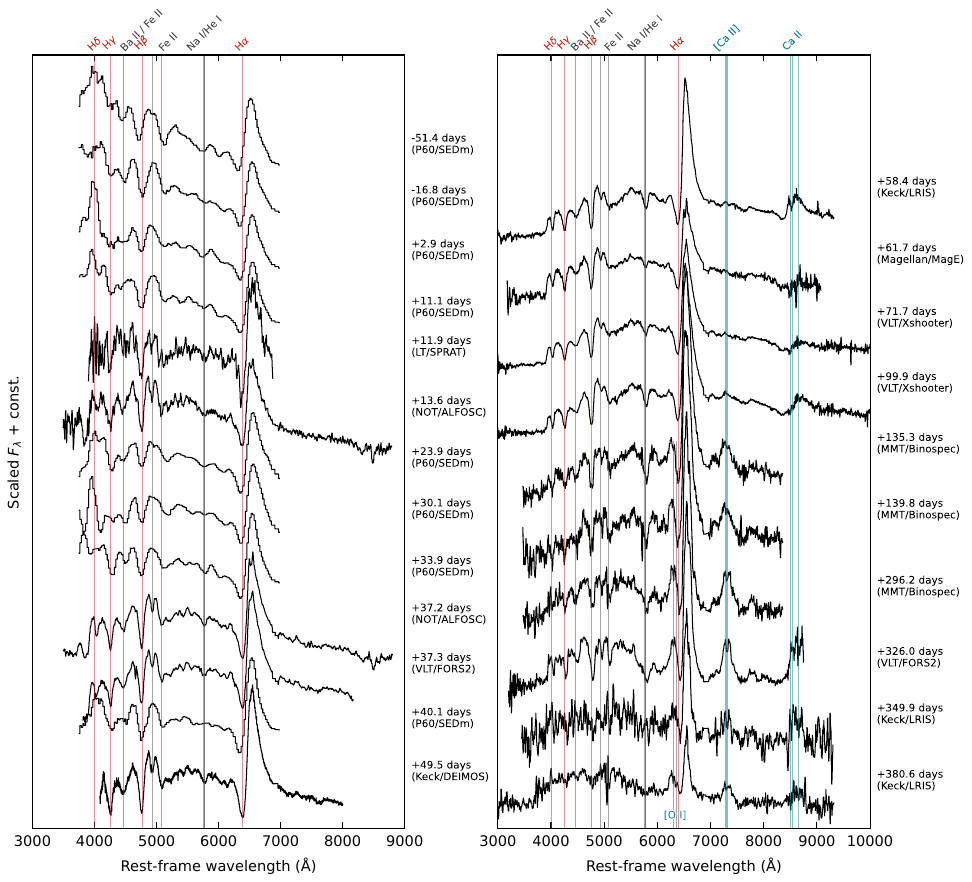}
    \caption{Spectroscopic sequence from 3000 to 10,000~\AA\ and from $-51.4$ to +380.6 rest-frame days after peak brightness (rebinned and smoothed with a Savitzky–Golay filter). The labels of all absorption features (red) are placed at the velocity coordinate $+6000\kms$ and emission features (blue) at zero velocity.
    }
    \label{fig:2023iop:spec:sequence_opt}
\end{figure*}

For spectra obtained up to $\sim 100$ days after maximum brightness, the phases of best-matching normal SNe~II are typically $\sim 16$ days after peak (range: 9--50 days), with no correlation between the template phase and the actual phase of \sn. At later epochs, the preferred matches are with SN~II spectra obtained at $\sim+74$ days. This behavior is noteworthy given that normal SNe~II generally reach maximum light within a few days of explosion and evolve substantially faster thereafter \citep{Rubin2016a,Gutierrez2017a, Martinez2022a, Hinds2025a, Ertini2026a}. Therefore, the discrepancy directly reflects \sn's extraordinarily slow spectroscopic evolution. A similar trend is found when restricting the comparison to SN\,1987A. In that case, the preferred matches correspond to spectra of SN\,1987A acquired $\sim 80$ days before maximum light, indicating the smaller extent of the hydrogen envelope in normal 87A-like SNe. 

\begin{figure*}[ht!]
    \centering
    \includegraphics[width=1\textwidth]{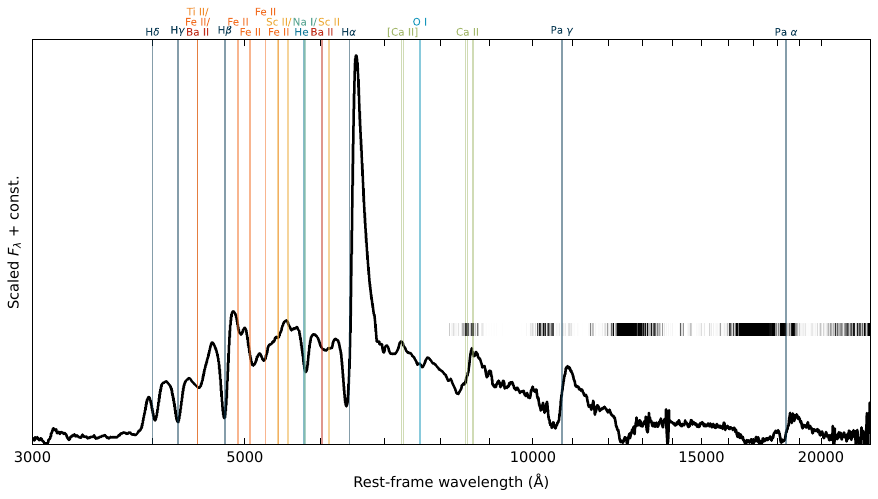}
    \caption{VLT/X-shooter spectrum on day~+71.7 and identification of the most prominent features. All features exhibit P-Cygni profiles, except for [\ion{Ca}{2}], which is pure emission. 
    The labels of the P-Cygni profiles are placed at a velocity of $6000\kms$ (approximately the ejecta velocity; Section~\ref{sec:spectroscopy:line_evolution}). 
    Some features 
    are formed at slower or faster portions of the ejecta (e.g., Balmer lines). 
    The line center of [\ion{Ca}{2}] is placed at zero velocity.
    The NIR suffers from strong atmospheric absorption; the opacity is indicated by the black vertical bars.
    }
    \label{fig:2023iop:spec:lineid}
\end{figure*}

\subsubsection{Spectroscopic Sequence}\label{sec:spectroscopy:sequence}

Figure~\ref{fig:2023iop:spec:sequence_opt} presents the optical spectral sequence for \sn, spanning $-51.4$ to $+380$ rest-frame days relative to peak light.\footnote{We omit displaying the epochs on days~+461.2 and +483.8 because they did not exhibit any SN features; they are discussed in Section~\ref{sec:redshift_host_attenuation}.} Already at the earliest epoch (day~$-51.4$), the spectrum exhibits well-developed P-Cygni profiles in the Balmer lines, while metal lines are present but comparatively weak. Over this extended interval, the spectra show a clear blue-to-red evolution driven by cooling, accompanied by a rich set of features that evolve very slowly and closely track \sn's gradual photometric evolution. Line profiles and relative strengths change only moderately over tens of days, indicating persistent photospheric conditions well beyond epochs at which most core-collapse SNe enter the nebular phase.

To facilitate line identification, we focus on the X-shooter spectrum from day~$+71.7$ in Figure~\ref{fig:2023iop:spec:lineid} and use \citet{Singh2019a} as a point of reference. At this phase, Balmer features remain prominent, while metal absorption lines from \ion{Fe}{2}, \ion{Ti}{2}, \ion{Sc}{2}, and \ion{Ba}{2} are relatively weak. Spectra obtained with Keck/LRIS, MagE, and X-shooter provide high sensitivity in the $u$ band. Essentially, they show no detectable flux blueward of 3500~\AA\ (rest frame), besides a minor contribution from the host galaxy. In addition, the Keck/DEIMOS, LRIS, MagE, and X-shooter data have high S/N and high spectral resolutions ranging from $R=1000$ to 8000. They reveal no evidence of narrow Balmer emission lines\footnote{The two NOT/ALFOSC spectra (days~$+13.6$ and $+37.2$) exhibit two sharp features around the peak of the H$\alpha$ P-Cygni profile. At \sn's redshift, the rest-frame H$\alpha$ line is observed at $\sim7234$~\AA, coinciding with the $\sim7200$~\AA\ H$_2$O telluric absorption complex. Under typical observing conditions at Observatorio del Roque de los Muchachos, this can result in measurable telluric absorption that could be misidentified as narrow, low-velocity absorption components. Observations of a telluric standard would have been needed to correct for this. The FORS2 observation on day~+326 was done in very good conditions (precipitable water vapor $\approx 2$~mm). That water vapor indeed affects this wavelength region is best seen in the high-resolution MagE and X-shooter spectra.} commonly associated with SNe~IIn. Both an excess flux below 3500~\AA\ or the presence of narrow Balmer emission would be indicative of ongoing CSM interaction \citep{Fransson2014a, Smith2017a}, but neither signature is observed. Together with the close resemblance to standard SNe~II (Section~\ref{sec:spectroscopy:template_matching}),  these observations show no direct spectroscopic evidence for ongoing CSM interaction during the observed epochs. This disfavors CSM interaction as a driver of \sn's high luminosity or unusually broad light curve, albeit the absorption-to-emission ratio of H$\alpha$ is unusually small (Section~\ref{sec:spectroscopy:line_evolution}).

We obtained two epochs of NIR spectroscopy on days~$+71.7$ and $+99.9$. No significant spectral evolution is observed between these epochs, and only a few weak features are detected. For this reason, we show only the day~$+71.7$ spectrum in Figure~\ref{fig:2023iop:spec:lineid}. The dominant transitions are Pa\,$\alpha$ and Pa\,$\gamma$, while Pa\,$\beta$ falls in a region strongly affected by atmospheric absorption.

On day~+99.9, the forbidden [\ion{Ca}{2}]\,$\lambda\lambda7291$, 7324  doublet becomes visible and subsequently strengthens. By day~+135.3, [\ion{O}{1}]\,$\lambda\lambda6300$, 6364 is also clearly detected. It may have been present at earlier phases, but blending with neighboring features (e.g., the P-Cygni profile of H$\alpha$) likely hinders its identification.

Intriguingly, the late-time spectra show the coexistence of nebular and photospheric signatures. Although forbidden emission from [\ion{O}{1}] and [\ion{Ca}{2}] is clearly established by day~+135, strong H$\alpha$ emission, H$\alpha$ absorption, and \ion{Fe}{2}\,$\lambda$5018 remain visible even in the spectrum at day~+326. This combination indicates that a substantial fraction of the ejecta remains optically thick while nebular emission becomes increasingly prominent, demonstrating that \sn evolves much more slowly toward the nebular phase than typical SNe~II.

\subsubsection{Line Evolution}\label{sec:spectroscopy:line_evolution}

Next, we zoom in on the evolution of the \ion{Fe}{2} and H$\alpha$ lines, which trace the bulk ejecta (\ion{Fe}{2}) and the fastest-moving H-rich material (H$\alpha$), respectively. Among the three strong \ion{Fe}{2} transitions between 4900 and 5200~\AA\ (rest frame), \ion{Fe}{2}\,$\lambda$5018 is the least blended and most straightforward to measure. To quantify the evolution of these lines, we normalize the spectra around each feature, apply a mild Savitzky-Golay smoothing filter, and measure the velocity-space location of the absorption minimum, the full width at half-maximum intensity (FWHM), and the EW (all in the rest frame). For H$\alpha$, we additionally measure the FWHM and EW of the emission component.

\begin{figure}
    \centering
        \includegraphics[width=1\columnwidth]{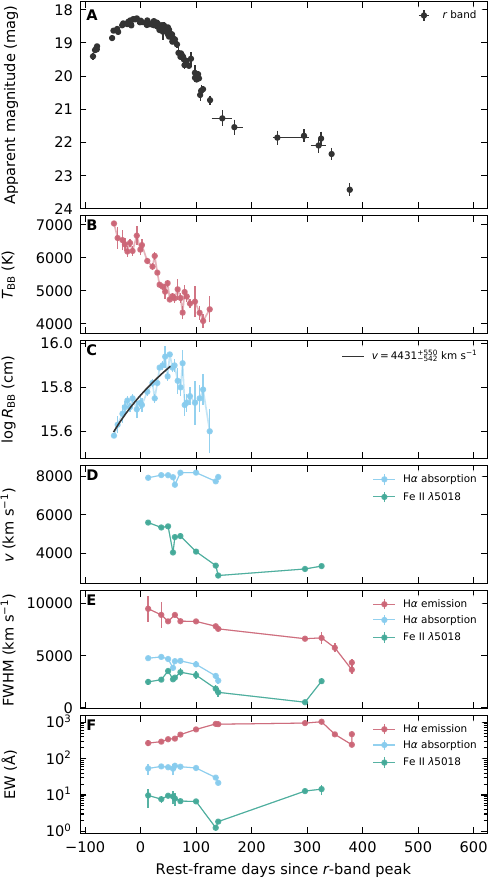}
    \caption{\sn's photometric and spectroscopic evolution versus rest-frame days since $r$-band peak. From top to bottom: the $r$-band light curve; the blackbody temperature and radius; the line center of the \ion{Fe}{2}\,$\lambda$5018 and H$\alpha$ P-Cygni absorption minima; the FWHM; and the EW of the H$\alpha$ emission and absorption components and of \ion{Fe}{2}\,$\lambda$5018. 
    }
    \label{fig:2023iop:spec:feature_evolution}
\end{figure}

Figure~\ref{fig:2023iop:spec:feature_evolution} places these measurements in the context of the light-curve evolution of \sn\ (panel 7.A), together with the evolution of the blackbody radius and temperature (panels 7.B and 7.C). The corresponding measurements are listed in Table~\ref{tab:line_params} of the Appendix. As discussed in Section~\ref{sec:lc:bolometric}, the photosphere of \sn\ cooled from $\sim 7000$~K on day~$-48$ to $\sim4500$~K by day~$+100$, while its radius expanded to $\sim 9\times10^{15}$~cm around day~$+57$ before receding into deeper layers of the ejecta, consistent with the behavior expected of a recombination-powered photosphere \citep{Arcavi2017a}.

\begin{figure*}
    \centering
    \includegraphics[width=1\textwidth]{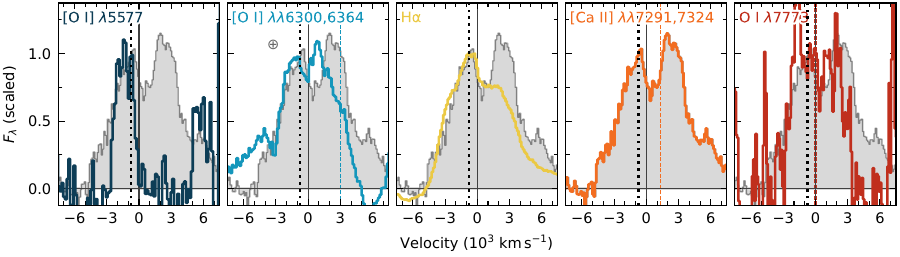}
    \caption{
    Comparison of the late-time line profiles in the day~+326 spectrum. The profiles are shown in velocity space relative to 5577.34, 6300.30, 6562.79, 7291.47, and 7774.17~\AA\ for [\ion{O}{1}]\,$\lambda$5577, [\ion{O}{1}]\,$\lambda\lambda$6300, 6364, H$\alpha$, [\ion{Ca}{2}]\,$\lambda\lambda$7291, 7324, and \ion{O}{1}\,$\lambda$7774, respectively. The [\ion{Ca}{2}] profile (gray) is repeated in every panel for reference. The location of the telluric B feature in the [\ion{O}{1}]\,$\lambda\lambda$6300,6364 panel is marked by the ``$\earth$'' symbol.
    }
    \label{fig:2023iop:spec:line_profiles}
\end{figure*}

Panel 7.D shows the evolution of the absorption-minimum velocities for \ion{Fe}{2} and H$\alpha$. The velocity of \ion{Fe}{2}\,$\lambda$5018 declines smoothly from $5600\,\kms$ on day~$+13.6$ to $3100\,\kms$ on day~$+139.8$, consistent with the photosphere receding into deeper ejecta layers. This evolution is broadly consistent with the expansion velocity inferred from the blackbody-radius evolution ($4300\pm200\kms$ measured from the data before the blackbody radius peaked) during the phase in which the blackbody radius increases approximately linearly. Beyond day~$+139.8$, the measured \ion{Fe}{2} velocity remains approximately constant at $\sim3300\kms$. The slight increase at late times may be an artifact of the lower S/N of the MMT/Binospec spectra. In contrast, the H$\alpha$ absorption velocity remains remarkably constant at 7600--$8200\,\kms$ between days~$+13.6$ and $+139.8$, showing no measurable deceleration.

Such a velocity offset is common in SNe~II and reflects H$\alpha$ forming at larger radii in the extended H envelope than metal lines such as \ion{Fe}{2}, which form closer to the photosphere and track its recession more closely \citep{Dessart2005a, Gutierrez2017a}. In those objects, however, both velocities decline with time. At later epochs, the H$\alpha$ absorption velocity can no longer be measured reliably because of the increasing contribution from [\ion{O}{1}]\,$\lambda\lambda6300$, 6364 emission (Figure~\ref{fig:2023iop:spec:sequence_opt}).

Although the velocities of H$\alpha$ and \ion{Fe}{2} evolve differently, both line widths decrease with time (panel 7.E). The same behavior is observed for the emission component of the H$\alpha$ P-Cygni profile. Because H$\alpha$ remains strongly in emission, its width can be tracked throughout the entire spectroscopic campaign. The width of the H$\alpha$ emission component closely follows the photometric evolution: it decreases between maximum light and approximately day~$+100$; remains nearly constant until day~$+326$; and then drops abruptly from $6600\pm100$ to $4300\pm300\kms$ (Appendix, Table~\ref{tab:line_params}). This transition coincides with a rapid fading of the light curve (Section~\ref{sec:lc:properties}). The dramatic narrowing of the H$\alpha$ emission profile occurs only during this late-time decline. By contrast, during the similarly rapid decline observed between days~$+50$ and $+100$, the FWHM decreases by only $500\pm300\kms$ (Appendix, Table~\ref{tab:line_params}).

Panel 7.F presents the EW evolution. The EW of the H$\alpha$ emission component increases monotonically from $268$ to $902$~\AA\ between days~$+13.6$ and $+135.3$, whereas the EW of the absorption component decreases from $54$ to $22$~\AA. Consequently, the absorption-to-emission EW ratio ($a/e$), a commonly used diagnostic of the Type~II SN Balmer-line diversity \citep{Gutierrez2017a, Csoernyei2026a}, drops from $0.20$ to $0.02$--$0.03$ over the same period. Thus, the H$\alpha$ profile is already emission-dominated at the earliest epoch and becomes progressively more so over time. In contrast, \ion{Fe}{2}\,$\lambda$5018 remains weak throughout the plateau phase, with EWs of only $7$--$10$~\AA, declining further to $1$--$2$~\AA\ by day~$+140$. Two later measurements (days~$+296$ and $+326$) suggest a restrengthening of \ion{Fe}{2}. However, comparing the profiles of \ion{Fe}{2}\,$\lambda$4924 and \ion{Fe}{2}\,$\lambda$5018 indicates that an additional line may contribute to the apparent strengthening of \ion{Fe}{2}\,$\lambda$5018. At even later epochs, the H$\alpha$ EW declines rapidly. Although an increasing contribution from the host-galaxy continuum may exaggerate this trend as the SN fades, and the exact EW measured in the day~+380 spectrum is sensitive to the details of the late-time data reduction, the simultaneous sharp decrease in the H$\alpha$ line width is largely insensitive to these effects. The concurrent declines in both EW and FWHM therefore suggest that the weakening of the line is at least partly intrinsic.

Taken together, the constant H$\alpha$ absorption velocity (Panel 7.D), the growing emission dominance of the profile (Panel 7.F), and the widening velocity offset from \ion{Fe}{2} (Panel 7.D) are difficult to reconcile with line formation in a single, homologously expanding photosphere. Previously studied 87A-like events show no comparable evolution \citep{Singh2019a}. The observations instead suggest that an increasing fraction of the H$\alpha$ emission arises in a region external to the recombining ejecta traced by \ion{Fe}{2}, such as CSM or a cool dense shell \citep{Chevalier1994a, Fransson2014a, Smith2017a, Nyholm2020a}. Discriminating among these configurations requires tailored radiative-transfer modeling.

\subsubsection{Line Profiles}\label{sec:spectroscopy:line_profiles}

The spectrum on day~+326 provides the highest-S/N view of the emission-line profiles at late times and offers a direct probe of the kinematic structure of the inner ejecta (Figure~\ref{fig:2023iop:spec:line_profiles}). Nebular emission-line profiles are known to be sensitive to the geometry and velocity structure of the emitting material \citep{Maeda2008a, vanBaal2023a, Fang2024a, vanBaal2024a}, and different ionic species can trace distinct regions of the ejecta \citep{Milisavljevic2010a,Jerkstrand2017b}.

The forbidden lines share a common kinematic signature. [\ion{O}{1}]\,$\lambda$5577, [\ion{O}{1}]\,$\lambda\lambda$6300, 6364 and \ion{O}{1}\,$\lambda$7774 are all single-peaked, with comparable widths. [\ion{O}{1}]\,$\lambda$5577 is blueshifted by $\sim1000\kms$ relative to the systemic velocity, as is the blue maximum of the emission complex at 7300~\AA. H$\alpha$ is also single-peaked, but remains optically thick at this phase, so its profile is shaped by radiative transfer rather than by the distribution of the emitting material. It shows a mild displacement toward redder velocities.

The spectral resolution does not permit a detailed decomposition of the individual components. The recurrence of similar offsets and widths nonetheless indicates that the line-forming regions are not symmetric about the systemic velocity, and that several transitions originate in related kinematic structures.

The 7300~\AA\ complex is the exception. Besides the blue maximum described above, it shows a second maximum near $+2600\kms$, separated from the first by $\sim3000\kms$. This component was absent up to day~+139.8 and fully developed by day~+296.2, when our coverage resumed after solar conjunction. No other feature in the spectrum shows a comparable second maximum. We tentatively attribute it to [\ion{Ni}{2}]\,$\lambda$7378 from stable $^{58}$Ni rather than to [\ion{Ca}{2}]\,$\lambda$7324 (Section~\ref{sec:discussion:niII}).

\subsection{Spectropolarimetry}\label{sec:polarimetry}

\begin{figure}
    \centering
    \includegraphics[width=1\linewidth]{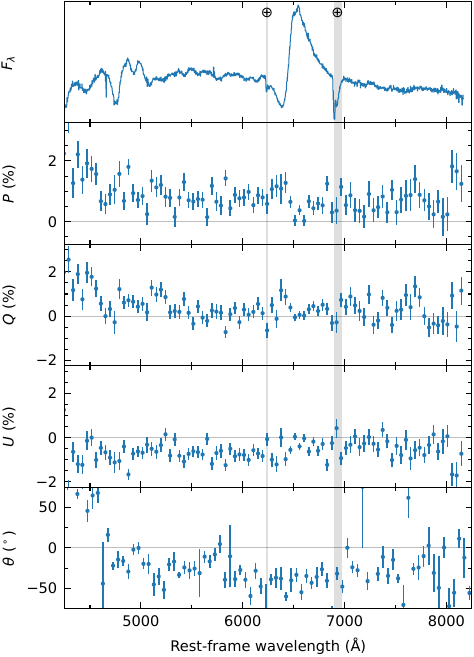}
    \caption{Spectropolarimetry of \sn on day~+37.3, obtained with FORS2 (50 \AA\ binning). The spectrum shows a high level of continuum polarization, 0.8--1\%, measured in line-free regions around H$\alpha$. 
    The gray vertical stripes and the ``$\earth$'' symbol mark the telluric A and B bands.
    }
    \label{fig:polarisation:spectrum}
\end{figure}

Spectropolarimetry probes the geometry of the photosphere and line-forming regions. Figure~\ref{fig:polarisation:spectrum} displays the spectropolarimetric observation obtained on day~+37.3.

Despite the modest S/N of the polarization spectrum, \sn exhibits significant continuum polarization. This interpretation is supported by the noticeable depolarization over the H$\alpha$ emission line. To first order, SN line emission is expected to be unpolarized, so it dilutes any underlying continuum polarization, causing the observed dip. We estimate the continuum polarization ($P$) from line‑poor wavelength intervals on either side of the H$\alpha$ P‑Cygni profile using a uniform 50~\AA\ spectral binning. In the 5800--6200~\AA\ (rest frame) interval we measure $P=0.83\pm0.09~\%$, while in the 7400--7800~\AA\ region we find $P=0.76\pm0.14~\%$. The inverse‑variance‑weighted mean of these wavelength intervals is $P=0.81\pm0.08~\%$. A measurement from a less-rebinned (25~\AA) spectrum yields a consistent but slightly higher value ($P=0.96\pm0.08~\%$), indicating that the inferred continuum polarization is not an artifact of the chosen binning scheme.

The observed global level of polarization is unlikely to be dominated by interstellar polarization (ISP), which produces a smooth, broad-band effect. Instead, the depolarization across the H$\alpha$ emission line can be used to constrain the ISP contribution. At the line peak, the spectrum is consistent with $Q\approx U \approx 0$, indicating a low combined ISP along the line of sight. This is supported by the low Galactic extinction $E(B-V)=0.044$~mag, corresponding to an expected Milky Way ISP of $9\times E(B-V)\leq0.4\%$ \citep{Serkowski1975a}. In addition, the \citet{Heiles2000a} catalog contains one star within $4^\circ$ of the SN position. It has a low polarization of only $P=0.070\pm0.035\%$, consistent with interstellar polarization. Together, these checks leave no evidence for a significant ISP contribution from either the Milky Way or \sn's host galaxy. Finally, the presence of line-locked depolarization across H$\alpha$ demonstrates that the continuum polarization is intrinsic to the SN and not caused by the lunar sky background, which would not reproduce such spectral features.

In addition to the depolarization observed across the H$\alpha$ emission component, the polarization spectrum shows an increase in polarization coincident with the absorption minimum of the P‑Cygni profile. This behavior indicates that the H‑rich ejecta are themselves aspherical. In this scenario, absorption by material along the line of sight preferentially removes unpolarized continuum photons, thereby enhancing the fractional contribution of scattered, and hence polarized, light originating from the asymmetric limb of the ejecta \citep[e.g.,][]{McCall1984a}. The resulting polarization peak at the absorption trough provides independent evidence that the line‑forming region shares the global asymmetry of the continuum-forming photosphere.

The measured continuum polarization of $P\approx0.8$--1.0~\% implies a substantial departure from spherical symmetry in the ejecta \citep{Hoeflich1991a, Jeffery1991a}. Canonical SNe~II are typically consistent with negligible continuum polarization during the photospheric phase, with polarization rising only once the photosphere recedes below the hydrogen envelope at the transition to the nebular phase \citep[e.g.,][]{Chornock2010a,Nagao2024a}. This timing depends on the thickness of the hydrogen envelope. For a more massive envelope, such as inferred for \sn under the PISN interpretation (Section~\ref{section:discussion:pisn}), this transition would be expected even later. \sn's significant continuum polarization already on day~+37.3 is therefore not straightforwardly explained by a receding photosphere alone. A bipolar explosion offers a more direct precedent: the Type~IIP SNe 2012aw and 2021yja show comparably high polarization already during the plateau phase, attributed to an energetic polar ejecta component that dominates before the bulk equatorial ejecta emerges at later times \citep{Nagao2024b}.

In contrast, 87A-like events are known to display significant polarization at comparable epochs. SN\,1987A exhibited continuum polarization of $\sim 0.6$~\% in the $R$ and $I$ bands between maximum light and several tens of days thereafter \citep{Jeffery1991a}, a behavior attributed to large‑scale asymmetry in the ejecta \citep{Hoeflich1991a, Wang2003a}. Similarly, the 87A‑like SN\,2018hna showed significant optical polarization during the photospheric phase \citep{Maund2021a}. The somewhat higher continuum polarization measured for \sn suggests an even more pronounced deviation from spherical symmetry. This could reflect a larger global axial ratio, or bipolar or lopsided structures imprinted on the hydrogen envelope by asymmetric energy deposition or shock propagation during the explosion \citep{Hoeflich1991a, Wang2008a}, similar to the bipolar geometry recently inferred for the Type~IIP SNe 2012aw and 2021yja \citep{Nagao2024b}. However, alternative interpretations involving asymmetric excitation or radiative-transfer effects without a strongly jet-driven explosion cannot be excluded \citep{Chugai1992a, Dessart2011a}. A comparably high continuum polarization ($\sim 0.8$\%) has, for instance, been attributed not to intrinsic ejecta asymmetry but to an aspherical circumstellar disk in the luminous, long-lived Type~II SN\,2021irp \citep{Reynolds2025b}.

\begin{figure}
    \centering
    \includegraphics[width=1\linewidth]{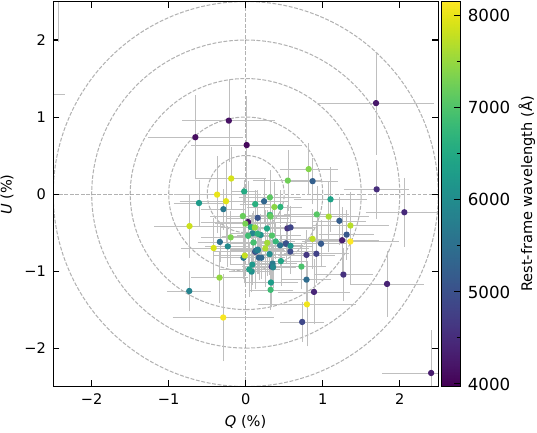}
    \caption{\sn in the Stokes $Q-U$ plane on day~+37.3 (binning 50~\AA). The points are color-coded by wavelength. Concentric circles show polarization degrees ranging from 0.5 to 2.5\%. The photosphere shows a clear departure from spherical symmetry, but the absence of a dominant axis does not indicate an axially symmetric geometry.
    }
    \label{fig:polarisation:quplane}
\end{figure}

While the continuum polarization amplitude clearly indicates significant asphericity in the photosphere, further constraints on the underlying geometry can be obtained from the distribution of points in the $Q-U$ plane (Figure~\ref{fig:polarisation:quplane}). Although the data are somewhat noisy, the measurements do not reveal a clear dominant axis indicative of simple axial symmetry, suggesting that the ejecta geometry may be more complex than a single coherent axisymmetric structure.

\subsection{Multiwavelength Limits}
\label{sec:multiwavelength_limits}

To probe nonthermal emission, circumstellar interaction \citep{Chevalier1982a, Chevalier1994a, Chevalier2017a}, and possible central engine activity \citep{Dexter2013a, Margalit2018a, Omand2018a}, as outlined in Section~\ref{section:discussion:roadmap}, we examined the X-ray and radio upper limits derived for \sn.
 
We find an unabsorbed X-ray flux limit of $<1.2\times10^{-13}~{\rm erg\,cm^{-2}\,s^{-1}}$ on day~+30.3 (0.3--10~keV; Appendix, Section~\ref{appendix:xrays}), corresponding to a luminosity of $<4\times10^{42}~{\rm erg\,s^{-1}}$ at $z=0.1020$. This limit is comparable to X-ray nondetections of H-poor SLSNe \citep{Levan2013a, Margutti2018a}, but lies more than two orders of magnitude above the X-ray luminosities seen in luminous SNe~IIn \citep{Chandra2015a}.
\citet{Alp2018a} reported that SN ejecta can remain optically thick to bound-free absorption across the whole 0.3--10~keV band for years to decades, so these data cannot test a central engine regardless of depth. An external shock is not screened by the ejecta, but at day~+30.3 the ejecta had reached only a few $\times10^{15}$~cm, well inside the CSM inferred from the late-time light curve, so no interaction signal was expected either.

Centimeter-wave follow-up observations with the AMI-LA on days~+84.4 and +102.6 yielded a nondetection with a root mean square (RMS) of 48~$\mu$Jy  (Appendix, Section~\ref{appendix:radio}). They translate to a $3\sigma$ spectral luminosity limit of $L_\nu < 4.4\times10^{28}~{\rm erg\,s^{-1}\,Hz^{-1}}$ at 15.5~GHz. 
The most radio-luminous SNe, such as the Type~Ic-BL SN\,PTF11qcj \citep[$\sim10^{29}~{\rm erg\,s^{-1}\,Hz^{-1}}$;][]{Corsi2014a}, reach luminosities above this limit, while the H-poor SLSN\,PTF10hgi \citep[$\sim10^{28}~{\rm erg\,s^{-1}\,Hz^{-1}}$;][]{Eftekhari2019a} sits below it. 
As in the X-rays, absorption rather than depth sets the constraint, though for different reasons in the two channels. For a central engine, the ejecta are optically thick to free-free absorption at centimeter wavelengths \citep{Mezger1967a, Murase2016a, Omand2018a, Margalit2018a}, and, given the probable large ejecta masses of extreme 87A-like events (Section~\ref{sec:discussion:ejecta_mass_diffusion}), they remain so for one to two decades (Section~\ref{section:discussion:roadmap}). An external shock is not screened by the ejecta, but CSM dense enough to power the light curve is itself optically thick to free-free absorption at 15.5~GHz at this epoch. Radio emission from strongly interacting SNe accordingly is expected to peak only after years \citep{Chevalier2017a}. Neither channel is therefore constrained at the levels of interest here.

\subsection{Host Galaxy}\label{sec:host_galaxy}

\begin{figure}
    \centering
    \includegraphics[width=1\columnwidth]{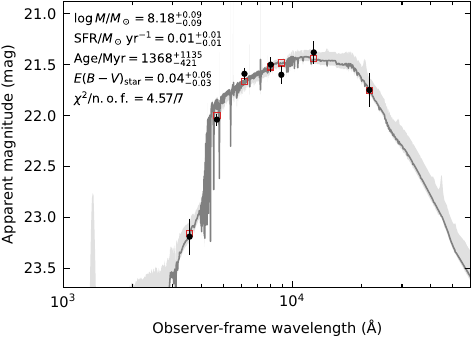}
    \caption{SED of the host galaxy from 1000 to 60,000~\AA\ (black dots). The solid line displays the best-fitting SED model. Red squares represent the model-predicted magnitudes. The fitting parameters are shown in the upper-left corner. The abbreviation ``n.o.f.'' stands for the number of filters. A sleeve indicates the $1\sigma$ credible region of 100 random draws.
    }
    \label{fig:host:sed}
\end{figure}

We model the observed host-galaxy SED (Table~\ref{tab:host_phot}) with the software package \program{Prospector}\footnote{\program{Prospector} uses the \program{Flexible Stellar Population Synthesis} (\program{FSPS}) code \citep{Conroy2009a} to generate the underlying physical model and \program{python-fsps} \citep{ForemanMackey2014a} to interface with \program{FSPS} in \program{python}. The \program{FSPS} code also accounts for the contribution from diffuse gas based on the \program{Cloudy} models of \citet{Byler2017a}. We use the dynamic nested-sampling package \program{dynesty} \citep{Speagle2020a} to sample the posterior probability distribution.
}
\citep{Johnson2021a}. We assume a Chabrier initial-mass function \citep[IMF;][]{Chabrier2003a}, a parametric star-formation history that rises linearly at early times and declines exponentially at late times [$t\, \exp(-t/t_{1/e})$], and attenuation following the prescription of \citet{Calzetti2000a}. The priors are identical to those adopted by \citet{Schulze2021a}.

Figure~\ref{fig:host:sed} shows the observed host-galaxy SED together with the best-fitting \textsc{Prospector} model. The observed photometry is well described by a star-forming dwarf galaxy with a stellar mass of $\log~(M_\star/M_\odot) = 8.18\pm0.09$, a star-formation rate of $0.01\pm0.01\,M_\odot\,{\rm yr}^{-1}$, and a stellar age of $1.4^{+1.1}_{-0.4}~\rm Gyr$.

The spectrum obtained on day~+623.8 exhibits nebular emission from H$\alpha$ and [\ion{O}{3}] $\lambda5007$ at a common redshift of $z=0.1020$. However, H$\alpha$ falls within a wavelength region strongly affected by telluric water-vapor absorption (Section~\ref{sec:spectroscopy:sequence}), which may suppress the observed flux and compromise estimates of the star-formation rate, attenuation, and gas-phase metallicity. 

\section{Discussion}\label{sec:discussion}

\subsection{Ejecta and Nickel Mass: Constraints from Photometric Timescales}

\subsubsection{Ejecta Mass}\label{sec:discussion:ejecta_mass_diffusion}

Before turning to detailed power-source models, we derive a basic estimate of \sn's total ejecta mass from its light-curve rise time. \sn's unusually slow rise may imply a long photon diffusion time through the homologously expanding ejecta. In the diffusion formalism of \citet{Arnett1982a}, this timescale is given by
\begin{equation}
t_{\rm d} \approx \left( \frac{2 \kappa M_{\rm ej}} {\beta c v} \right)^{1/2},
\end{equation}
where $\kappa$ is the effective opacity, $M_{\rm ej}$ is the ejecta mass, $v$ is a characteristic ejecta velocity, and $c$ is the speed of light. The dimensionless constant $\beta \approx 13.8$ depends on the ejecta density structure \citep{Arnett1982a}. If radiative diffusion primarily governs the rise, $t_{\rm rise} \approx t_{\rm d}$, the ejecta mass is given by
\begin{equation}
M_{\rm ej} \approx \frac{\beta c v t_{\rm rise}^{2}} {2\kappa}.
\end{equation}
The ejecta mass then scales as
\begin{align}
M_{\rm ej} &\approx 12.8~M_{\odot}\nonumber\\
& \times \left( \frac{t_{\rm rise}}{90~{\rm day}} \right)^{2} \left( \frac{v}{6000\kms} \right)  \left( \frac{\kappa}{0.3~{\rm cm^{2}~g^{-1}}} \right)^{-1} .
\end{align}
Adopting the lower limit on the rise time of $\sim88$~days (Section~\ref{sec:lc:properties}; Table~\ref{tab:lc:properties}) and assuming 
$v=6000\kms$ (Section~\ref{sec:spectroscopy:line_evolution}; Appendix Table~\ref{tab:line_params}), and an effective gray opacity $\kappa = 0.3~\rm cm^{2}\,g^{-1}$ (consistent with values used in semianalytic models for both regular SNe~II and 87A-like SNe, e.g., \citealt{Nagy2016a}), yields an ejecta mass of $>12\,M_\odot$.

\subsubsection{Nickel Mass}\label{sec:discussion:ni_mass}

In SNe powered by the decay chain $^{56}{\rm Ni}\rightarrow {^{56}{\rm Co}}\rightarrow{^{56}{\rm Fe}}$, the peak luminosity is determined by the amount of radioactive nickel. \sn resembles SNe\,1987A and 2018hna in light-curve shape but exceeds their peak luminosity by $\sim3.7$--4.0~mag (Table~\ref{tab:87A:comparison}). If this offset were entirely due to radioactive decay, and adopting as calibrators the nickel yields of SNe\,1987A and 2018hna\footnote{\citet{Singh2019a} reported a nickel yield of $0.087~M_\odot$, assuming a distance modulus of 30.52. In this paper, we assume a distance modulus of 30.11 (Table \ref{tab:87A:comparison}), which reduces SN\,2018hna's nickel yield by $\sim31\%$.} ($0.07$ and $0.06\,M_\odot$, respectively; \citealt{Bouchet1991a,Suntzeff1991a,Singh2019a}), the implied $^{56}$Ni mass for \sn is
\begin{equation}
M_{\rm Ni}({\rm \sn}) \approx M_{\rm Ni,ref} \times 10^{0.4\,\Delta m},
\end{equation}
where $\Delta m$ is the peak-magnitude offset relative to each calibrator. This gives $M_{\rm Ni}({\rm \sn}) \approx 2.2\,M_\odot$ for both calibrators.

The initial post-maximum decline resembles the end of the recombination phase in SNe~II (Section~\ref{section:discussion:87A:lc}). Afterward, \sn faded by only $0.2$~mag\,(100~day)$^{-1}$ between days~+150 and +325, markedly slower than the $\sim1$~mag\,(100~day)$^{-1}$ expected for pure $^{56}$Co decay, before dropping sharply thereafter (Section \ref{sec:lc:properties}). Radioactive decay alone therefore did not power \sn throughout this interval. The last detection of \sn, on day~+377.1, nonetheless provides an independent, and potentially tighter, upper bound on the nucleosynthesized nickel mass. By this epoch all $^{56}$Ni has decayed into $^{56}$Co, whose $e$-folding time is $\tau_{\rm Co}=111.3$~days \citep{Nadyozhin1994a}, so the decay luminosity reduces to
\begin{equation}
L_{\rm decay}(t) \approx M_{\rm Ni}\times1.33\times10^{43}~{\rm erg\,s^{-1}}\,M_\odot^{-1}\times e^{-t/\tau_{\rm Co}},
\label{eq:nickel:tail}
\end{equation}
where $t$ is measured from the explosion rather than from maximum light.

\sn's rise of $>88$~days (Section~\ref{sec:lc:properties}) places the last detection at $>465$~days after explosion, and attributing the entire observed luminosity to radioactive decay then gives $M_{\rm Ni}\lesssim2\,M_\odot$, comparable to the peak-based bound. This remains conservative for two reasons. The explosion may have preceded the first detection by considerably longer, and the luminosity adopted here does not capture the full bolometric output, since the infrared contribution is unconstrained by our data at these epochs (Section~\ref{sec:lc:bolometric}).

Equation~\ref{eq:nickel:tail} further assumes complete $\gamma$-ray trapping.
\citet{Jerkstrand2016a} reported that SNe~II become optically thin to $\gamma$-rays $\sim1$~yr after explosion, and PISNe only after $\sim2$~yr, owing to their more massive ejecta. At $>1.2$~years after explosion, \sn lies within this transition even for a very massive progenitor, so trapping is unlikely to be complete. This would raise the nickel mass derived from the tail, relaxing the bound further.

The upper end of these nickel-mass estimates is inconsistent with a neutrino-driven explosion and instead requires the PISN channel \citep{Fowler1964a, Barkat1967a, Rakavy1967a}, which we discuss in Section~\ref{section:discussion:pisn}. A nickel yield of a few tenths of a solar mass, or lower, cannot be excluded. Radioactive decay would then not be the primary power source at any phase, and another energy source would be required.

\subsection{Oxygen Mass: Constraints from Late-time Spectroscopy}
\label{sec:mass_constraints}
\label{subsec:omass}

Estimating the core oxygen mass from the nebular [\ion{O}{1}]\,$\lambda\lambda6300,6364$ doublet is a commonly used method for constraining the progenitor mass \citep{Uomoto1986a,Li1992a,Jerkstrand2014a}.
The observed doublet ratio contains key information regarding the optical depth of the line-forming gas. In the optically thin limit, the flux ratio is $F_{6300}/F_{6364} = 3$, whereas optically thick emission drives this value toward unity. 

From our spectrum at day~+326, we measure $F_{6300}/F_{6364} \approx 2.3$. This suggests that the observed doublet ratio lies near the boundary between the optically thin and thick regimes. However, residual H$\alpha$ absorption and dust formation may partially suppress the 6364~\AA\ component, making the intrinsic ratio uncertain. We therefore evaluate the oxygen mass under both regimes separately.

\paragraph{Optically thin regime}

In the optically thin limit, the neutral oxygen (\ion{O}{1}) mass can be derived from the combined luminosity of the [\ion{O}{1}]\,$\lambda\lambda6300,6364$ doublet using the analytical approximation of \citet{Uomoto1986a},
\begin{equation}
M_{\rm O\,I} \simeq 10^{-4} \left( \frac{L_{6300+6364}} {10^{38}~{\rm erg~s^{-1}}} \right) \exp\!\left( \frac{2.28\times10^4~{\rm K}} {T} \right) M_\odot,
\label{eq:M_O_thin}
\end{equation}
which reproduces detailed NLTE (non-local thermal equilibrium) spectral calculations to within $\sim 30\%$ under optically thin conditions \citep{Jerkstrand2014a,Jerkstrand2015a,Schweyer2025a}.

To constrain the oxygen temperature $T$, one can use the flux ratio between the [\ion{O}{1}]\,$\lambda5577$ and [\ion{O}{1}]\,$\lambda\lambda6300,6364$ transitions. Assuming both of these transitions to be in LTE with respect to the ground-state, one obtains Equation~2 of \citet{Jerkstrand2014a},
\begin{equation}
\frac{L_{5577}}{L_{6300,6364}} \approx 38\, \frac{\beta_{5577}} {\beta_{6300,6364}} \exp\!\left( -\frac{25790~{\rm K}} {T} \right),
\label{eq:OI_temp}
\end{equation}
where $\beta$ denotes the Sobolev escape probability [$\beta=(1-\exp(-\tau_\lambda))/\tau_\lambda$ and $\tau_\lambda$ being the Sobolev optical depth]. The lower level of [\ion{O}{1}]\,$\lambda\lambda6300,6364$ is in the ground state, whereas the lower level of [\ion{O}{1}]\,$\lambda5577$ is an excited state. This difference implies $\beta_{5577} > \beta_{6300,6364}$ \citep{Jerkstrand2017b}.
Since we are evaluating in the optically thin limit, we assume $\beta_{6300,6364} \approx 1$, so it follows that also 
$\beta_{5577} \approx 1$. Our line luminosity measurements (Table~\ref{tab:nebular:line_luminosities}) then yield a gas temperature of $T \approx 4250$~K.

\begin{table}
\caption{Line luminosities of nebular lines in the day +326 spectrum}
\centering
\begin{tabular}{lc}
\hline\hline
Species & Luminosity \\
& $(10^{39}\,\rm erg\,s^{-1})$\\
\hline
{[\ion{O}{1}]}\,$\lambda$5577 &$1.15 \pm 0.07$ \\
{[\ion{O}{1}]}\,$\lambda$6300 &$9.23 \pm 0.08$ \\
{[\ion{O}{1}]}\,$\lambda$6364 &$3.98 \pm 0.07$ \\
\hline
\end{tabular}
\label{tab:nebular:line_luminosities}
\end{table}

Crucially, we deem it unlikely that [\ion{O}{1}]\,$\lambda5577$ can be approximated as being in LTE for our +326\,d spectrum. As discussed by \citet{Jerkstrand2014a}, this transition has a large critical electron density (above which the lines form in LTE), $n_{e,\rm crit} = \beta_{5577} \times$ 10$^{8}$ cm$^{-3}$, with [\ion{O}{1}]\,$\lambda\lambda6300,6364$ having $n_{e,\rm crit} = \beta_{6300,6364} \times$ 3 $\times$ 10$^{6}$ cm$^{-3}$. This can be compared to, for example, PISN models by \citet{Jerkstrand2016a}, where O-rich zones have $n_{e} \approx 10^{6}$ -- 10$^{7}$ cm$^{-3}$ at 400 days post explosion, or to the CCSN models of \citet{Barmentloo2026a}, 
which have a similar range of $n_{e}$. If [\ion{O}{1}]\,$\lambda5577$ is out of LTE, this leads to an underestimate of the oxygen temperature, so that $T \approx 4250$~K should be understood as a strict lower limit, and thus in turn any estimate of the oxygen mass for the optically thin limit as an upper limit. Keeping this in mind, Equation~\ref{eq:M_O_thin} yields for $T \gtrsim 4250$~K a mass of $M^{\rm thin}_{\rm O\,I} \lesssim 2.8~M_\odot$. Here we do assume that [\ion{O}{1}]\,$\lambda\lambda6300,6364$ is in LTE, which from the discussion above is plausible, but also not guaranteed. 

\paragraph{Optically thick regime}

If the [\ion{O}{1}] doublet is optically thick, the emergent line luminosity no longer scales linearly with oxygen mass. Instead, a strict lower limit is obtained by setting the Sobolev optical depth \citep{Sobolev1957a} of the doublet to at least unity,
\begin{equation}
\tau_{\rm S} = \frac{1}{8\pi} \frac{g_k}{g_l} A_{\lambda} \lambda^{3} n_{l} t \gtrsim 1,
\label{eq:sobolev_OI}
\end{equation}
where $g_k$ and $g_l$ are the statistical weights of the upper and lower levels, $A_{\lambda}$ is the Einstein spontaneous emission coefficient, $n_{l}$ is the lower-state population density, and $t$ is the elapsed time since explosion. This expression is valid as long as $n_{u} \ll n_{l}$, which holds provided $E_{u}-E_{l} \gg kT$. Assuming the lower state is in LTE, which is well justified given that it lies within the ground-state term, the population density is
\begin{equation}
    n_{l} = \frac{M_{\rm ion}}{\mu m_{p}} \frac{1}{V_{\rm emit}} \frac{g_{l}}{Z(T)} \exp(-E_{l}/kT),
    \label{eq:nl}
\end{equation}
where $\mu$ is the atomic weight of oxygen, $Z(T)$ is the partition function, $E_{l}$ is the energy of the lower level above the ground state, and $V_{\rm emit} \equiv {4\pi}/{3}\,(v_{\rm O}t)^{3} f_{\rm O}$ is the emitting volume under the assumption of homologous expansion, with $f_{\rm O}$ the oxygen filling factor and $v_{\rm O}$ the characteristic velocity derived from the [\ion{O}{1}] profile. Because both lower levels lie within the ground-state term ($E_{l}/k \lesssim 230$~K), the Boltzmann factor is close to unity at the temperatures of interest, and we set it to unity in what follows.

The two doublet components share the same upper level but differ in their transition probabilities, with $A_{6300} \approx 3\,A_{6364}$, so that $\tau_{6364} < \tau_{6300}$ at any given oxygen mass. A genuinely optically thick doublet requires both components to satisfy $\tau_{\rm S} \gtrsim 1$; therefore, we impose this condition on the weaker $\lambda6364$. Evaluating Equations~\ref{eq:sobolev_OI} and \ref{eq:nl} for $\lambda6300$ alone would give $M_{\rm O} > 151\,f_{\rm O}\,M_\odot$, whereas imposing the condition on $\lambda6364$ yields
\begin{equation}
    M_{\rm O}^{\rm thick} > 455
    \left(\frac{v_{\rm O}}{4000\,\kms}\right)^{3}
    \left(\frac{t}{420\,\rm day}\right)^{2}
    f_{\rm O}\ M_\odot .
    \label{eq:M_O_thick}
\end{equation}
If the 6364~\AA\ component is only marginally optically thick, this limit decreases by roughly the doublet ratio.

We estimate $v_{\rm O}$ using two different methods: first, a double-Gaussian decomposition of the [\ion{O}{1}]\,$\lambda\lambda6300,6364$ profile, and second, the half-width at half-maximum intensity measured from the red wing of the 6364~\AA\ line. The second method may suffer from residual H$\alpha$ absorption and dust formation, whereas the first can be affected by line blending along the blue wing. These two approaches yield characteristic velocities of  $\sim 4000\kms$ and $5900\kms$, respectively, corresponding to $M^{\rm thick}_{\rm O} > 455\,f_{\rm O}\,M_\odot$ and $M_{\rm O}^{\rm thick} > 1459\,f_{\rm O}\,M_\odot$.

For oxygen masses typical of ordinary CCSNe ($M_{\rm O} \approx 1\,M_\odot$; \citealt{Jerkstrand2011a, Jerkstrand2014a, Jerkstrand2015a, Ergon2022a}), the doublet remains optically thick only for $f_{\rm O} \lesssim 0.002$ at $v_{\rm O} = 4000\kms$, or $f_{\rm O} \lesssim 0.0007$ at $v_{\rm O} = 5900\kms$. For the oxygen masses predicted by H-rich PISN models suitable to explain \sn's early light curve ($M_{\rm O} \approx 50\,M_\odot$; Table~\ref{tab:pisn:models}; \citealt{Dessart2013a}), the corresponding requirements are $f_{\rm O} \lesssim 0.11$ and $f_{\rm O} \lesssim 0.03$.

For comparison, hydrodynamic simulations show that Rayleigh-Taylor instabilities and nickel-bubble expansion fragment the metal core into dense, high-velocity clumps \citep{Hammer2010a}. Nebular spectral modeling of magnetar-powered SNe requires the most extreme clumping, $f_{\rm O}\lesssim0.01$, reflecting the strongly compressed inner ejecta produced by sustained central energy injection \citep{Jerkstrand2017a}, whereas modeling of ordinary CCSNe adopts $f_{\rm O}\approx 0.01$--0.1 \citep{Jerkstrand2011a, Jerkstrand2014a, Jerkstrand2015a, Ergon2022a}. By contrast, hydrodynamic simulations of pair-instability explosions find negligible mixing, with no significant redistribution of $^{56}$Ni and hence no nickel-bubble-driven clumping \citep{Chen2020a}, so that 1D models reproduce the ejecta structure well \citep{Joggerst2011a, Whalen2014a}. Such largely unmixed ejecta imply comparatively high filling factors, $f_{\rm O}\gtrsim0.1$.

\subsection{Stable [\ion{Ni}{2}]\,$\lambda$7378}\label{sec:discussion:niII}

\begin{figure}
    \centering
    \includegraphics[width=1\columnwidth]{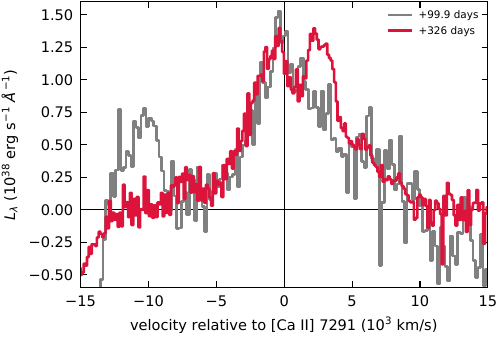}
    \caption{Emergence of the red component of the 7300~\AA\ complex. Continuum-subtracted profiles on days~+99.9 and +326, shown in velocity space relative to [\ion{Ca}{2}]\,$\lambda$7291. The blue halves are indistinguishable, and the profiles diverge only redward of $\sim+1000\kms$. A distinct second maximum, absent at day~+99.9, appears near $+2600\kms$ at day~+326. We attribute it to stable [\ion{Ni}{2}]\,$\lambda$7378, blueshifted by $\sim900\kms$ (Section~\ref{sec:discussion:niII}).}
    \label{fig:caII:epochs}
\end{figure}

Figure~\ref{fig:caII:epochs} compares the [\ion{Ca}{2}] region around 7300~\AA\ on days~+99.9 and +326. The blue side of the line complex remains unchanged between the two epochs. Both profiles rise identically from $-5000\kms$ to the peak near the [\ion{Ca}{2}]\,$\lambda$7291 rest velocity and begin to diverge only redward of $\sim1000\kms$. By day~+326, a distinct emission component has emerged near $+2600\kms$ that is absent at day~+99.9.

This feature is unlikely to be [\ion{Ca}{2}]\,$\lambda$7324 from material on the far side of the SN. Such material would contribute to both components of the doublet. Its [\ion{Ca}{2}]\,$\lambda$7291 emission would appear at approximately $+1300\kms$, nearly coincident with the expected position of [\ion{Ca}{2}]\,$\lambda$7324 ($+1333\kms$). However, the observed profile exhibits a local minimum at this velocity and shows no excess flux relative to day~+99.9.

At day~+326, [\ion{Ca}{2}] is one of the nebular features least affected by radiative-transfer effects. In models of late-time core-collapse SN ejecta, the [\ion{Ca}{2}] doublet becomes effectively transparent at an early stage, with its emission tracing the full ejecta volume to within a few percent \citep{vanBaal2026a}. Although that model assumes a lower ejecta mass than expected for \sn (Sections~\ref{sec:discussion:ejecta_mass_diffusion}, \ref{section:discussion:lightcurve_modeling}), it also adopts a lower expansion velocity, such that the ejecta densities may not differ substantially. Unlike transitions at shorter wavelengths, [\ion{Ca}{2}] should sample the receding hemisphere nearly as effectively as the approaching one.

No other feature in the day~+326 spectrum shows a comparable double-horn structure. [\ion{O}{1}]\,$\lambda$5577, [\ion{O}{1}]\,$\lambda\lambda$6300, 6364 and \ion{O}{1}\,$\lambda$7774 are all single-peaked at widths comparable to [\ion{Ca}{2}] (Figure~\ref{fig:2023iop:spec:line_profiles}), and some share the $\sim1000\kms$ blueshift noted above. This suggests that the red horn is unlikely to arise from calcium. [\ion{Ni}{2}]\,$\lambda$7378, arising from stable $^{58}$Ni, provides a more plausible identification. It lies at $+3551\kms$ relative to [\ion{Ca}{2}]\,$\lambda$7291, so the observed peak corresponds to a blueshift of $\sim900\kms$, which is similar to the blueshifted component observed in, e.g., H$\alpha$ and the \ion{O}{1} features.

To estimate the luminosity of the red horn, we aligned the blue flanks of the continuum-subtracted line profiles at days~+99.9 and +326 and subtracted the earlier spectrum from the later one. Integrating the residual flux between $+1000$ and $+5000\kms$ yields $\sim10^{-16}~{\rm erg\,s^{-1}\,cm^{-2}}$, corresponding to a luminosity of $\sim3\times10^{39}~{\rm erg\,s^{-1}}$. The uncertainty is dominated by the absolute flux calibration and is estimated to be at the level of several tens of percent.

The critical electron density of [\ion{Ni}{2}]\,$\lambda$7378 is $\sim1.5\times10^{6}~{\rm cm^{-3}}$ \citep{Barmentloo2026b}, above which the level populations approach local thermodynamic equilibrium. At 400 days, electron densities at the core reach $10^{7}$--$10^{8}~{\rm cm^{-3}}$ in PISN models \citep{Jerkstrand2016a} but only $\sim10^{6}~{\rm cm^{-3}}$ in ordinary core-collapse SN models \citep{Barmentloo2026b}. LTE is a reasonable approximation for the PISN scenario, but not for the core-collapse SN one. In either scenario, [\ion{Ni}{2}]\,$\lambda$7378 should be optically thin by $\gtrsim400$ days after explosion according to the aforementioned PISN and CCSN models. In the optically thin LTE limit, the  [\ion{Ni}{2}]\,$\lambda$7378 luminosity is given (through e.g., equation 42 in \citealt{Jerkstrand2017b}) by
\begin{equation}
L_{7378} \approx \left(\frac{M_{\rm Ni\,II}}{0.1~M_\odot}\right) \frac{\exp(-19500~{\rm K}/T)}{Z(T)} \times 10^{43}~{\rm erg\,s^{-1}},
\label{eq:niII}
\end{equation}
where $Z(T)$ is the partition function and $T$ the gas temperature. The observed luminosity implies $M_{\rm Ni\,II}\approx0.017~M_\odot$ at $T=5000$~K, a temperature typical of massive PISN models (see e.g., figure 5 in \citealt{Jerkstrand2016a}), and $\approx0.17~M_\odot$ at $T=3000$~K. 

Stable nickel yields of a few $\times10^{-2}~M_\odot$ are reached in the PISN grid of \citet{Heger2002a} for helium-core masses of $100~M_\odot$ and above. Model grids \citep{Limongi2018a, Woosley2019a, Ertl2020a} and observations \citep{Aitken1988a, Jerkstrand2015b, Sollerman2021a, Barmentloo2026b} of ordinary core-collapse explosions synthesize a few $\times10^{-3}~M_\odot$ of $^{58}$Ni, an order of magnitude less. The core-collapse supernova interpretation is further disfavored because LTE is unlikely to hold at these lower densities, making Equation~\ref{eq:niII} a strict lower limit on the required nickel mass. The temperature and yield constraints reinforce one another. The higher temperatures of massive PISN models lower the required nickel mass to a value those models supply. The cooler core-collapse models raise it beyond anything they produce.

At later epochs, the temperature in the Ca-rich zones 
in PISN models is found (see e.g., figure 5 in \citealt{Jerkstrand2016a}) to be $\sim$ 1000 K lower than in the Ni-rich zones ($T_{\mathrm{Ni}} \sim$ 5000 K for He130 model and $T_{\mathrm{Ni}} \sim$ 4000 K for He100 model)\footnote{Models in the grid of \citet{Heger2002a} are labeled by the mass of the bare helium core in solar masses, so He100 and He130 denote cores of $100$ and $130~M_\odot$.}. With almost the same temperature dependence ($L_{\mathrm{[Ca\,II]}} \propto \exp(-19700 \,\mathrm{K} / T )$), one then expects $L_{[\mathrm{Ni \,II}]}$/$L_{[\mathrm{Ca \,II}]}$ to increase by a factor $\sim$ 3 -- 5 (for He130 and He100, respectively) compared to early times, where the higher and more similar zone temperatures likely lead to comparable exponentials. This explains the absence/weaker contribution of [\ion{Ni}{2}] in the early spectrum, and its arrival in the late spectrum.

Section~\ref{section:discussion:pisn} shows that a hydrogen-rich pair-instability model of a $190~M_\odot$ star, R190NL\footnote{The \citet{Dessart2013a} model is labeled by zero-age main-sequence mass rather than by core mass, with R190NL denoting a red-supergiant progenitor of $190~M_\odot$.}, reproduces \sn's early light curve and spectra. Its oxygen core mass of $\sim79~M_\odot$ is almost identical to that of the He100 model. Because stable $^{58}$Ni is synthesized in the innermost burning layers, the two models are expected to produce comparable amounts, despite their somewhat different $^{56}$Ni yields ($\sim2.6$ and $\sim5~M_\odot$, respectively). The stable nickel mass inferred above is comfortably within the range expected for R190NL. If \sn is a PISN, the late-time luminosity excess must have a separate origin, with circumstellar interaction as a leading candidate (Section~\ref{section:discussion:late_time}).

If correct, identifying the red horn as [\ion{Ni}{2}]\,$\lambda$7378 would provide direct evidence for the production of a substantial mass of stable nickel, strongly supporting a PISN interpretation of \sn. However, this interpretation presently rests on a single spectral feature. Confirmation will require either detecting additional [\ion{Ni}{2}] transitions, which would require NIR/MIR observations at late times, or dedicated nebular radiative-transfer modeling tailored to \sn.

\subsection{Light-curve Modeling}\label{section:discussion:lightcurve_modeling}
\subsubsection{Pair-instability Supernova Scenario}\label{section:discussion:pisn}

We now turn our attention to the light-curve modeling. The broad, luminous light curve is a hallmark of PISN models. We compare the observations of \sn with PISN models of H-rich stars with a zero age main-sequence mass of $190~M_\odot$ from \citet{Dessart2013a} that are based on the calculations of \citet{Heger2002a}. The properties of the blue- and red-supergiant progenitors (B190NL and R190NL), differing in the mass of the H envelope, are summarized in Table~\ref{tab:pisn:models}. These models assume zero metallicity (i.e., a Population-III star) and include nonlocal energy deposition (i.e., $\gamma$-ray transport). Using the publicly available models\footnote{\href{https://zenodo.org/records/5524918}{https://zenodo.org/records/5524918}}, we compute the light curves in the ZTF $g$, $r$, and $i$ bands at the redshift of \sn.

\begin{figure*}[t!]
    \centering
    \includegraphics[width=1\textwidth]{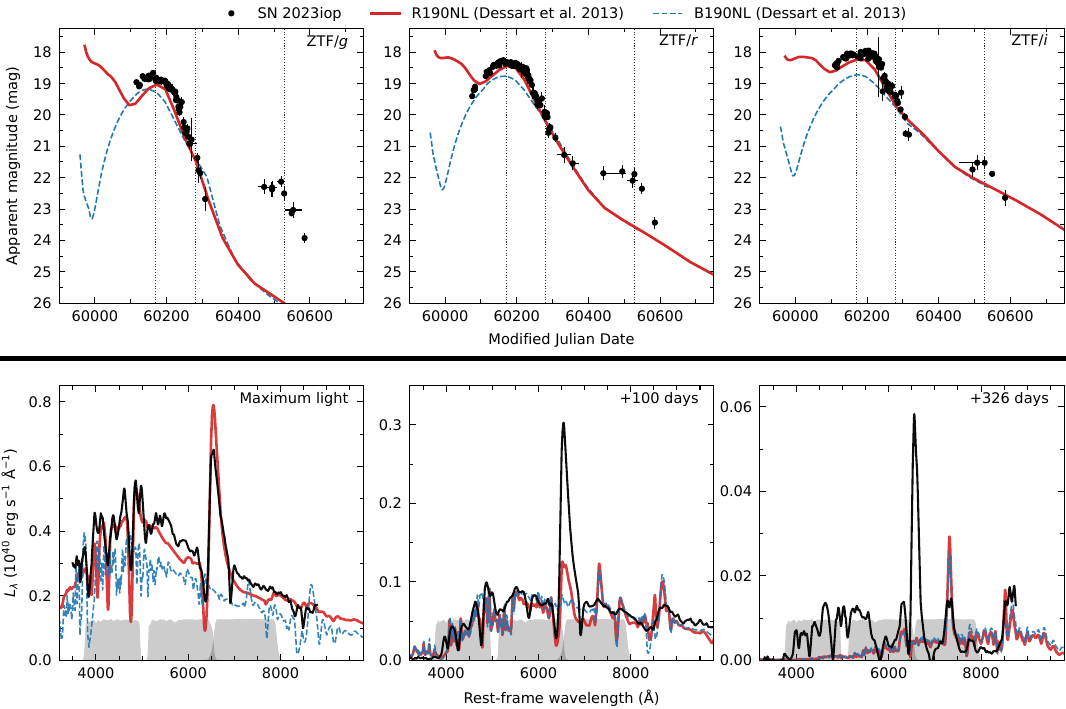}
    \caption{Comparison of the multiband light curves and spectra of \sn with H-rich PISN models.
    \textit{Upper panels:} The R190NL model reproduces the observed light curves well up to $+150$ days relative to peak light, in contrast to the B190NL model. 
    At later epochs, the models fade by $\sim$1 mag per 100 days ($^{56}$Co decay). By contrast, \sn exhibits a long-lasting plateau and remains several magnitudes brighter than all models. This excess is strongly wavelength-dependent, being most pronounced in the $g$ band and least pronounced in the $i$ band. The vertical lines in each panel indicate the spectroscopic epochs shown below.
    \textit{Lower panels:} Near peak brightness, the observed spectrum is essentially indistinguishable from R190NL, whereas B190NL already lacks H features owing to its receded photosphere. At day~+100, R190NL continues to provide a good overall match, but differences begin to emerge. By day~+326, the divergence becomes pronounced, particularly in the $g$ band, mirroring the photometric excess.
    For guidance, the ZTF $gri$ filter transmission functions are shown in light gray.
    }
    \label{fig:discussion:pisne}
\end{figure*}

Up until 150 days after peak brightness (Figure~\ref{fig:discussion:pisne}, top panels), the R190NL model provides a very good match to both the rise and peak luminosity of \sn, whereas B190NL is systematically $\sim0.5$~mag fainter at peak in all bands. The agreement with the R190NL model extends to the spectral evolution (Figure~\ref{fig:discussion:pisne}, bottom panels). At maximum light, the observed spectrum is very similar to that of R190NL. We note that this model predicts stronger Balmer features than are observed, but also a low absorption-to-emission ratio for H$\alpha$ that is qualitatively similar to that observed for \sn. This good spectroscopic correspondence persists to +100 days, although H$\alpha$ is significantly stronger in the data, suggesting that \sn may retain a more massive H envelope than assumed in this PISN model. The B190NL model fails to reproduce the prominent H features because its H envelope is less extended and the photosphere has already receded into the inner metal-rich layers.

\begin{table}
\caption{Properties of PISN models \citep{Dessart2013a}.}\label{tab:pisn:models}
\centering
\small
\setlength{\tabcolsep}{3pt}
\begin{tabular}{lcccccccc}
\hline\hline
Name & $M_i$  & $M_f$ & $R_*$ & $M({\rm H})$ &$M({\rm O})$& $M({\rm Ni})$ & $E_{\rm kin}$\\
     & ($M_\odot$) & ($M_\odot$) & ($R_\odot$) & $(M_\odot)$& $(M_\odot)$& $(M_\odot)$ & (B)\\
\hline
B190NL & 190 & 133.9 &  186 &  5.7 & 48.1 & 3.0 & 34.5\\     
R190NL & 190 & 164.1 & 4044 & 23.6 & 48.1 & 2.6 & 33.2\\
\hline
\end{tabular}
\tablecomments{$M_i$ is the zero-age main-sequence (ZAMS) mass, and $M_f$ the mass of the star at the time of explosion.
}
\end{table}

At later phases, however, the agreement breaks down completely. Beyond $\sim200$ days, the R190NL model follows the expected $^{56}$Co decay rate, declining at $\sim1$~mag per 100~days. In contrast, \sn fades much more slowly, by merely $\approx0.2$~mag per 100~days (Section~\ref{sec:lc:properties}). By day~+326, the discrepancy reaches $\sim3.5$~mag in $g$, 1.7~mag in $r$, and $\sim0.7$~mag in $i$. Such a slow decline and blue excess are incompatible with a purely radioactively powered PISN, in which the luminosity directly tracks the exponentially declining energy deposition from $^{56}$Co.

The spectral evolution reinforces this conclusion. By day~+326, the R190NL model has transitioned to a red, nebular phase, as expected once the ejecta become optically thin. In stark contrast, \sn remains blue and distinctly photospheric, with strong H$\alpha$ emission. This behavior cannot be reconciled with this PISN model, which predicts a transition to the nebular phase on comparable timescales. The continued presence of a photosphere and strong H emission naturally explains the excess flux in the $g$ and $r$ bands and points to an additional energy source beyond radioactive decay, which we discuss in the next section.

The PISN models used here were computed in the absence of circumstellar interaction, even though their progenitors can lose several tens of $M_\odot$ before explosion (Table~\ref{tab:pisn:models}). A late-time excess has also been reported for the H-poor PISN candidate SN\,2018ibb \citep{Schulze2024a} and the H-rich candidate SN\,2023vbw \citep{Hiramatsu2026a}. In both cases, it was attributed to CSM interaction rather than taken as evidence against a PISN origin. The excesses differ in spectral shape between these objects, so we draw no direct analogy. A suite of PISN models that include ejecta-CSM interaction would be needed to establish whether the excess in \sn could arise within the PISN channel.

\subsubsection{Late-time Emission}\label{section:discussion:late_time}

\begin{figure}
    \centering
    \includegraphics[width=1\columnwidth]{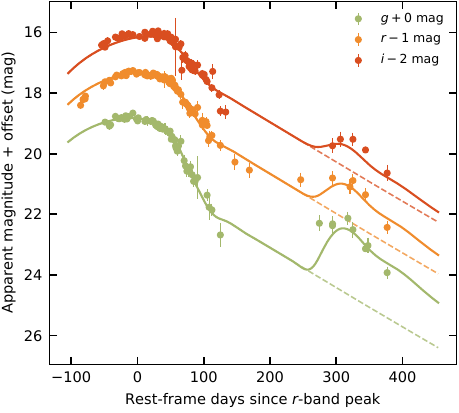}
    \caption{
    Two-stage phenomenological modeling of the $gri$ light curves. The photometry is plotted with the indicated magnitude offsets. Dashed curves show the radioactive-decay and recombination baseline, while solid curves denote the baseline plus the additive blackbody component. The latter describes the timing and color dependence of the late excess and is not interpreted as a physical blackbody.}
    \label{fig:discussion:radioactive_blackbody}
\end{figure}

We used two descriptions of the late-time excess to model and better understand the origin of the late-time behavior of \sn. The first is a deliberately phenomenological fit to the $gri$ photometry, shown in Figure~\ref{fig:discussion:radioactive_blackbody}. We fit the light curve in two stages with \program{Redback}. First, we establish a baseline at late times by tackling the first peak, fitting it with a radioactive-decay and recombination model~\citep{Nagy2016a} via \program{SNmix}~\citep{Sarin2026b} and sampling the posterior via \program{Bilby}~\citep{Ashton2019a}. Thereafter, we use this baseline as a prior and add a blackbody component, that we then fit to the $gri$ photometry. 

We stress that this analysis is deliberately simple. We are not inferring both components together but using the physically well-motivated radioactive component to anchor the extra phenomenological component. The additional component begins at $269.7$ rest-frame days after the $r$-band peak and reaches its maximum at $295.8$ days. At that maximum, its effective luminosity, temperature, and radius are $8.2\times10^{41}~{\rm erg\,s^{-1}}$, 8500~K, and $4.7\times10^{14}$~cm, respectively. These quantities describe the timing and color of the excess, but should not be interpreted as measurements of a real blackbody component. As highlighted by the spectrum (Figure~\ref{fig:2023iop:spec:sequence_opt}), the observed late-time SED is not well described by a blackbody. Nevertheless, it helps set some rough constraints to understand what the emission component could be.

The late-time light curve shares several characteristics with the light curve near peak, including a sudden drop in brightness (Figure~\ref{fig:2023iop:lc}). Such a ``replay'' could occur if interstellar or circumstellar dust scatters supernova light from an earlier epoch back into the line of sight \citep{Chevalier1986a}. If \sn originates from an extremely massive progenitor, a scattering screen from substantial pre-explosion mass loss is not a contrived scenario. In fact, a light echo was discovered in the H-poor SLSN iPTF16eh 
\citep{Lunnan2018a}.

To test this quantitatively, we model the observed late-time spectrum as a combination of the residual PISN-model spectrum and a rescaled, dust-scattered copy of the near-peak spectrum,
\begin{equation}
f_{\rm late}(\lambda) = f_{\rm floor}(\lambda) + a\,f_{\rm incident}(\lambda)\,S\!\left(\frac{\lambda}{1+z},\theta\right),
\end{equation}
where $f_{\rm late}$ is the observed late-time spectrum, $f_{\rm incident}$ is the near-peak spectrum being scattered, $S(\lambda,\theta)$ is a tabulated Milky Way dust-scattering function \citep{Sugerman2003a}, $\theta$ is the scattering angle, $f_{\rm floor}$ is the late-time PISN spectrum from \citet{Dessart2013a} shown in Figure~\ref{fig:discussion:pisne}, $a$ is a scaling factor, and $z$ is the redshift of \sn.

This model reproduces \sn's late-time brightness and color reasonably well. The scattering angle has no significant impact on the shape of the spectrum, but only affects the absolute flux level, which the scaling factor $a$ absorbs. However, the fitted spectrum shows a clear mismatch to the data. \sn's spectroscopic evolution also remains smooth throughout, with no discontinuity of the kind a light echo would produce, replaying the spectrum at maximum light \citep{Rest2011a}. This is best seen in H$\beta$, whose velocity declines smoothly from $\sim5900\kms$ near maximum light to $\sim4100\kms$ at day~+326. This gradual change is also evident in the overall spectral similarity between days~+139.8 and +326, despite the $\sim$186-day gap separating them (Figure~\ref{fig:2023iop:spec:sequence_opt}). We therefore rule out a light echo as the origin of \sn's late-time emission.

Interpreting the expanding blackbody component in the context of CSM interaction is also challenging. Producing such a hot, quasithermal component at large radii would require finely tuned conditions, and it is difficult to reconcile this scenario with the lack of clear spectroscopic signatures of interaction (e.g., narrow Balmer emission lines). While weak or geometrically confined interaction cannot be excluded a priori \citep{Sollerman2021a}, invoking it in this context appears contrived. Furthermore, the inferred temperature is higher than typically expected for emission associated with dust or reprocessed radiation. This skepticism is reinforced by comparison with SN\,2021irp, a luminous, long-lived Type~II SN whose excess luminosity and $\sim0.8$\% continuum polarization have been attributed to interaction with aspherical, disk-like CSM of mass $\gtrsim2\,M_\odot$ \citep{Reynolds2025a, Reynolds2025b}. Unlike \sn, however, SN\,2021irp shows clear interaction signatures, including narrow P~Cygni components, an \ion{Fe}{2} pseudocontinuum, and a lack of absorption lines. The absence of any such features in \sn's spectra suggests that, if CSM interaction contributes to its late-time emission, the interacting material must be more deeply hidden than in SN\,2021irp.

Given the difficulty of identifying the origin of the emission spectroscopically or via photometry, we instead explore a different question, whether CSM interaction is viable even at a bolometric level. We first construct a restricted pseudobolometric light curve directly from the 24 flux-calibrated, Milky-Way-extinction-corrected spectra. We integrated each spectrum over the common observer-frame interval $4650$--$9000$~\AA\ (rest-frame $4220$--$8168$~\AA), spanning phases from $-52$ to $+380$ days. We use this reconstruction of the bolometric luminosity rather than from the photometry as it better highlights the late-time plateau and excess, and it can be more easily restricted to energy over a specific wavelength range. 

\begin{figure}
    \centering
    \includegraphics[width=1\columnwidth]{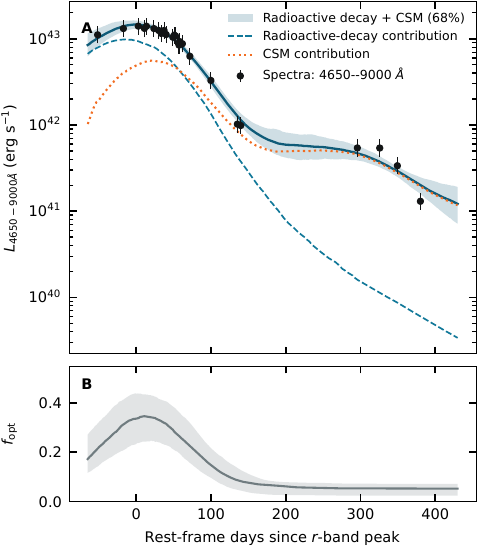}
    \caption{
    Joint modeling of the spectroscopic pseudobolometric light curve. Panel A shows the luminosity integrated over the common observer-frame interval $4650$--$9000$~\AA\ (black points). The solid curve and shaded region are the median and 68\% posterior-predictive interval of the radioactive-decay plus CSM model. Dashed and dotted curves represent the median radioactive-decay/recombination and interaction contributions, respectively, after both have been mapped into the observed interval by the same $f_{\rm opt}(t)$. Panel B shows the median and 68\% interval of this single time-dependent optical fraction.}
    \label{fig:discussion:radioactive_csm}
\end{figure}

We fitted these luminosities jointly with a radioactive-decay/recombination component as above and the interaction power generated by the SN ejecta interacting with a static CSM density profile \program{redback-csm}~\citep{Sarin2026a}. The CSM density was represented by a static 24-node penalized spline to provide maximum flexibility in CSM structure, but we stress that the priors are built to target a smooth power-law density profile. We then assume that only a time-dependent fraction $f_{\rm opt}(t)$, of the radioactive and interaction power is visible in the $4650$--$9000$~\AA\ range. A low floor plus a Gaussian enhancement was adopted for this fraction, allowing the optical share to be small at both early and late phases, and largest near the optical peak. The $^{56}$Ni mass was restricted to $<5~M_\odot$, motivated by the pair instability models. We assume a conservative 0.10~dex luminosity uncertainty in the likelihood. The data, posterior-predictive light curve, component decomposition, and inferred optical fraction are shown in Figure~\ref{fig:discussion:radioactive_csm}.

This model reproduces the restricted pseudobolometric light curve with small scatter. The inferred ejecta mass is $60^{+43}_{-22}~M_\odot$, the kinetic energy is $18^{+12}_{-8}\times10^{51}$~erg, and the nickel mass is $3.1^{+1.5}_{-1.7}~M_\odot$ (median and 68\% credible interval). The corresponding characteristic velocity is $5370^{+640}_{-510}\kms$. The spherical-equivalent CSM mass is much less tightly constrained, $7^{+20}_{-5}~M_\odot$, because it is strongly degenerate with the radiative efficiency, which we constrain to $0.24^{+0.22}_{-0.13}$, and with the fraction of the luminosity emerging in the fitted wavelength interval. The inferred CSM occupies radii of $\sim 10^{16}$ to $7\times10^{16}$~cm. The fraction of the total luminosity emerging in optical rises from a floor of $0.05^{+0.02}_{-0.03}$ to $0.36\pm0.09$ near $+13^{+22}_{-21}$ days, with a characteristic width of $58^{+14}_{-9}$ days.

The SN parameters in the joint fit are broadly compatible with the energetic and nickel-rich explosion suggested by the R190NL PISN comparison ($E_{\rm kin}=33.2\times10^{51}$~erg and $M_{\rm Ni}=2.6~M_\odot$), although the ejecta mass is substantially lower than that of the R190NL model ($164~M_\odot$; Table~\ref{tab:pisn:models}). This difference could be driven by the uncertain explosion epoch as well as systematic uncertainty and/or different assumptions in our modeling, such as differences in density structure and opacity.

We note that the calculation above assumes a spherical system. For an equatorial band with half-opening angle $\theta$, a useful first-order conversion is $f_\Omega=\Omega/(4\pi)=\sin\theta$. Holding the radial density fixed within the occupied solid angle gives $M_{\rm CSM,true}\approx f_\Omega M_{\rm CSM,sph}$ and $\epsilon_{\rm fit}\approx f_\Omega\epsilon_{\rm int}$. The $30^\circ$--$50^\circ$ half-opening angle inferred for the disk-like CSM of SN\,2021irp \citep{Reynolds2025a,Reynolds2025b} would thus turn our median spherical-equivalent mass into $\sim 3.3$--$5.0~M_\odot$, while requiring an intrinsic conversion efficiency of $\sim 0.32$--0.48. These scalings are illustrative because changing the geometry also changes the shock dynamics, optical depth, viewing-angle dependence, and emergent spectrum. Likewise, if a smaller fraction of the interaction power lies in our restricted optical interval than inferred here, more bolometric shock power --- and generally more CSM at fixed efficiency --- is required. The change is not linear because CSM mass, efficiency, and spectral redistribution are covariant.

The red color near peak brightness (Section~\ref{sec:lc:properties}) independently disfavors a dominant early interaction component, although our joint fit was not explicitly forced to make the early CSM luminosity vanish and cannot predict colors because it uses a single $f_{\rm opt}(t)$. At late times, the absence of narrow lines or other unambiguous interaction signatures (Sections~\ref{sec:lc:properties}, \ref{sec:spectroscopy:sequence}) requires any interaction to be hidden, optically thick, geometrically confined, or sufficiently kinematically broadened. The continuum polarization (Section~\ref{sec:polarimetry}) makes a nonspherical configuration plausible. Moreover, the absence of a significant change in the optical-NIR colors across the rapid decline (Section~\ref{sec:lc:properties}) argues against newly formed dust being its principal cause. A finite equatorial CSM structure, analogous in broad geometry to that proposed for SN\,2021irp but spectroscopically more deeply buried, could instead produce a long interaction-powered phase followed by a rapid decline when the shock reaches its outer edge. Two-dimensional radiation-hydrodynamic simulations of ejecta colliding with a circumstellar disk show that the dissipation is concentrated where the ejecta reach the disk rather than in the surrounding wind \citep{Suzuki2019a}. A tenuous wind would then be shocked too weakly to produce narrow lines, and the disk emission would emerge only once the overlying ejecta become optically thin. The present light-curve fit shows that such a scenario is energetically viable, and the configuration above offers a route to the absence of spectral signatures, though neither establishes the geometry.

\subsubsection{Fallback Accretion Powering}\label{section:discussion:fallback}

We next consider whether fallback accretion onto a central compact remnant can account for \sn's light curve. In this scenario, the initial explosion fails to unbind the star. A substantial fraction of the stellar material then falls back onto the newly formed compact object, and the accretion energy released at the center powers the observed emission \citep{Dexter2013a, Moriya2018c}.

We follow the approach of \citet{Moriya2018c}, who applied it to the extreme 87A-like SN OGLE-73, by using their 40 and 60~$M_\odot$ progenitors. Light curves are computed with the one-dimensional (1D) radiation-hydrodynamics code \program{STELLA} \citep{Blinnikov1998a, Blinnikov2000a, Blinnikov2006a}, with the accretion power deposited as thermal energy just above the mass cut. The models contain no $^{56}$Ni, so the entire light curve, including the tail, is powered by fallback accretion alone. In the models investigated here, an initial thermal energy $E_{\rm in}=3\times 10^{50}~\mathrm{erg}$ is released at the stellar center. Following the initial energy input, the fallback accretion energy $\dot{E}_{\rm acc}$ is injected from the center as
\begin{equation}
\dot{E}_{\rm acc} = \eta\,\dot{M}\,c^{2},
\end{equation}
where $\eta$ is an efficiency for converting the accretion energy to thermal energy, $\dot{M}(t)$ is the accretion rate which asymptotically approaches $\dot{M}\propto t^{-5/3}$, and $c$ is the speed of light. The accretion efficiencies are adjusted to match the luminosity of \sn, and are $\eta=4\times 10^{-3}$ ($40\,M_\odot$) and $\eta=5\times10^{-3}$ ($60\,M_\odot$).

\begin{figure}
    \centering
    \includegraphics[width=1\columnwidth]{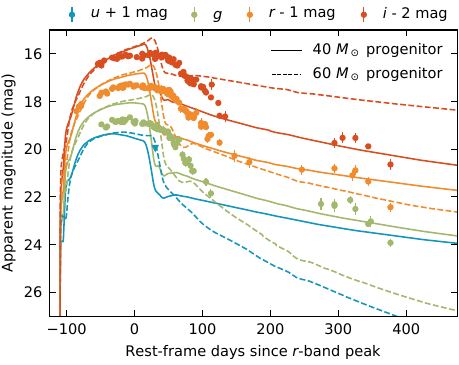}
    \caption{   
    Comparison of \sn's $ugri$ light curves (detections $\bullet$; nondetections $\blacktriangledown$) with fallback-accretion-powered models for progenitors of $40\,M_\odot$ (solid) and $60\,M_\odot$ (dashed). The models contain no $^{56}$Ni, so the entire light curve, including the tail, is powered by accretion onto the central remnant. Both models reproduce the extended rise and the luminous maximum, but the abrupt decline at the end of the recombination phase occurs earlier and more sharply than observed, most conspicuously in $u$ and $g$. At late times the $40\,M_\odot$ model is too bright in all bands by a band-dependent amount, while the $60\,M_\odot$ model is too faint in $g$ and too bright in $i$, indicating that the predicted colors, and not merely the luminosity scale, disagree with the data.
    }
    \label{fig:discussion:fallback}
\end{figure}

Figure~\ref{fig:discussion:fallback} compares \sn's $ugri$ light curves with the two models. Both reproduce the broad morphology of the observation. The rise lasts more than 100~days, and the tail fades slowly without any radioactive contribution. They fail, however, in two respects.

First, the abrupt decline at the end of the recombination phase occurs too early and is far too sharp. The models drop at day~$\sim+30$ ($40\,M_\odot$) and day~$\sim+60$ ($60\,M_\odot$). \sn instead steepens gradually from day~$\sim+50$ onward, with no comparable discontinuity. The mismatch is most severe in the blue, where the model $u$ and $g$ light curves fall by several magnitudes across the transition. \citet{Moriya2018c} encountered the same behavior for OGLE-73. They attributed it to the heating source being purely central. Recombination in the H-rich envelope terminates abruptly, and the photosphere then recedes rapidly.

Outward leakage of accretion energy through collimated ejecta would smooth this transition, as would outward mixing of $^{56}$Ni. Neither is included in the models. Our $u$-band limit constrains both. Any process that delays the rapid recession of the photosphere keeps the emergent spectrum blue longer. It therefore raises the predicted $u$-band flux at the epoch probed by our nondetection. Smoothing the transition enough to match the observed decline would push the models above that limit.

Second, the models do not reproduce the observed late-time colors. The $40\,M_\odot$ model is too bright at late times in every band, but by a band-dependent amount. The $60\,M_\odot$ model is simultaneously too faint in $g$ and too bright in $i$. The accretion efficiency $\eta$ scales the bolometric output without altering the emergent SED. Adjusting $\eta$ could therefore absorb a uniform luminosity offset, but not a color mismatch. The multiband comparison provides a constraint that the $r$-band light curve alone does not.

The models also fade smoothly at late times and do not reproduce the sharp decline observed from day~$\sim+325$ (Section~\ref{sec:lc:properties}). \citet{Moriya2018c} noted that energy released at the center may eventually push back the infalling material and suppress the accretion. The light curve would then fade faster than the models predict. They invoked this effect to account for a possible late decline in OGLE-73 near day~$+400$. An abrupt late fade is therefore a predicted feature of the scenario, even though the models computed here do not capture it.

Fallback accretion in this spherically symmetric, centrally heated form thus does not provide a satisfactory description of \sn. Our $u$-band limit constrains the two remedies usually invoked for the sharpness of the recombination drop. Departures from spherical symmetry remain a possible escape route. A spread of line-of-sight transition epochs might smear the drop while raising the blue flux less than a uniform smoothing would, though we cannot test this with the available models. We therefore regard fallback accretion as disfavored rather than excluded.

\label{subsec:moriya_fallback}

\subsection{Comparison with H-rich SLSNe}\label{section:discussion:SLSN-II}

Owing to its peak absolute magnitude of $M_r\approx-20.14$~mag (Table~\ref{tab:lc:properties}), \sn also belongs to the broader class of H-rich SLSNe. Most display narrow (a few hundred $\kms$) or intermediate-width ($\sim1100\kms$) H$\alpha$ profiles, generally attributed to ongoing interaction between the ejecta and dense CSM \citep{Gal-Yam2017a, Moriya2017a}. \citet{Kangas2022a} found that only a small fraction of H-rich SLSNe instead show broad Balmer lines ($>5000\kms$). \sn, with its broad and persistently regular Balmer P-Cygni profiles (Section~\ref{sec:spectroscopy:sequence}), belongs to this rarer subgroup.

\begin{figure}
\includegraphics[width=\columnwidth]{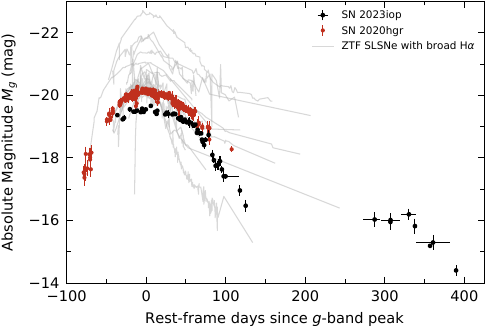}
\caption{Rest-frame absolute $g$-band light curve of \sn (black) compared with the 14 H-rich SLSNe with broad H$\alpha$ from ZTF \citep[gray;][]{Kangas2022a}, spanning $z = 0.057$--0.40. \sn is the faintest object at peak ($M_g \approx -19.8$ mag), yet it has one of the broadest light curves. Highlighted is SN\,2020hgr that is photometrically similar but spectroscopically different from \sn. Horizontal error bars on \sn reflect the width of its photometric time bins.
}
\label{fig:lc:2023iop_vs_Kangas}
\end{figure}

To place \sn in this context, we compare its rest-frame $g$-band light curve with the ZTF broad-H$\alpha$ SLSN~II sample of \citet{Kangas2022a} (Figure~\ref{fig:lc:2023iop_vs_Kangas}). Two features stand out. First, with a peak absolute magnitude of $M_{g,\rm peak}\approx-19.8$ mag, \sn is the faintest object in the comparison. Second, its light-curve width places it at the opposite extreme of the distribution. Its half-flux rise time is $t_{1/2,\rm rise}>36$ rest-frame days,\footnote{\citet{Kangas2022a} reported only the rise time from 50\% peak flux for their sample. We therefore measured the light-curve parameters of the comparison objects using the same method as for \sn.  The resulting 50\%-flux rise times agree with those reported by \citet{Kangas2022a}.} a lower limit because the $g$-band monitoring began well after explosion, later than the first $r$-band observation. The well-sampled $r$-band light curve gives $t_{1/2, \rm rise}=75$ rest-frame days (Table~\ref{tab:lc:properties}). Because the $g$- and $r$-band peaks are separated by only 12.6 rest-frame days, small compared with the rise timescale itself, the true $g$-band rise time is expected to be comparable, and thus longer than that of any object in the \citet{Kangas2022a} sample. Its half-flux decline time, $t_{1/2,\rm decline}\approx66$ rest-frame days, is roughly twice the sample median of $\sim34$ days, and essentially tied with the longest in the sample, SN\,2020yue ($\sim66.5$ days). \sn is therefore not a scaled-down version of a typical SLSN~II. 

The closest photometric analog within the broad-line sample is SN\,2020hgr \citep{Kangas2022a}, also the slowest-evolving object. Its early-time light curve resembles that of \sn near maximum brightness, but its published coverage is limited to phases around peak brightness, precluding any comparison with \sn's extended plateau and slow spectroscopic evolution (Sections~\ref{sec:spectroscopy:sequence}, \ref{section:discussion:87A:lc}). Despite their photometric similarity, the two objects differ spectroscopically. SN\,2020hgr exhibits a persistent UV excess and an anomalously weak H$\alpha$ absorption component, which led \citet{Kangas2022a} to conclude that it is an interaction-powered event. \sn shows neither signature. Its $u$-band and \swift/UVOT photometry reveal no UV excess at peak (Section~\ref{sec:lc:properties}), and its H$\alpha$ profile retains a strong, regular P-Cygni absorption trough throughout the photospheric phase (Section~\ref{sec:spectroscopy:line_evolution}). A further difference emerges in the strength of forbidden emission lines at late times. Such lines are essentially absent across the broad-line SLSN~II sample of \citet{Kangas2022a}. Even the weak forbidden emission lines reported in SN\,2020hgr remain far weaker than the well-developed [\ion{Ca}{2}]\,$\lambda\lambda7291$,7324 and [\ion{O}{1}]\,$\lambda\lambda6300$,6364 emission that strengthens in \sn from day~$+135.3$ onward (Section~\ref{sec:spectroscopy:sequence}). The single broad-line SLSN with a comparable light curve therefore appears to be powered differently from \sn.

\sn is thus doubly unusual. It belongs to the rare broad-line tail of the H-rich SLSN population, and even within that tail its combination of luminosity, rise time, and late-time spectral behavior has no close counterpart. Together with the comparison to the 87A-like population in Section~\ref{section:discussion:87A}, this places \sn at the extreme end of the 87A-like class in both luminosity and timescale. Its SLSN classification, by contrast, follows from peak luminosity alone and does not carry the interaction-powered origin that characterizes most events of that class.

\subsection{Comparison with Other SN\,1987A-like Objects}\label{section:discussion:87A}

\sn's light curve is reminiscent of the Type~II SN 1987A. To place \sn in context, we compare its light curve and spectra with those of regular 87A-like SNe and with the small set of previously identified extreme 87A-like events. As a reference, we use the volume-limited sample of 87A-like SNe from the ZTF Census of the Local Universe survey \citep{De2020a, Sit2023a}, with ZTF light curves from Miller et al. (in prep.) and spectra from \citet{Singh2019a} and \citet{Sit2023a} retrieved from \textsc{WISeREP}. We also include SN\,1987A itself, whose photometry and spectroscopy \citep{Hamuy1990a, Pun1995a} are likewise available on \textsc{WISeREP}. To extend the comparison toward the most luminous events, we add the extreme 87A-like SNe OGLE-73 \citep{Terreran2017a} and 2020faa \citep{Yang2021a, Salmaso2023a}. Although not formally an 87A-like SN, we also include the extraordinary Type~II SN iPTF14hls \citep{Arcavi2017a, Sollerman2019a} as a further point of comparison.
 
All light curves and spectra were corrected for Milky Way extinction and, where applicable, for host-galaxy attenuation. For nearby objects such as SNe~1987A and 2018hna, redshift-based distance estimates are unreliable. We instead adopt the distance moduli from \citet{Pietrzynski2019a} and \citet{Karachentsev2013a}, respectively. For each object, we derive characteristic light-curve properties following the methodology outlined in Section~\ref{sec:lc:properties}.

Recently, \citet{Hiramatsu2026a} reported a further extreme 87A-like event, SN\,2023vbw ($M_{\rm peak}\approx-19.2$~mag), which they interpret as a pair-instability explosion. It is $\sim0.9$~mag fainter at peak than \sn and radiated $3\times10^{50}$~erg, so \sn remains the most luminous and most energetic member of the class. A detailed comparison
of the two events is deferred to Skoglund et al. (in prep.) and does not affect the conclusions drawn here.

\begin{figure*}
    \centering
    \includegraphics[width=1\textwidth]{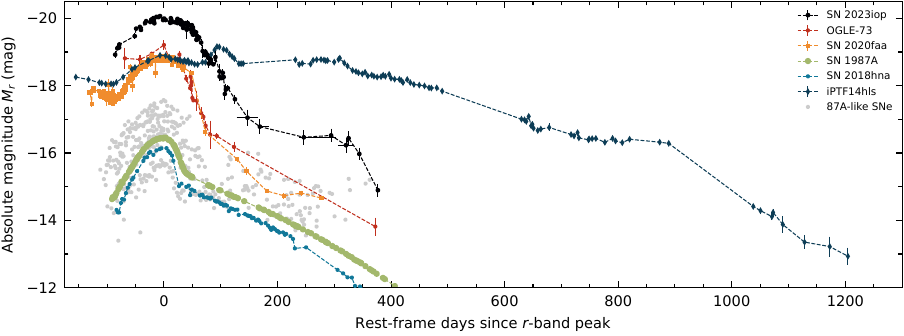}
    \caption{\sn's light curve compared with 87A-like SNe from the volume-limited ZTF Census of the Local Universe survey, together with SN~1987A and the two most extreme 87A-like events, OGLE-73 and SN~2020faa. \sn is $\sim3.7$~mag more luminous at peak than typical 87A-like SNe, and exhibits a significantly broader light curve. Even relative to the extreme 87A-like events, \sn remains $\sim$1~mag more luminous and has a broader light curve. 
    }
    \label{fig:87A-like:lc}
\end{figure*}

\begin{table*}
%\centering
\small
\caption{Light-curve parameters of the SN\,1987A-like comparison objects shown in Figure~\ref{fig:87A-like:lc}.}
\label{tab:87A:comparison}
\hspace{-2.25cm}
\begin{tabular}{lcccccccc}
\hline\hline
Object & $z$ & $\mu$ & $M_{\rm peak}$ & $(g-r)_0$ & 
       $\tau_{1/2}$ & $\tau_{1/e}$ & $\tau_{1/10}$  \\
       &     & (mag) & (mag)          & (mag)      
       & (day) & (day) & (day) \\
\hline
SN\,2018hna     & 0.0024 & $30.11$ & $-16.19$ & $0.59$ & $44.8^{+1.7}_{-6.9}$/$16.7^{+0.7}_{-0.5}$ & $55.2^{+0.2}_{-0.4}$/$19.8\pm1.2$ & $>76.8$/$178.9\pm1.2$ \\
SN\,1987A   & 0.001 & $18.48$ & $-16.45$ & $1.08$      & $45.0\pm0.4$/$30.2\pm0.3$ & $56.1\pm0.4$/$40.2^{+0.6}_{-0.5}$ & $>80.9$/$199.8\pm2.3$ &  \\
SN\,2020faa     & 0.0389 & 36.25   & $-18.88$ & $0.33$  & $59.6\pm0.2$/$58.9^{+2.0}_{-1.6}$ & $68.0\pm0.2$/$53.3^{+2.1}_{-1.9}$ & $>120.6$/$69.2^{+34.3}_{-1.7}$ &  \\
OGLE-73        & 0.1225 & 38.86   & $-19.23$ & $0.37$ & $>57$/$39.3\pm0.9$ & $>57$/$46.7\pm0.7$ & $>57$/$89.3^{+4.8}_{-3.6}$ \\
\hline
iPTF14hls      & 0.0344 & 35.97   & $-18.80$& $0.56$ & $70.5\pm0.5$/$395.8\pm1.2$ & $>142.5$/$466.9^{+39.5}_{-3.5}$ & $>142.5$/$741.6^{+7.2}_{-227.1}$\\
\hline
\textbf{\sn} & \textbf{0.1019} &  \textbf{38.43} & $\mathbf{-20.14}$ & \textbf{0.44} & $\mathbf{75.1^{+1.3}_{-1.1}}$/$\mathbf{66.4\pm0.5}$ & $\mathbf{>85.8}$/$\mathbf{77.2\pm0.7}$ & $\mathbf{>85.8}$/$\mathbf{131.7^{+3.8}_{-3.1}}$\\
\hline
\end{tabular}
\tablecomments{
All timescales are measured from the observed $R$/$r$ data and corrected for time dilation. The first and second values represent the rise and the decline timescales, respectively. The values of \sn are the same as in Table \ref{tab:lc:properties} and are reported here again for easy direct comparison.
}
\end{table*}

\subsubsection{Light Curve}\label{section:discussion:87A:lc}

Figure~\ref{fig:87A-like:lc} compares the $r$-band absolute-magnitude light curve of \sn with those of the comparison sample. With $M_r = -20.1$~mag (Table~\ref{tab:87A:comparison}), \sn is the most luminous object in the sample by a substantial margin: it exceeds the peak luminosities of OGLE-73 ($M_R = -19.2$~mag) and SN~2020faa ($M_r = -18.9$~mag), the most luminous extreme 87A-like SNe known previously, by $\sim0.9$ and $1.3$~mag, respectively. Regular 87A-like events span $M \approx -15.6$ to $-17.4$~mag, so \sn outshines objects such as SN\,2018hna ($M_r = -16.2$~mag) and SN\,1987A ($M_R \approx -16.5$~mag) by $\sim3.7$--4.0~mag, not only at peak brightness but throughout most of its evolution.

The rest-frame $(g-r)_0$ color of \sn at $r$ maximum, measured from the spectrum on day $+13.6$, is $g-r = 0.44$~mag. The same procedure yields $g-r = 0.36$~mag for SN\,2020faa ($-8$~days) and $g-r = 0.33$~mag for OGLE-73 ($-16.1$~days; Table~\ref{tab:87A:comparison}). Since H-rich SNe cool and redden after maximum brightness, these pre-maximum measurements are likely somewhat bluer than the true peak colors. For comparison, SN\,1987A had $g-r \approx 1.08$~mag at maximum light, while SN\,2018hna reached $g-r \approx 0.59$~mag ($-9.6$~days). The color of \sn thus lies between the regular and extreme 87A-like subclasses. 

The post-maximum evolution reveals a qualitative distinction between the regular and extreme 87A-like events. Regular objects decline from peak light and settle onto a linear (in magnitudes) tail consistent with the $^{56}$Co$\to^{56}$Fe decay rate of $0.98$~mag\,(100~day)$^{-1}$, as exemplified by SNe\,1987A and 2018hna. In contrast, SNe\,2020faa and 2023iop first decline gradually to $\sim$50\% of their peak flux and then drop much more rapidly. Between the epochs of $1/e$ and $1/10$ peak flux, the decline rates are $2.6$~mag\,(100~day)$^{-1}$ for \sn, $3.4$~mag\,(100~day)$^{-1}$ for OGLE-73, and $8.8$~mag\,(100~day)$^{-1}$ for SN~2020faa. No regular 87A-like SN shows such a steep drop. It resembles instead the end of the recombination phase in Type~IIP SNe \citep[e.g.,][]{Zampieri2017a}, although the preceding light curves are concave (87A-like) rather than plateau-like. After the drop, however, the light curves of \sn and SN~2020faa flatten again and fade far more slowly than $^{56}$Co decay would allow (\sn: $0.2~{\rm mag\,(100~day)^{-1}}$; Section~\ref{sec:lc:properties}).

At $\sim376$~days after peak light, the brightness of \sn dropped sharply by 3~mag\,(100~day)$^{-1}$ (Section~\ref{sec:lc:properties}), a behavior without counterpart among the regular 87A-like SNe. Whether OGLE-73 and SN\,2020faa did the same is unknown, as neither was observed at these epochs. A comparably rapid late-time drop, following an extended slow-decline phase, has been observed in the CSM-interaction-powered luminous Type~II SN\,2021irp, which steepened to $2.7$~mag\,(100~day)$^{-1}$ after $\sim250$~days \citep{Reynolds2025a}. There, the transition was attributed to either a sharp decrease in the CSM-interaction power or the onset of dust formation and reprocessing \citep{Reynolds2025a}. 

A further intriguing comparison object is the peculiar Type~II SN~iPTF14hls \citep{Arcavi2017a, Sollerman2019a}. Although not formally classified as an 87A-like event, its early photometric evolution closely resembles that of the extreme 87A-like SNe OGLE-73, 2020faa, and 2023iop (Figure~\ref{fig:87A-like:lc}). All four objects exhibit unusually broad light curves and long diffusion timescales compared to ordinary SNe~II, pointing to exceptionally massive H-rich ejecta, an additional energy source, or both. The later evolution tells a different story. While SNe\,2023iop and 2020faa, and likely OGLE-73, decline relatively rapidly after the end of their broad maximum-light phase, iPTF14hls maintained a high luminosity for several hundred days and rebrightened repeatedly \citep{Arcavi2017a, Sollerman2019a}. A possible interpretation is that the early-time similarity reflects a shared ejecta configuration, a massive H envelope with a long diffusion timescale, while a second parameter governs the late-time evolution, for instance the strength and duration of central-engine activity \citep{Dessart2018b}, fallback accretion \citep{Moriya2018c}, or CSM interaction \citep{Andrews2018a}. iPTF14hls and the extreme 87A-like events would then be neighbors in a broader family of massive H-rich explosions rather than products of a single explosion or progenitor channel.

\begin{figure*}
    \centering
    \includegraphics[width=1\textwidth]{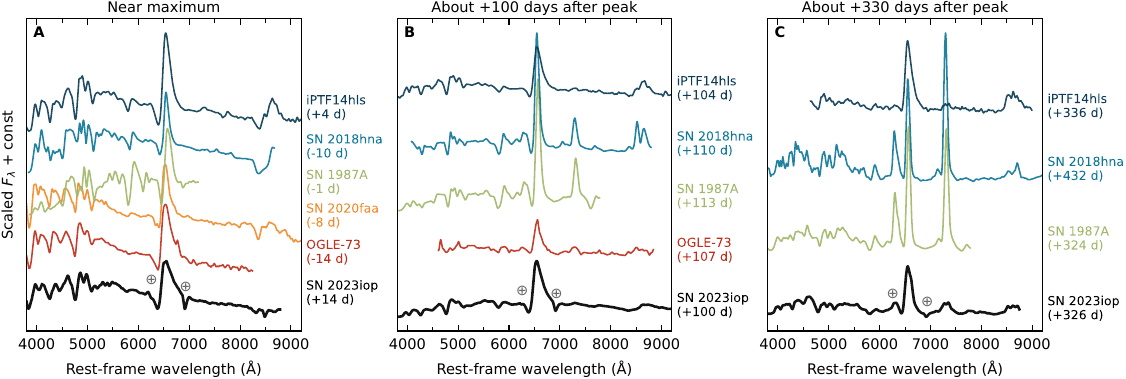}
    \caption{
    \sn's spectra compared with 87A-like SNe and the peculiar Type~II SN iPTF14hls near maximum light (panel A), $+100$ days (panel B), and $+330$ days (panel C). Spectra are smoothed with a Gaussian ($\sigma=15$~\AA) and normalized over 5500--6200~\AA, to allow determining which objects have continuum- or line-dominated spectra.
    \textit{Panel A:} At peak light, all objects resemble normal SNe~II, but the extreme 87A-like events (SNe 2020faa, OGLE-73) and iPTF14hls are bluer, show weaker metal absorption, higher velocities, and lower H$\alpha$ EWs than the regular 87A-like SNe (SNe 1987A, 2018hna). \sn broadly follows the extreme class.
    \textit{Panel B:} Differences strengthen with time. While regular 87A-like SNe develop nebular features ([\ion{Ca}{2}]\,$\lambda\lambda7291$, 7324, \ion{Ca}{2} NIR triplet) and strong H$\alpha$ emission, \sn remains largely photospheric and shows little evidence for deceleration, even in contrast with the extreme 87A-like SNe.
    \textit{Panel C:} Even at $\sim$330 days, \sn remains predominantly photospheric, with weak nebular lines ([\ion{O}{1}]\,$\lambda\lambda6300$, 6364, [\ion{Ca}{2}], \ion{Ca}{2} NIR) and 50\% weaker H$\alpha$ emission than the regular 87A-like SNe, which by this phase exhibit strong nebular spectra. iPTF14hls behaves similarly, retaining an H$\alpha$ absorption component and a well-defined emission feature near 8600~\AA. \sn's spectra were not corrected for telluric absorption lines; their location is marked by ``$\earth$.''
    }
    \label{fig:87A-like:spectra}
\end{figure*}

\subsubsection{Spectrum}\label{section:discussion:87A:spec}

Figure~\ref{fig:87A-like:spectra} compares the spectral evolution of \sn with that of regular and extreme 87A-like SNe. All spectra were scaled to the same continuum level between 5500 and 6200~\AA\ (rest frame), enabling a direct comparison of the continuum shapes and the relative strengths of the absorption and emission features. Near maximum light, all objects show the characteristic appearance of a Type~II SN, with broad Balmer P-Cygni profiles superposed on a blue continuum. Nevertheless, clear differences separate the regular and extreme members of the class. The spectra of OGLE-73, SN\,2020faa, and \sn are systematically bluer than those of SNe~1987A and 2018hna, indicating higher photospheric temperatures, reduced line blanketing by iron-group elements, or both.
 
In addition, the metal absorption features, particularly those of Fe-group species, are noticeably weaker and less distinct in the extreme 87A-like objects (Section~\ref{section:discussion:87A:metallicity}). Their spectra appear smoother, with absorption troughs that are both shallower and more heavily blended, consistent with the larger expansion velocities of the extreme events (see below). A further difference is the systematically lower H$\alpha$ absorption-to-emission ratio in OGLE-73, SN\,2020faa, and \sn compared to SNe 1987A and 2018hna. In all of these respects, the extreme events differ from their regular counterparts, and \sn sits firmly within the extreme group.
 
The clearest difference is the pace of the spectral evolution. By $\sim 100$ days after maximum light, SNe 1987A and 2018hna have already developed conspicuous nebular features, including strong [\ion{Ca}{2}] emission and a markedly strengthened H$\alpha$ emission component \citep{Pun1995a, Singh2019a}. In contrast, \sn remains largely photospheric at comparable epochs, with persistent Balmer and \ion{Fe}{2} P-Cygni profiles and only weak signs of a transition toward the nebular phase. OGLE-73 and SN\,2020faa occupy an intermediate regime, evolving more slowly than regular 87A-like SNe but more rapidly than \sn. The spectroscopic sequence mirrors the photometric one (Section~\ref{section:discussion:87A:lc}). Both point to a continuum of progressively longer diffusion and recombination timescales, plausibly set by the mass of the H envelope \citep{Zampieri2003a, Dessart2013a, Pumo2023a}.

By $\sim330$ days (Figure~\ref{fig:87A-like:spectra}, panel C), the contrast has widened further. SNe\,1987A and 2018hna are fully nebular by then, and their spectra are dominated by strong, relatively narrow emission lines with little underlying continuum. \sn instead retains a well-defined continuum with only weak [\ion{O}{1}], [\ion{Ca}{2}], and \ion{Ca}{2} NIR emission, and its H$\alpha$ emission is $\sim50\%$ weaker than in the regular 87A-like events. Nearly 1\,yr after maximum light, \sn has therefore not yet reached the state that SN\,1987A attained within $\sim100$ days. iPTF14hls evolved even more slowly, remaining largely photospheric at a comparable phase. 

\begin{figure}
    \includegraphics[width=1\columnwidth]{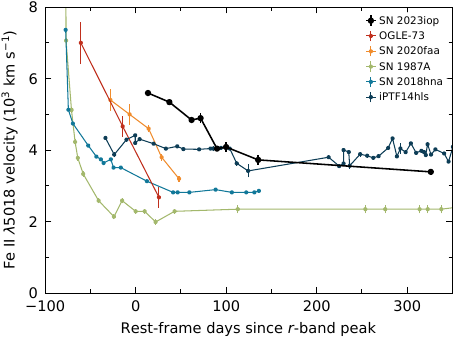}
    \caption{
    Evolution of the \ion{Fe}{2}\,$\lambda$5018 photospheric velocity of \sn compared with 87A-like SNe and iPTF14hls. The regular events (SNe 1987A and 2018hna) reach higher velocities at early phases but decelerate rapidly. The more luminous events evolve more slowly and retain higher velocities at comparable phases, consistent with more massive ejecta slowing the inward recession of the photosphere in velocity space. iPTF14hls is a clear outlier; it holds a nearly constant $\sim4000\kms$ for more than 400 days despite being far less luminous than the extreme 87A-like events, so its flat velocity evolution does not follow the same luminosity trend.
 }
 \label{fig:87A-like:velocity}
\end{figure}

The slow spectral evolution is also reflected in the velocities. Figure~\ref{fig:87A-like:velocity} shows that the photospheric velocity inferred from \ion{Fe}{2}\,$\lambda$5018 declines rapidly in the regular 87A-like SNe, whereas the evolution is substantially slower in the more luminous events. \sn exhibits the most gradual velocity evolution in the sample, maintaining comparatively high velocities at phases when SNe 1987A and 2018hna have already decelerated significantly. Similar behavior has been noted for OGLE-73 and SN\,2020faa \citep{Terreran2017a, Yang2021a}. It is expected if the extreme objects carry more massive ejecta, because the photosphere then recedes more slowly in velocity space and the nebular phase begins later \citep{Arnett1980a, Dessart2013a, Pumo2023a, Pumo2025a}.

While the velocity evolution points toward unusually massive ejecta, the spectropolarimetry constrains their geometry. The continuum polarization of \mbox{$P \approx 0.8$--1.0\%} measured on day~+37 shows that the photosphere departed significantly from spherical symmetry at an epoch when ordinary SNe~II are essentially unpolarized (Section~\ref{sec:polarimetry}).

\begin{figure}
\includegraphics[width=\columnwidth]{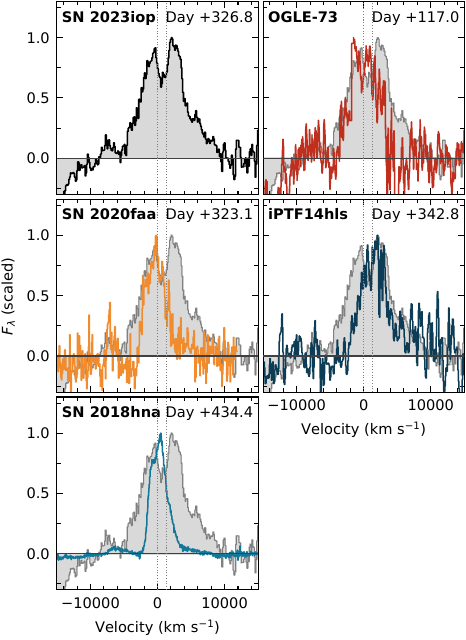}
\caption{Continuum-subtracted [\ion{Ca}{2}]\,$\lambda\lambda7291$, 7324 profiles near rest-frame day~+326, scaled to a common peak and shown in velocity space relative to 7291~\AA\ (dotted lines mark the two components of the doublet). The \sn profile (gray) is shown in every comparison panel for reference. Among the comparison objects,
only \sn shows a second maximum. 
OGLE-73 displays a broad, flat-topped morphology that may hint at similar substructure, although the S/N is insufficient to assess this robustly. SN\,2020faa and iPTF14hls exhibit a single-peaked profile, similar to the regular 87A-like SN\,2018hna. 
}
\label{fig:spec:CaII_profiles}
\end{figure}

Figure~\ref{fig:spec:CaII_profiles} compares the late-time [\ion{Ca}{2}]\,$\lambda\lambda$7291,7324 profiles of \sn\ with those of other 87A-like events and iPTF14hls. Among the comparison sample, only \sn\ exhibits a second maximum in this region, offset from the primary peak by $\sim3000\kms$.
OGLE-73 shows a broad, flat-topped profile that may qualitatively resemble that of \sn, although the S/N is insufficient to determine whether similar substructure is present. By contrast, SN\,2020faa and iPTF14hls\footnote{The red half of the line profile is affected by the telluric A band. This hinders interpreting whether the skewness of the profile is intrinsic or introduced by the telluric correction.} exhibit largely single-peaked profiles, similar to the more typical 87A-like SN\,2018hna. If the red horn in \sn\ is indeed [\ion{Ni}{2}]\,$\lambda$7378 rather than [\ion{Ca}{2}] (Section~\ref{sec:discussion:niII}), its absence in the comparison objects suggests that they did not synthesize enough stable nickel to produce a detectable line. The inferred nickel mass of a few $\times10^{-2}~M_\odot$ is reached only in the most massive pair-instability models, and all objects in the comparison sample are substantially less luminous than \sn. The appearance of the red horn exclusively in the most luminous event thus supports both the line identification and \sn's status as an outlier. 
iPTF14hls shows no analogous feature \citep[Figure~\ref{fig:spec:CaII_profiles};][]{Arcavi2017a,Sollerman2019a}, indicating that slow spectroscopic evolution and the production of detectable stable nickel are not necessarily linked.

\subsubsection{Metallicity}\label{section:discussion:87A:metallicity}

\citet{Dessart2014a} showed that the EW of \ion{Fe}{2}\,$\lambda$\,5018 can serve as a proxy for the natal iron abundance and hence for the metallicity of the progenitor at the time of its formation. To compare the absorption strengths at a common epoch, we measured the EW of \ion{Fe}{2}\,$\lambda$\,5018 at a photospheric temperature of 6000~K, adopting the temperature evolution from \citet{Arcavi2017b}, \citet{Terreran2017a}, \citet{Singh2019a}, and \citet{Salmaso2023a}, together with our measurements from Section \ref{sec:lc:bolometric}. 

The resulting rest-frame EWs define the following sequence, from lowest to highest: OGLE-73 ($6.0\pm0.5$~\AA), SN\,2018hna ($7\pm1$~\AA), \sn ($8\pm1$~\AA), SN\,1987A ($13\pm1$~\AA), and SN\,2020faa ($16\pm4$~\AA). iPTF14hls spans a broader range of $10$--$20$~\AA\ \citep[taken from][]{Arcavi2017b}, overlapping with the upper half of this sequence. All of these values are substantially lower than that of SN\,2023ixf \citep[35--40~\AA; taken from][]{Singh2024a}, which exploded in a solar-metallicity environment \citep{VanDyk2024a}. This suggests that the 87A-like events in our comparison sample span a range of metal-poor environments, with SN\,2018hna, \sn, and OGLE-73 occupying the lowest-metallicity end.

To place \sn on an absolute metallicity scale, we compiled oxygen-abundance measurements of \ion{H}{2} regions near the sites of SNe\,1987A, 2018hna, 2020faa, iPTF14hls, and OGLE-73 \citep{Mattila2010a, Arcavi2017b, Terreran2017a, Singh2019a, Yang2021a}. These studies find metallicities of $\sim0.1$--$0.6~Z_\odot$, following the same ordering as the \ion{Fe}{2}\,$\lambda$\,5018 EWs. This correspondence indicates that the progenitor of \sn formed in a markedly metal-poor environment. Section~\ref{sec:host_galaxy} presents an independent, host-mass-based metallicity estimate for \sn.

Reduced line-driven mass loss at low metallicity allows stars to retain more-massive H envelopes until core collapse \citep{Smith2014a, Vink2022a}, providing a plausible physical link between these metal-poor environments and \sn's observed properties.

\subsubsection{Host Galaxies}

\begin{figure}
    \centering
    \includegraphics[width=1\linewidth]{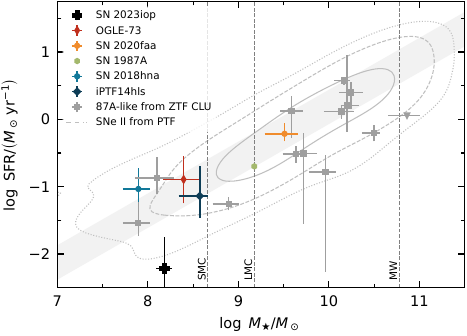}
    \caption{Host galaxy of \sn shown in the context of hosts of 87A-like SNe from the volume-limited ZTF Census of the Local Universe survey, together with SN\,1987A and the two extreme 87A-like events OGLE-73 and SN\,2020faa, and regular SN~II from the Palomar Transient Factory (66, 90, 95\%-ile contours). The gray band indicates the star-formation-rate vs. stellar-mass relation for star-forming galaxies. \sn occurred in one of the least massive (equivalent to a very low metallicity as indicated by the top axis) host galaxies and in a system with comparatively low recent star formation among those known to host this SN subclass.
    }
    \label{fig:host:mass_vs_sfr}
\end{figure}

Supernova host galaxy studies have proven to be powerful tools for constraining SN progenitor populations \citep{Lunnan2014a, Leloudas2015a, Galbany2016a, Perley2016a, Chen2017b, Taggart2021a, Pessi2023b}. To place the host of \sn into context, we compare it with hosts of 87A-like SNe from the volume-limited ZTF Census of the Local Universe survey \citep{Sit2023a} and with the extreme 87A-like events OGLE-73 and SN\,2020faa (Figure~\ref{fig:host:mass_vs_sfr}). We also compare these systems with the host-galaxy population of ordinary SNe~II from the Palomar Transient Factory study of \citet{Schulze2021a}. The hosts of regular 87A-like SNe broadly overlap with the ordinary SN~II host distribution, covering the mass range of sub-SMC \citep[$4.6\times10^8~M_\odot$;][]{McConnachie2012a} and MW-like galaxies \citep[$6\times10^{10}~M_\odot$;][]{Licquia2015a}. The extreme 87A-like events, however, occupy its low-mass tail. The host galaxy of iPTF14hls falls in the same low-mass region, considerably less massive than the typical hosts of nearby SNe~II.

With a stellar mass of only $\sim10^{8.2}\,M_\odot$, the host of \sn is among the least massive galaxies in our comparison sample. It is more than two orders of magnitude less massive than the median host of the known 87A-like population. It is also a factor of 30 less massive than the common hosts of nearby SNe~II  \citep{Schulze2021a}. Through the galaxy mass--metallicity relation \citep{Tremonti2004a}, this low mass implies a significantly lower progenitor metallicity than typically inferred for ordinary SNe~II. Those SNe predominantly occur in approximately solar-metallicity environments \citep{Schulze2021a}. This independent, mass-based estimate is consistent with the spectroscopic and nebular metallicity constraints presented in Section~\ref{section:discussion:87A:metallicity}.

Although the current sample remains small, the emerging picture is that extreme 87A-like events preferentially occur in metal-poor host galaxies, and that metallicity plays an important role in shaping their properties. As argued in Section~\ref{section:discussion:87A:metallicity}, reduced mass loss in such environments would let massive stars retain more of their H envelope until core collapse, which could explain the defining traits discussed in Sections~\ref{section:discussion:87A:lc} and \ref{section:discussion:87A:spec}, including broad light curves, slow velocity evolution, and delayed transition to the nebular phase.

Taken together, this evidence places \sn at an even more extreme end of the 87A-like parameter space than OGLE-73 and SN\,2020faa, with a higher luminosity and longer characteristic timescales than either object. However, it remains unclear whether the regular and extreme 87A-like events form a single physical sequence or merely occupy a similar region of observational parameter space. One possibility is that the 87A-like light-curve morphology can arise from multiple channels, including explosions that differ substantially in progenitor structure and powering mechanism. In this context, iPTF14hls is particularly interesting: although it shares several properties with the extreme 87A-like events during the first few hundred days, its subsequent evolution suggests that additional physical ingredients, such as prolonged central-engine activity or CSM interaction, may play an important role \citep[e.g.,][]{Andrews2018a, Wang2018a, Woosley2018a, Sollerman2019a}.

\subsection{A Multiwavelength Roadmap for Extreme 87A-like Supernovae}\label{section:discussion:roadmap}

The growing sample of extreme 87A-like events includes objects with very similar early-time photometric and spectroscopic properties but potentially different physical origins. Pair-instability explosions, central-engine activity, fallback accretion, and hidden CSM interaction can all reproduce broad light curves and long-lived photospheric spectra, so the power source cannot be inferred from optical data alone. The most discriminating diagnostics may instead arise in the UV, IR, radio, or X-rays, on timescales ranging from days after explosion (shock breakout, UV excess) to years (dust formation, CSM interaction, radio transparency, accretion-powered X-rays). Below we outline observational tests for each scenario. They are primarily intended for newly discovered extreme 87A-like events that future surveys will find, but some of them can still be applied to \sn.

The importance of such diagnostics is illustrated by \sn itself. Its luminosity plateau followed by a rapid decline would remain undetected in individual Rubin/LSST images\footnote{\url{https://rubinobservatory.org/for-scientists/rubin-101/key-numbers}} at $z \gtrsim 0.3$ (single-image depths: $g \approx 24.5$, $r \approx 24.0$, $i \approx 23.4$, $z \approx 22.9$~mag), or in \textit{James Webb Space Telescope} (\jwst) transient surveys beyond $z \gtrsim 2$ at COSMOS-like depths \citep[$m_{\rm lim}\approx28$~mag;][]{Fox2026a} and $z \gtrsim 6$ even in JADES \citep[$m_{\rm lim}\approx30$~mag;][]{DeCoursey2025a}. At such distances, \sn would be classified as a PISN candidate from its peak luminosity, rise time, and early spectra alone. If it is not a PISN, such misclassifications would bias inferred PISN rates across cosmic time \citep{Briel2024a,Gabrielli2024a}, with ramifications for massive-star and galaxy evolution \citep{Heger2002a,Heger2003a} and the interpretation of the LIGO black-hole mass function \citep{LIGO2026a, Tong2026a}.

\paragraph{(i) PISN Diagnostics~}~
PISNe are most directly identified through their exceptionally large iron-group yields, best probed in the nebular phase. The thermonuclear disruption of very massive helium cores synthesizes up to tens of solar masses of $^{56}$Ni, and once the ejecta turn optically thin the spectra should be dominated by neutral and low-ionization emission from intermediate-mass elements (\ion{Si}{1}, \ion{S}{1}, \ion{Ca}{1}--{\small II}) and iron-group elements ([\ion{Fe}{2}--{\small III}], [\ion{Co}{2}], [\ion{Ni}{2}]), whose integrated luminosities trace the iron-group mass, its distribution, and mixing \citep{Jerkstrand2016a}. At $\gtrsim1~\rm yr$, cooling shifts to the NIR, producing prominent 2--5~$\mu$m features \citep{Jerkstrand2016a} analogous to those seen in Type Ia SNe \citep{Kwok2023a, Blondin2023a}. \jwst/NIRSpec \citep[1--5~$\mu$m;][]{Jakobsen2022a} and MIRI \citep[5--28~$\mu$m;][]{Wells2015a} cover these features over a wide redshift range. Since core-collapse explosions yield $\ll0.5~M_\odot$ of iron-group material and only rarely $\sim 1~M_\odot$ \citep{Anderson2019a, Taddia2019a, Srinivasaragavan2024a}, the strength of these features cleanly separates the two channels. Their absence would argue against a PISN origin.

Dust provides a complementary test. PISN models predict $\sim1$--$30~M_\odot$ of dust ($\sim15$--30\% of the progenitor mass), mainly silicates and oxides, forming within a few years \citep{Schneider2004a}. These predicted dust masses exceed those inferred for nearby core-collapse SNe, typically $\sim0.01$--$1~M_\odot$, by one to several orders of magnitude \citep{Matsuura2011a, Szalai2019a, Shahbandeh2023a, Subrayan2026a}. The predicted yields are upper limits and are sensitive to mixing, molecule formation, and reverse-shock destruction. Nevertheless, dust masses $\gtrsim1\,M_\odot$ at $\gtrsim100~\rm K$ are readily detectable with MIRI in the not-so-distant Universe, providing an unambiguous discriminator against core-collapse or magnetar-powered explosions \citep{Omand2019a, Gomez2024b, Chen2026b}.

Early-time UV observations offer another test. Radiation-hydrodynamic calculations predict exceptionally luminous shock-breakout flashes from H-rich PISNe \citep[for a review, see][]{Katz2012a, Kasen2011a, Chen2014a}. These flashes peak in the extreme UV and soft X-rays, reach luminosities of $\sim10^{45}$--$10^{46}~\rm erg~s^{-1}$, and last for several hours in the progenitor rest frame. They are substantially more luminous than the shock-breakout emission expected from ordinary core-collapse SNe. High-cadence UV facilities such as ULTRASAT \citep{Shvartzvald2024a} and UVEX \citep{Kulkarni2011a} could therefore directly test a pair-instability origin.

\paragraph{(ii) Magnetar Diagnostics~}~
Whereas the PISN scenario is primarily constrained through its nucleosynthetic products and radioactive-energy deposition, models powered by a compact remnant are more naturally tested through signatures of prolonged energy injection into the ejecta. Sustained magnetar input maintains high ionization in the inner ejecta and enhances the excitation of Ne, Ar, and S, whose strongest transitions lie in the IR. The [\ion{Ne}{2}]\,12.8\,$\mu$m fine-structure line is predicted to be the brightest MIR feature, with $L \approx 10^{39}$--$10^{40}\,\rm erg\,s^{-1}$ in magnetar-powered SNe, and its strength is set by the magnetic field, spin-down timescale, and ejecta mass, making it a direct probe of the central engine \citep{Omand2023a, Dessart2024a, Omand2026a}. It is not expected to be prominent in purely radioactive explosions, where ionization is lower, and energy deposition is confined to iron-group-rich regions \citep{Jerkstrand2016a}. MIRI Medium Resolution Spectroscopy \citep[MRS;][]{Wells2015a} is uniquely suited to detect this line in nearby PISN-like transients. However, MIRI/MRS is split into four channels, and Channel 4 ($\lambda > 17.98\,\mu\rm m$) is significantly less sensitive than Channels 1--3; redshifting [\ion{Ne}{2}]\,12.8\,$\mu\rm m$ into Channel 4 imposes an effective redshift limit of $z \approx 0.4$ for this diagnostic method. Combined with NIRSpec and optical nebular data, its detection or absence discriminates between magnetar, radioactive, and interaction powering.

A magnetar also predicts a secondary, engine-driven shock breakout. The inflated high-pressure bubble drives a radiative shock through the ejecta that emerges days after explosion as a UV/optical transient with luminosities of $\sim10^{43}$--$10^{44}~\rm erg~s^{-1}$ and durations of several days \citep{Kasen2016a, Chen2026a}. It may appear as a distinct first peak or as a short-lived UV excess preceding the main light-curve maximum. High-cadence UV monitoring is therefore among the most direct probes of central-engine activity.

\paragraph{(iii) Fallback Accretion Diagnostics~}~
Fallback accretion resembles magnetar-powered models but distinctively predicts long-lived emission from the accretion flow, whose luminosity tracks $\dot{M} \propto t^{-5/3}$ \citep{Chevalier1989a} or an advection-dominated plateau \citep{Dexter2013a} independent of the environment. An interaction luminosity instead follows the CSM density profile. A steady wind ($\rho_{\rm csm} \propto r^{-2}$) gives $L_{\rm X} \propto t^{-1}$ \citep{Immler2006a}, whereas shells and density cutoffs imprint abrupt rebrightenings or drops \citep{Chevalier1994a, Dwarkadas2005a, vanMarle2010a, Margutti2017a}.

Detecting the accretion flow is difficult. The engine lies beneath the ejecta, where its emission is absorbed by the K-shell electrons of the metals \citep{Alp2018a}. Hydrogen and helium contribute little at these energies, and a fallback explosion ejects little iron, so the opacity per unit mass falls well below that of the metal-rich compositions for which it is usually computed. Scaling the relation of \citet{Andreoni2022a} to our ejecta masses therefore overestimates the optical depth. Even so, only emission above $\sim10$--30~keV escapes within the first few years, while the soft band stays opaque for decades. Because an advection-dominated flow carries most of its energy inward rather than radiating it, fallback models predict only $L_{\rm X}\approx10^{39}~\rm erg\,s^{-1}$ \citep{Dexter2013a, Moriya2018c}, so even a \nustar-class detection would require an event within $\sim10$~Mpc. An external shock can also be heavily absorbed early, but by its own CSM, whose column falls rapidly as the shock overruns it \citep{Chandra2015a, Grefenstette2023a}. Soft X-rays therefore emerge within months for an interacting SN and only after decades for a buried engine.

Radio observations are more promising, though the same competition applies. Free-free absorption scales steeply with frequency and time ($\tau_{\rm ff}\propto\nu^{-2.1}\,t^{-5}$; \citealt{Mezger1967a, Murase2016a, Margalit2018a, Omand2018a}), so the dense, still-ionized ejecta clear from high frequencies downward. For $M_{\rm ej}=12$--$60~M_\odot$, they become transparent after $\approx2$--4~yr in the ALMA bands \citep[84--950~GHz;][]{Wootten2009a} but only after one to two decades at centimeter wavelengths. Early ALMA monitoring is the only route to catching an engine as it emerges, at the cost of observing where a synchrotron spectrum has already declined. For a source as luminous as PTF10hgi \citep[$L_\nu \lesssim 10^{28}~\rm erg\,s^{-1}\,Hz^{-1}$ at 6~GHz;][]{Eftekhari2019a, Law2019a} with a spectrum falling as $\nu^{-0.7}$, a deep observation near the transparency epoch reaches $\sim100$~Mpc, an order of magnitude beyond the hard X-ray test, and further if the self-absorption turnover lies in the millimeter bands. Centimeter facilities instead provide population coverage. The DSA-2000 \citep{Hallinan2019a} will image the sky north of $\delta = -30^\circ$ every four months at $\sim2~\mu\rm Jy\,beam^{-1}$ RMS (0.7--2~GHz), and the ngVLA \citep[$\lesssim0.4~\mu\rm Jy\,beam^{-1}$ per hour at $\sim8$~GHz;][]{Selina2018a} adds multifrequency coverage of the turnover.

Radio emission is less specific, however. Its brightness depends on uncertain jet energetics and microphysics, and CSM interaction \citep{Chevalier1982a,Chevalier1994a,Chevalier2017a} or a magnetar-born pulsar-wind nebula \citep{Omand2018a,Margalit2018a} can produce comparable luminosities. The spectral turnover helps, since a buried source clears from high frequencies downward whereas an external shock is limited by synchrotron self-absorption \citep{Chevalier1982b}. Distinguishing the two engines (i.e., black hole and magnetar) is harder, as both lie behind the same screen. A pulsar-wind nebula is intrinsically radio-bright, radiating from radio to $\gamma$-rays, whereas an advection-dominated flow radiates poorly and emits in the radio only if it launches a jet. A firmer separation may come from the optical light curve, through an abrupt accretion turn-off \citep{Dexter2013a, Moriya2018c} or periodic modulation from a companion, as in SN\,2022jli \citep{Chen2024a}.

\paragraph{(iv) CSM Diagnostics~}~
CSM interaction converts ejecta kinetic energy into radiation at external shocks and can in principle supply both the high peak luminosity and the late-time flux excess. The difficulty lies in its concealment. If the CSM is confined and rapidly engulfed, or the interaction proceeds in a buried disk or torus, the narrow emission lines that define SNe~IIn are suppressed, and the light curve alone cannot separate interaction from radioactive or engine powering \citep{Smith2017a, Nyholm2020a, Salmaso2023a}. Hidden interaction nevertheless leaves subtler imprints. These include narrow ($\sim10^{2}\kms$) Balmer features from the unshocked CSM, intermediate-width ($\sim10^{3}\kms$) components with electron-scattering wings \citep{Chugai2001a, Huang2018a, Fransson2014a}, boxy flat-topped late-time H$\alpha$ from a cold dense shell \citep[e.g., iPTF14hls;][]{Andrews2018a}, and weaker P-Cygni absorption troughs, which occur when continuum light from outside the line-forming region dilutes the absorption feature beneath it \citep{Branch2000a}. Detecting or excluding these signatures requires $R \gtrsim 5000$ ($\lesssim60\kms$) and high S/N, since the features are weak perturbations on a bright continuum.

Interaction also reshapes the SED. The shock sustains a hot continuum that resists the reddening of a recombination-cooled photosphere and produces a persistent blue and UV excess. \citet{Kangas2022a} identified such excesses in some H-rich SLSNe, most notably SN\,2020hgr, whose otherwise ordinary SN~II spectra show an extremely weak H$\alpha$ absorption component. Together, the UV excess and the diluted absorption identify it as a primarily interaction-powered event despite the absence of narrow SN~IIn-like lines. UV and $u$-band photometry around maximum brightness, combined with the strength of the H$\alpha$ trough, thus tests for hidden interaction even when optical spectra appear unremarkable. This diagnostic also separates interaction from engine powering. An internal engine heats the ejecta below the line-forming layer, so the P-Cygni absorption remains strong, whereas external interaction dilutes the absorption and imprints the CSM velocity field on the line profiles. If optically thick CSM dominates the luminosity, the shock breakout itself is also modified, with a larger breakout radius, a lower effective temperature, and UV emission that lasts days rather than hours and evolves more slowly in color \citep{Chevalier2017a,Nakar2010a,Sapir2011a,Zimmerman2024a}.

For \sn, these CSM tests return negative results. The $u$-band and \swift/UVOT photometry around maximum light show no UV excess (Section~\ref{sec:lc:properties}), and the high-resolution, high-S/N spectra reveal strong, regular H$\alpha$ P-Cygni absorption (Figure~\ref{fig:2023iop:spec:sequence_opt}), in marked contrast to SN\,2020hgr, with no narrow or intermediate-width features at any epoch. We find no evidence that CSM interaction dominates during the epochs probed by our optical and UV data.  Any interaction scenario for the late-time excess must therefore operate without producing narrow lines or absorption filling. Radio and X-ray observations constrain extended, wind-like material, but a shock confined within the ejecta is screened as effectively as an embedded engine, so nondetections do not exclude it.

Several of these diagnostics will soon be available for \sn. Approved \jwst\ observations \citep{Schulze2026a} will obtain deep MIRI imaging and spectroscopy, directly testing the pair-instability (dust formation) and magnetar ([\ion{Ne}{2}] line) scenarios. In X-rays, our upper limit on the unabsorbed flux ($<1.2\times10^{-13}~{\rm erg\,cm^{-2}\,s^{-1}}$, 0.3--10~keV; corresponding to $L_{\rm X}<4\times10^{42}~{\rm erg\,s^{-1}}$ at $z=0.1020$; Section~\ref{sec:multiwavelength_limits}) is more than three orders of magnitude above the fiducial fallback-accretion luminosity discussed above, so this test remains inconclusive. Moreover, the ejecta remain optically thick to bound-free absorption across this band at the epoch observed (Section~\ref{sec:multiwavelength_limits}), so a nondetection is expected regardless of depth.

In the radio, our 15.5~GHz limit ($L_\nu < 4.4 \times 10^{28}~\rm erg\,s^{-1}\,Hz^{-1}$; Section~\ref{sec:multiwavelength_limits}) lies a factor of $\sim4$ above the PTF10hgi detection \citep{Eftekhari2019a}. Given the possibly large ejecta masses and low expansion velocities of extreme 87A-like events, however, the free-free optical depth at centimeter wavelengths remains high for one to two decades \citep{Mezger1967a,Murase2016a,Margalit2018a,Omand2018a}, so a deeper observation at this frequency would not test an embedded engine. The higher frequencies accessible to ALMA are far less affected (see above) and become transparent within a few years of explosion. \sn may now be entering the epoch at which such a search becomes feasible. Centimeter monitoring with the VLA and, eventually, systematic coverage with DSA-2000 will supply the long-term light curve once those frequencies clear.

\section{Conclusions}\label{sec:conclusion}

We have presented optical, NIR, UV, radio, X-ray, and spectropolarimetric observations of \sn, a luminous H-rich SN at $z=0.1020$. Its broad light curve, slow spectral evolution, and delayed transition toward the nebular phase make it the most extreme 87A-like supernova known. Our main conclusions are as follows.

\begin{itemize}[leftmargin=*, nosep, topsep=0.5em]
\item \sn is the most extreme 87A-like transient currently known. It reached a peak luminosity of $M_r \approx -20.1$ mag, $\geq0.9$~mag brighter than OGLE-73 and SN\,2020faa, and exhibited an exceptionally broad light curve with a rise time $\gtrsim88$~rest-frame days.
\item \sn's peak luminosity also places it among the H-rich SLSNe. Broad Balmer lines are rare in that class, and \sn is one of the few such objects. It is the faintest of them at peak but has one of the broadest light curves, and its well-developed late-time forbidden lines have no counterpart among them. It does not show the interaction signatures that characterize most H-rich SLSNe.
\item Despite its extraordinary luminosity, the optical spectra closely resemble those of regular Type~II SNe. However, the spectral evolution is much slower than in both normal SNe~II and classical 87A-like events, with strong photospheric features remaining visible several hundred days after maximum light.
\item The combination of broad light curves, slow velocity evolution, delayed nebular transition, and host-galaxy properties points toward an unusually massive H-rich envelope as a key ingredient of \sn and, possibly, the broader population of extreme 87A-like transients.
\item The light-curve rise of $\gtrsim88$~rest-frame days implies an ejecta mass of $M_{\rm ej}>12~M_\odot$. 
The peak luminosity requires $\approx2.2~M_\odot$ of $^{56}$Ni, compatible with the tail-based upper limit and with the yield from the joint interaction fit. These numbers exceed the yields attainable in neutrino-driven explosions, but both the
peak- and tail-based values assume radioactive decay dominates the luminosity, and are therefore upper limits. 
\item An emission feature emerging near 7300~\AA\ after day~+140 is tentatively identified as [\ion{Ni}{2}]\,$\lambda$7378 from stable $^{58}$Ni. The implied mass of $\approx0.02~M_\odot$ is reached only in the most massive PISN models and exceeds
ordinary core-collapse supernova yields by an order of magnitude. For the $^{58}$Ni/$^{56}$Ni ratios of $\approx0.01$--0.02 predicted by these models, it corresponds to $\approx1$--$2~M_\odot$ of $^{56}$Ni, consistent with the light curve modeling.
\item The continuum polarization of 0.8--1.0\% measured on day~+37.3 indicates a substantial departure from spherical symmetry at the photospheric phase.
\item The host galaxy is a low-mass dwarf system, placing \sn within the metal-poor tail of the SN~II host-galaxy population. Together with OGLE-73, SN\,2020faa, and iPTF14hls, this suggests that metallicity may play a role in producing the most slowly evolving H-rich explosions. A metal-poor environment is also a prerequisite for the pair-instability channel \citep{Langer2007a}, since retaining a helium core of $\gtrsim65~M_\odot$ requires weak line-driven winds.
\item Existing H-rich pair-instability models reproduce much of the early light-curve evolution but not the prolonged photospheric phase or the late-time luminosity, whose excess is most pronounced in the blue. These models are computed without any interaction between the ejecta and the material their progenitors are expected to shed. Hidden, geometrically confined interaction could supply the missing energy without producing the narrow lines that are absent from the spectra at every epoch. The late-time mismatch therefore does not by itself exclude a pair-instability origin.
\item Fallback accretion reproduces the broad light-curve morphology but does not match late-time colors, and our $u$-band limit constrains the modifications that would otherwise smooth its recombination transition. A light echo is excluded by the smooth spectroscopic evolution.
\item More broadly, \sn demonstrates that markedly different physical mechanisms can produce similar optical observables. The extreme 87A-like events therefore need not share a common powering mechanism. Pair-instability explosions, hidden circumstellar interaction, and central-engine activity may all occur within the population, and \sn itself appears to require more than one of them.
\item The features that identify \sn as unusual are the first to be lost with distance. Its late-time plateau and subsequent rapid decline lie below the single-image depth of Rubin/LSST beyond $z\approx0.3$, and below \jwst\ survey limits beyond $z\approx2$--6, leaving only the peak luminosity, rise time, and early spectra, from which \sn would be classified as a PISN candidate without reservation. Yet it is the late-time data that reveal a plateau no powering mechanism reproduces and that require an additional energy source. Distant PISN candidates will therefore appear cleaner than they are, biasing the PISN rate inferred across cosmic time and the constraints that rate places on massive-star evolution and on the black-hole mass function.
\end{itemize}

The high $^{56}$Ni mass, the large ejecta mass, and the metal-poor host each point toward a very massive progenitor, and the tentative detection of stable nickel strengthens that case. We consider a pair-instability explosion the most plausible interpretation of \sn. In that case, the late-time excess must come from an additional energy source, most likely circumstellar interaction. The identification of the nickel line rests on a single feature, however, and needs independent confirmation. It also remains unclear whether the extreme 87A-like events are a continuous extension of the classical population or a distinct class of H-rich explosions. Answering these questions requires a larger sample with comparably comprehensive coverage and the multiwavelength diagnostics outlined in Section~\ref{section:discussion:roadmap}: early UV coverage of the shock breakout, MIR searches for dust and fine-structure lines, and radio and X-ray monitoring of the putative engine. For \sn itself, the approved \jwst\ observations will provide the first of these tests.

\begin{acknowledgments}

We thank Irene Salmaso for sharing the spectra of SN\,2020faa, and Yuxin Dong and Alexa C. Gordon for performing some of the Keck observations.

\indent K.C. acknowledges the support from the National Science and Technology Council, Taiwan, under grant No. NSTC 115-2112-M-001-001- and the Academia Sinica, Taiwan, under a career development award under grant No. AS-CDA-111-M04.

\indent M.W.C. acknowledges support from the National Science Foundation with grant numbers PHY-2117997, PHY-2308862 and PHY-2409481.

\indent P.C. acknowledges support from the Zhejiang Provincial Top-Level Research Support Program.

\indent T.-W.C. acknowledges financial support from the Yushan Fellow Program of the Ministry of Education, Taiwan (MOE-111-YSFMS-0008-001-P1), and from the National Science and Technology Council, Taiwan (NSTC grant 114-2112-M-008-021-MY3).

\indent The research group of A.V.F. at U.C. Berkeley received financial assistance from the Christopher R. Redlich Fund, as well as donations from Gary and Cynthia Bengier, Clark and Sharon Winslow, Alan Eustace and Kathy Kwan, Timothy and Melissa Draper, Briggs and Kathleen Wood, Sanford Robertson, Ellyn and Alan Seelenfreund (W.Z. is a Bengier-Winslow-Eustace Specialist in Astronomy, T.G.B. is a Draper-Wood-Seelenfreund Specialist in Astronomy, Y.Y. was a Bengier-Winslow-Robertson Fellow in Astronomy),
and numerous other donors.

\indent A.G-Y's research is supported by ISF, IMOS and BSF grants, as well as the André Deloro Institute for Space and Optics Research, the Center for Experimental Physics, a WIS-MIT Sagol grant, the Norman E Alexander Family M Foundation ULTRASAT Data Center Fund, and Magnus Konow in honor of his mother Olga Konow Rappaport;  A.G-Y is the incumbent of the The Arlyn Imberman Professorial Chair.

\indent J.I.G.H. acknowledges financial support from the Spanish Ministry of Science, Innovation and Universities (MICIU) project PID2023-149982NB-I00.

\indent K.H. would like to thank the Caltech Presidential Postdoctoral Fellowship program and President Rosenbaum.

\indent A.J. and A.S. acknowledge support from the Knut and Alice Wallenberg Foundation through project ``Gravity Meets Light'', and the Swedish National Research Council through grants 2018-03799 and 2026-05262.

\indent T.K. acknowledges support from the Research Council of Finland project 360274.

\indent C.L. is supported by DoE award \#\,DE-SC0025599.

\indent K.M. acknowledges support from JSPS KAKENHI grant No. JP24KK0070, JP24K00682, and JP23H04894. 

\indent F.P. acknowledges support from the Spanish Ministerio de Ciencia, Innovaci\'on y Universidades (MICINN) under grant number PID2022-141915NB-C21.

\indent M.P. acknowledges support from a UK Research and Innovation Fellowship (UKRI1062).

\indent A.R. acknowledges INAF project Supporto Arizona \& Italia.

\indent N.S is supported by the Kavli foundation.

\indent Y.Y.'s research is partially supported by the Tsinghua University Dushi Program.

Funded by the European Union (ERC, project number 101042299, TransPIre). Views and opinions expressed are however those of the author(s) only and do not necessarily reflect those of the European Union or the European Research Council Executive Agency. Neither the European Union nor the granting authority can be held responsible for them.

% ZTF
Based on observations obtained with the Samuel Oschin Telescope 48-inch and the 60-inch Telescope at the Palomar Observatory as part of the Zwicky Transient Facility project. ZTF is supported by the National Science Foundation under Grants No. AST-1440341, AST-2034437, and currently Award \#2407588. ZTF receives additional funding from the ZTF partnership. Current members include Caltech, USA; Caltech/IPAC, USA; University of Maryland, USA; University of California, Berkeley, USA; Cornell University, USA; Drexel University, USA; University of North Carolina at Chapel Hill, USA; Institute of Science and Technology, Austria; National Central University, Taiwan; German Centre for Astrophysics, Germany; and OKC, University of Stockholm, Sweden. Operations are conducted by Caltech's Optical Observatory (COO), Caltech/IPAC, and the University of Washington at Seattle, USA.

% ZTF FP
The ZTF forced-photometry service was funded under the Heising-Simons Foundation grant \#12540303 (PI: Graham).

% Fritz
The Gordon and Betty Moore Foundation, through both the Data-Driven Investigator Program and a dedicated grant, provided critical funding for SkyPortal.

% Keck
The authors wish to recognise and acknowledge the very significant cultural role and reverence that the summit of Maunakea has always had within the Native Hawaiian community. We are most fortunate to have the opportunity to conduct observations from this mountain. Some of the data presented herein were obtained at Keck Observatory, which is a private 501(c)3 non-profit organization operated as a scientific partnership among the California Institute of Technology, the University of California, and the National Aeronautics and Space Administration. The Observatory was made possible by the generous financial support of the W. M. Keck Foundation.

% ATLAS
This work has made use of data from the Asteroid Terrestrial-impact Last Alert System (ATLAS) project. The ATLAS project is primarily funded to search for near earth asteroids through NASA grants NN12AR55G, 80NSSC18K0284, and 80NSSC18K1575; byproducts of the NEO search include images and catalogs from the survey area. This work was partially funded by Kepler/K2 grant J1944/80NSSC19K0112 and HST GO-15889, and STFC grants ST/T000198/1 and ST/S006109/1. The ATLAS science products have been made possible through the contributions of the University of Hawaii Institute for Astronomy, the Queen's University Belfast, the Space Telescope Science Institute, the South African Astronomical Observatory, and The Millennium Institute of Astrophysics (MAS), Chile.

% CIERA
\indent W. M. Keck Observatory, MMT Observatory, and Zwicky Transient Facility access was supported by Northwestern University and the Center for Interdisciplinary Exploration and Research in Astrophysics (CIERA).

% ESO
\indent Based in part on observations made with ESO Telescopes at the La Silla Paranal Observatory under program IDs 111.24ME, 112.26W6, 113.26U4, and 115.281K.

% LBT
The LBT is an international collaboration among institutions in the United States and Europe. At the time data were acquired for this research, LBT Corporation Members were the University of Arizona on behalf of the Arizona Board of Regents; Istituto Nazionale di Astrofisica, Italy; LBT Beteiligungsgesellschaft, Germany, representing the Max-Planck Society, the Leibniz Institute for Astrophysics Potsdam, and Heidelberg University; and The Ohio State University, representing The Ohio State University, University of Notre Dame, University of Minnesota, and University of Virginia.

% LT
The Liverpool Telescope is operated on the island of La Palma by Liverpool John Moores University in the Spanish Observatorio del Roque de los Muchachos of the Instituto de Astrofisica de Canarias with financial support from the UK Science and Technology Facilities Council.

% NOT
\indent Based in part on observations made with the Nordic Optical Telescope, owned in collaboration by the University of Turku and Aarhus University, and operated jointly by Aarhus University, the University of Turku, and the University of Oslo, representing Denmark, Finland, and Norway (respectively), the University of Iceland, and Stockholm University at the Observatorio del Roque de los Muchachos, La Palma, Spain, of the Instituto de Astrofisica de Canarias. The NOT data were obtained under program IDs 67-001, 67-415, 68-008, and 68-202.

% Swift
This work made use of data supplied by the UK \swift Science Data Centre at the University of Leicester.

% GALEX
Some of the data presented in this paper were obtained from the Mikulski Archive for Space Telescopes (MAST) at the Space Telescope Science Institute. STScI is operated by the Association of Universities for Research in Astronomy, Inc., under NASA contract NAS5-26555.

% PS1
The Pan-STARRS1 Surveys (PS1) and the PS1 public science archive have been made possible through contributions by the Institute for Astronomy, the University of Hawaii, the Pan-STARRS Project Office, the Max-Planck Society and its participating institutes, the Max Planck Institute for Astronomy, Heidelberg and the Max Planck Institute for Extraterrestrial Physics, Garching, The Johns Hopkins University, Durham University, the University of Edinburgh, the Queen's University Belfast, the Harvard-Smithsonian Center for Astrophysics, the Las Cumbres Observatory Global Telescope Network Incorporated, the National Central University of Taiwan, the Space Telescope Science Institute, the National Aeronautics and Space Administration under Grant No. NNX08AR22G issued through the Planetary Science Division of the NASA Science Mission Directorate, the National Science Foundation Grant No. AST-1238877, the University of Maryland, Eotvos Lorand University (ELTE), the Los Alamos National Laboratory, and the Gordon and Betty Moore Foundation.

% SDSS
Funding for the SDSS and SDSS-II has been provided by the Alfred P. Sloan Foundation, the Participating Institutions, the National Science Foundation, the U.S. Department of Energy, the National Aeronautics and Space Administration, the Japanese Monbukagakusho, the Max Planck Society, and the Higher Education Funding Council for England. The SDSS Web Site is http://www.sdss.org/. The SDSS is managed by the Astrophysical Research Consortium for the Participating Institutions. The Participating Institutions are the American Museum of Natural History, Astrophysical Institute Potsdam, University of Basel, University of Cambridge, Case Western Reserve University, University of Chicago, Drexel University, Fermilab, the Institute for Advanced Study, the Japan Participation Group, Johns Hopkins University, the Joint Institute for Nuclear Astrophysics, the Kavli Institute for Particle Astrophysics and Cosmology, the Korean Scientist Group, the Chinese Academy of Sciences (LAMOST), Los Alamos National Laboratory, the Max-Planck-Institute for Astronomy (MPIA), the Max-Planck-Institute for Astrophysics (MPA), New Mexico State University, Ohio State University, University of Pittsburgh, University of Portsmouth, Princeton University, the United States Naval Observatory, and the University of Washington.

% SEDM
SED Machine is based upon work supported by the National Science Foundation under Grant No. 1106171.

\end{acknowledgments}

\facilities{
AMI-LA, 
ATLAS,
Magellan:Baade (IMACS),
Magellan:Clay (MagE),
Keck:I (LRIS),
Keck:II (DEIMOS),
LBT (MODS, LUCI),
Liverpool:2m (IO:O),
MMT (Binospec, MMIRS),
NEOWISE,
NOT (ALFOSC, NOTCam),
PO:1.5m (SEDm),
VLT:Antu (FORS2),
VLT:Kueyen (X-shooter), 
Swift (UVOT, XRT),
WISE,
ZTF
}

\bibliographystyle{aasjournalv7}

\appendix

\section{Observations and Data Reduction}\label{appendix:observations}
\subsection{Photometry}\label{obs:photometry}
\subsubsection{Observations with All-sky Surveys}\label{sec:obs:surveys}

\paragraph{\textbf{Asteroid Terrestrial-impact Last Alert System (ATLAS)}~}~ATLAS is a network of robotic 0.5\,m $f/2$ Wright-Schmidt telescopes designed for rapid, all-sky monitoring \citep{Tonry2018b}. During the period relevant to this work, ATLAS operated sites at Haleakal\={a} and Mauna Loa (Hawaii, USA), Sutherland (South Africa), and El Sauce (Chile), providing coverage of both hemispheres.

Each ATLAS unit has a field of view of $\sim30~\mathrm{deg}^2$. Observations are obtained using two custom broad optical filters: cyan ($c$; $\sim4200$--6500\,\AA) and orange ($o$; $\sim5600$--8200\,\AA; \citealt{Tonry2018b}). Individual exposures are 30\,s in duration, typically reaching a $5\sigma$ limits of $\sim 19.5$--19.8\,mag (AB) under dark conditions. The ATLAS survey strategy prioritises cadence over depth. Each field is generally observed multiple times per night, with revisit times on the order of 30--60\,min, and the full accessible sky is typically covered every 1--2 nights \citep{Tonry2018b}.

ATLAS data are processed using an automated reduction and image-subtraction pipeline, including photometric and astrometric calibration against external reference catalogues, described by \citet{Stalder2017a}. We ran the ATLAS Forced Photometry service\footnote{\href{https://fallingstar-data.com/forcedphot}{https://fallingstar-data.com/forcedphot}} \citep{Shingles2021a} on all relevant images. The data were clipped and binned using the \program{Python} script \program{plot\_atlas\_fp.py}\footnote{\url{https://gist.github.com/thespacedoctor/86777fa5a9567b7939e8d84fd8cf6a76}}. We used a bin size of 2 days for all data. Measurements with a significance of $<3\sigma$ were removed from the final light curve.

\paragraph{\textbf{40\,cm Wide-field Infrared Survey Explorer}~}~
The field of \sn was also observed during the {\it Near-Earth Object Wide-field Infrared Survey Explorer (NEOWISE)} all-sky MIR survey, the reactivation mission of the {\it WISE} space telescope, in the $W1$ (3.4 $\mu$m) and $W2$ (4.6 $\mu$m) bands \citep{Wright2010a, Mainzer2014a}. We retrieved time-resolved coadded images of the field created as part of the unWISE project \citep{Lang2014a, Meisner2018a}.

\paragraph{\textbf{Zwicky Transient Facility (ZTF)}~}~ZTF is a wide-field optical survey operating on the 1.2\,m Samuel Oschin Schmidt Telescope at the Palomar Observatory (California, USA). ZTF is operated as a public-private partnership, combining a public time-domain survey with a suite of private and collaboration-led programs. The telescope is equipped with a 600-megapixel mosaic camera covering $47~\mathrm{deg}^2$ \citep{Dekany2020a}, enabling rapid synoptic coverage of the northern sky \citep{Bellm2019a}.

Observations were performed in the ZTF $g$ and $r$ bands as part of the public survey, with additional $i$-band imaging obtained through a private survey program. Each exposure had a duration of 30\,s, reaching a median $5\sigma$ point-source depth of $g \approx 20.6$\,mag and $r \approx 20.4$\,mag (AB) under nominal conditions \citep{Bellm2019a, Masci2019a}. During the 2023--2024 observing seasons, the public survey strategy revisited most extragalactic fields approximately at a two-night cadence.

ZTF data were processed through the real-time image-subtraction pipeline at the Infrared Processing and Analysis Center (IPAC), which performs bias subtraction, flat-fielding, astrometric calibration, and photometric zeropoint determination, followed by image differencing using reference images \citep{Masci2019a}. Transient candidates are identified on difference images and distributed via public-alert streams \citep{Masci2019a, Patterson2019a, Mahabal2019a}. These candidates were vetted and managed using the \emph{Fritz} Marshal system \citep{Coughlin2023a}, which is based on the SkyPortal Framework \citep{vanderWalt2019a}.

We reprocessed the ZTF data of \sn using the ZTF Forced Photometry service \citep{Masci2023a}. These were cleaned and calibrated as described by \citep{Masci2023a}. To boost the S/N, we combined same-night observations in each filter.

\subsubsection{Targeted Observations}

\paragraph{\textbf{8.2\,m ESO Very Large Telescope~}}~
Between July and October 2024, we obtained photometry in $gRIz$ with the FOcal Reducer/low dispersion Spectrograph 2 \citep[FORS2;][]{Appenzeller1998a} at the ESO 8.2\,m Very Large Telescope at Paranal Observatory (Chile). These data suffered from progressively increasing contamination by the host galaxy. To remove the host contribution, we took a final set of deep $gRIz$ photometry in June 2025. These reference images were obtained under similar observing conditions, but their integration times were a factor of 2 longer to ensure a reliable host subtraction. We reduced the data with the FORS2 pipeline v5.6.5\footnote{\href{https://www.eso.org/sci/software/pipelines/fors}{https://www.eso.org/sci/software/pipelines/fors}} and the ESO workflow engine ESOReflex \citep{Freudling2013a}. The world coordinate system was calibrated with the software package \program{astrometry.net} \citep{Lang2010a}.

\paragraph{\textbf{8.4\,m Large Binocular Telescope~}}~
We performed NIR observations in January 2026 (after the SN had faded below the host level) with the Large Binocular Telescope (LBT), located on Mount Graham (Arizona, USA). The LBT consists of two comounted 8.4\,m primary mirrors, providing a large collecting area and excellent image quality. Imaging was obtained using the LUCI instruments (LUCI1 and LUCI2; \citealt{Seifert2003a}), which provide imaging and spectroscopic capabilities in the NIR. Observations were acquired in the $J$ and $K_s$ bands using standard dither patterns to enable accurate sky subtraction. The data were processed using the data-reduction pipeline developed at INAF -- Osservatorio Astronomico di Roma \citep{Fontana2014a}, which includes bias subtraction and flat-fielding, bad-pixel and cosmic-ray masking, astrometric calibration, sky subtraction, and coaddition. Field stars from the \gaia DR3 catalog \citep{Gaia2016a, Gaia2023a} were used for astrometric calibration.

\paragraph{\textbf{2.0\,m Liverpool Telescope~}}~The Liverpool Telescope is a fully robotic 2.0\,m telescope located at the Observatorio del Roque de los Muchachos on La Palma (Canary Islands, Spain). Between June 2023 and August 2024, optical imaging was performed using the IO:O instrument \citep{Steele2004a}. Observations were obtained in Sloan-like $g$, $r$, $i$, and $z$ filters. Reduced images were downloaded from the LT archive and processed using a custom image-subtraction and analysis software (K. Hinds, K. Taggart, et al., in prep.). Image stacking and alignment were performed using \program{SWarp} \citep{Bertin2002a} where required. Image subtraction was done using a pre-explosion reference image in the appropriate filter from the Pan-STARRS Data Release 1 (DR1) \citep{Chambers2016a}. Photometry was measured using PSF-fitting techniques based on the methods of \citet{Fremling2016a}. 

\paragraph{\textbf{6.5\,m Magellan Telescope}~}~Optical imaging with the Magellan Baade 6.5\,m telescope was obtained using the Inamori-Magellan Areal Camera and Spectrograph (IMACS; \citealt{Dressler2006a}), located at Las Campanas Observatory, Chile. IMACS was operated in imaging mode with Sloan-like $g$ and $r$ filters. The data were reduced following standard optical imaging procedures, including bias subtraction, flat-field correction, and astrometric calibration.

\paragraph{\textbf{6.5\,m MMT~}}~Late-time photometry was acquired in the $g$ and $r$ bands with Binospec \citep{Fabricant2019a} on the 6.5\,m MMT telescope at Mount Hopkins (Arizona, USA), as well as in the $J$ band with the MMT and Magellan Infrared Spectrograph \citep[MMIRS;][]{McLeod2012a}. The data were reduced using the Pipeline for Optical/infrared Telescopes in Python for Reducing Images (\program{POTPyRI})\footnote{\url{https://github.com/CIERA-Transients/POTPyRI}}, which automatically conducts standard image-reduction procedures including bias subtraction, flat-field correction, and astrometric calibration.

The $g$- and $ r$-band observations suffered from tracking loss. This led to an elongated and deformed point-spread function. Due to the this, we omit these data from further analysis.

\paragraph{\textbf{Neil Gehrels Swift Observatory}~}~ We acquired two epochs with the 30\,cm Ultraviolet/Optical Telescope \citep[UVOT;][]{Roming2005a} aboard the \swift\ satellite \citep{Gehrels2004a} between 14 and 17 September 2023 in $w2$, $m2$, $w1$, $u$, $b$, $v$. The science-ready data were retrieved from the \swift\ archive\footnote{\href{https://www.swift.ac.uk/swift_portal}{https://www.swift.ac.uk/swift\_portal}}. \sn was not detected in these data. Since the data were taken only 3 days apart and \sn evolves extremely slowly, we coadded all images in each filter using \program{fappend} and \program{uvotimsum} in \program{HEAsoft}\footnote{\href{https://heasarc.gsfc.nasa.gov/docs/software/heasoft/}{https://heasarc.gsfc.nasa.gov/docs/software/heasoft/}} version 6.30 \citep{Heasoft2014a}.

\paragraph{\textbf{2.56\,m Nordic Optical Telescope~}}~The 2.56\,m Nordic Optical Telescope (NOT) is located at the Observatorio del Roque de los Muchachos on La Palma, Canary Islands, Spain. Optical imaging was acquired with the Andaluc\'{i}a Faint Object Spectrograph and Camera\footnote{\href{https://www.not.iac.es/instruments/alfosc}{https://www.not.iac.es/instruments/alfosc}} (ALFOSC), using standard Sloan-like $g$, $r$, and $i$ filters, in January and November 2024. Data reduction was performed with \program{PyNOT}\footnote{\href{https://github.com/jkrogager/PyNOT}{https://github.com/jkrogager/PyNOT}}, which applies standard procedures, including bias subtraction, flat-field correction, and coadding data. Strong winds, dust, and clouds affected the January observations. In addition, a mismatch between the ALFOSC science images and the VLT template images affects the image subtraction. Therefore, we do not report photometry for those data.

NIR imaging at the NOT was obtained using the NOTCam instrument \citep{Djupvik2010a} in the $J$, $H$, and $K_s$ bands between September 2023 and the end of June 2024. The data were reduced using standard NIR techniques, including dark subtraction, flat-fielding, sky subtraction, and coaddition.

\paragraph{\textbf{Palomar 60-inch Telescope}~}~Additional optical imaging was obtained with the robotic Palomar 60-inch (1.52\,m) telescope (P60; \citealt{Cenko2006a}), located at Palomar Observatory (California, USA), using the Spectral Energy Distribution Machine \citep[SEDM;][]{Blagorodnova2018a} in imaging mode. Images were obtained in Sloan-like $g$ and $r$. The data were reduced using the data-reduction pipeline \program{FPipe} \citep{Fremling2016a}.

\subsubsection{Supernova Photometry}

The ZTF and ATLAS forced-photometry light curves were calibrated following \citet{Masci2019a} and \citet{Smith2020a}, respectively. To increase the S/N of the ATLAS data, we combined same-night observations in each filter with inverse-variance weighting, using \program{plot\_atlas\_fp.py}\footnote{\url{https://gist.github.com/thespacedoctor/86777fa5a9567b7939e8d84fd8cf6a76}}, and imposed an S/N threshold of 5. The ZTF data were combined with a dynamic binning scheme (also using inverse-variance weighting): (1) $\rm min(\Delta t) = 1$~day and $\rm min(S/N) = 10$ between MJD 60000 and 60100; (2) $\rm min(\Delta t) = 1$~day and $\rm min(S/N) = 5$ between MJD 60100 and 60300; and (3) $\rm min(\Delta t) = 20$~days and $\rm min(S/N) = 4$ between MJD 60300 and 60700, where $\Delta t$ is the width of the time bin. The higher S/N threshold in the first interval accounts for the elevated noise of the earliest observations, obtained as part of the ZTF twilight survey \citep{Bellm2019b}. The coarser binning in the third interval is needed to recover \sn, which is not detected in individual images at late times; the exact choice of binning parameters does not affect the overall shape of the late-time light curve.
 
Photometry from targeted ground-based optical imaging was extracted with the \textsc{AutoPhOT} pipeline \citep{Brennan2022d}. We first refined the world coordinate system of each image with \textsc{Astrometry.net} \citep{Lang2010a} and constructed an effective-PSF (ePSF) model from bright, isolated stars using \program{Photutils} \citep{Bradley2024a}. Zeropoints were calibrated against stars from the ATLAS All-Sky Stellar Reference Catalog REFCAT2 \citep[][]{Tonry2018a}. Science and reference images were aligned with \textsc{SWarp} \citep{Bertin2002a}, and the SN flux was isolated through image subtraction with the saccadic fast Fourier transform algorithm \citep[SFFT;][]{Hu2022a}. Whenever \sn was not detected at the $\geq3\sigma$ level, we determined the limiting magnitude of the image via artificial-source injection, as detailed by \citet{Brennan2022d}.

At the time of the NIR observations with NOTCam, \sn was several magnitudes brighter than its host galaxy, making host contamination negligible. We therefore measured the SN brightness directly, without image subtraction, using the aperture-photometry tool of \citet{Schulze2018a}, which is based on \program{Source Extractor} \citep{Bertin1996a}. The instrumental magnitudes were calibrated against stars from the Two Micron All Sky Survey \citep[2MASS;][]{Skrutskie2006a} Point Source Catalog \citep{Cutri2003a}, measured in the same way.

For the MMIRS NIR data, we extracted the photometry using a different method for each epoch. For the first epoch, we performed image subtraction with \program{HOTPANTS} v5.11 \citep{Becker2015a}, using the LBT image as the reference and measuring the residual flux by aperture photometry. In the second epoch, the image quality was substantially better than that of the LBT reference, so subtraction would have degraded it. We instead recovered the SN flux by forward modeling the science frame, following an approach similar to that of \program{GALFIT} \citep{Peng2010a}. We modeled the image as the sum of a PSF-convolved S\'ersic host galaxy, a point source at the fixed transient position, and a constant sky background. An empirical PSF was built for each epoch from unsaturated field stars with \program{Source Extractor} \citep{Bertin1996a} and \program{PSFEx} \citep{Bertin2011a}. We fitted all model parameters simultaneously by nonlinear least squares and derived the uncertainties from the covariance matrix scaled by the reduced $\chi^2$. The photometry was calibrated with the same stellar sample and PSF model. We regarded any epoch with a point-source flux uncertainty above 50\% as unreliable, since such values indicate that the host and transient were not sufficiently resolved.

\sn is not detected in the UVOT images. We measured the brightness at the position of \sn with the \swift\ tool \program{uvotsource}, using a source aperture of radius $10''$ and a substantially larger background region, and calibrated the photometry with the calibration files from September 2020.
 
For the MIR data, we performed image subtraction on the NEOWISE images with a custom pipeline \citep{De2020b} based on the \program{ZOGY} algorithm \citep{Zackay2016a}, using the full-depth coadds from the \wise and {\it NEOWISE} missions (obtained during 2010--2014) as reference images. Photometric measurements were then obtained through forced PSF photometry at the position of \sn on the subtracted images.
 
\begin{figure*}
    \centering
    \includegraphics[width=1\textwidth]{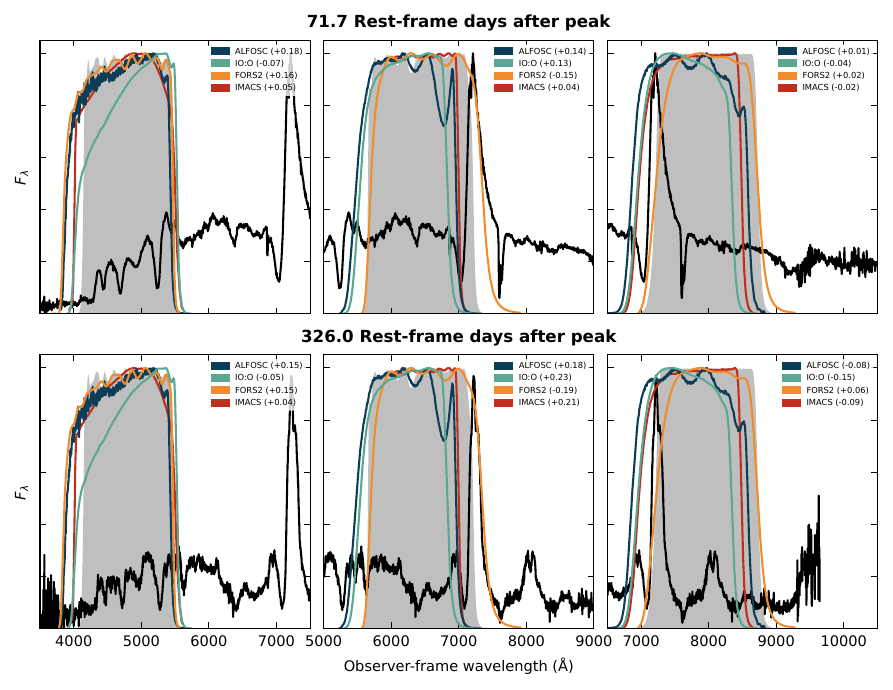}
    \caption{Spectra of \sn at 71.7 and 326.0 days after peak light. Overlaid are the filter transmissions of similar but not identical filters (left/center/right: \textit{g}, \textit{r}, \textit{i}). Since different instruments were used at various epochs and \sn's evolution is slow, we apply a color term ($X_{\rm Instrument} - X_{\rm ZTF}$ in magnitudes) for each of the instruments (values are reported in parentheses; mag units).
    }
    \label{fig:spec:colour_terms}
\end{figure*}

Strong spectral features can introduce time-dependent color terms between similar but not identical filters and, if uncorrected, add non-negligible systematic scatter to the light curves \citep[e.g.,][]{Stritzinger2002a}. Appendix Figure~\ref{fig:spec:colour_terms} illustrates this issue: it shows the spectra at $+71.7$ and $+326.0$~days, overlaid with the response curves of the filters used during the follow-up campaign. A comparison of the ZTF $r$ and $i$ bands with the other filters shows that individual filters can either cover or miss particularly strong spectral features. However, since each instrument was used only during a limited time span, over which the spectral features changed little, we can assume a constant color term (also known as $s$ correction) for each instrument.
 
Where needed, all measurements were converted to the AB system using the offsets from \citet{Blanton2007a}, \citet{Breeveld2011a}, and \citet{Cutri2012a}.

Appendix Tables~\ref{tab:SN2023iop:photometry} and \ref{tab:SN2023iop:photometry:wise} summarize the final photometric measurements. Nondetections are reported as $3\sigma$ upper limits. This table also presents the photometry after applying the $s$ correction and correcting for Milky Way extinction, assuming the \citet{Cardelli1989a} attenuation law, a total-to-selective extinction ratio of $R_V=3.1$, and the \citet{Schlafly2011a} Milky Way extinction map. The value of the Milky Way extinction along the line of sight of \sn is $E(B-V) = 0.05$~mag. All photometric measurements will also be made available through \textsc{WISeREP}.

\subsubsection{Host-galaxy Photometry}

After the SN had faded, we observed the SN host galaxy with the ESO VLT and the LBT from the $g$ to the $K$ bands. To measure the galaxy flux and preserve the intrinsic galaxy colors, we selected the deepest image and defined a circular aperture encompassing the entire galaxy. Next, we adjusted the aperture size according to the seeing conditions for each image and measured the brightness using the aforementioned aperture-photometry tool developed by \citet{Schulze2018a}. The photometry was calibrated against stars in the Sloan Digital Sky Survey (SDSS) DR9 \citep{Ahn2012a} and stars from the 2MASS Point Source Catalog \citep{Cutri2003a}. The NIR measurements were converted to the AB system using the values reported by  \citet{Blanton2007a}. The final photometry is listed in Appendix Table~\ref{tab:host_phot}. These measurements are not corrected for Milky Way extinction, but the correction was applied to all derived measurements reported in this paper.

\subsection{Spectroscopy}\label{obs:spectroscopy}

\paragraph{\textbf{8.2\,m ESO Very Large Telescope}~}
~We acquired two medium-resolution spectra with the X-shooter instrument \citep{Vernet2011a} on 1 November and 2 December 2023 (days~+71.7 and +99.9). All observations were performed in nodding mode and with 1\farcs0/0\farcs9/0\farcs9 wide slits (UVB/VIS/NIR). The integration time was 3600 s for each epoch. The data were reduced following \citet{Selsing2019a}. In brief, we first removed cosmic rays with the tool  \program{astroscrappy}\footnote{\url{https://github.com/astropy/astroscrappy}}, which is based on the cosmic-ray removal algorithm by \citet{vanDokkum2001a}. Afterward, the data were processed with the X-shooter pipeline v3.3.5 and the ESO workflow engine \program{ESOReflex} \citep{Goldoni2006a, Modigliani2010a}. The UVB and VIS-arm data were reduced in stare mode to boost the S/N. The individual rectified, wavelength-calibrated, and flux-calibrated two-dimensional (2D) spectra files were coadded using tools developed by J. Selsing\footnote{\href{https://github.com/jselsing/XSGRB_reduction_scripts}{https://github.com/jselsing/XSGRB\_reduction\_scripts}}. The NIR data were reduced in nodding mode to ensure good sky-line subtraction. In the third step, we extracted the 1D spectra of each arm in a statistically optimal way using tools by J. Selsing. Finally, the wavelength calibration of all spectra was corrected for barycentric motion. The spectra of the individual arms were stitched by averaging the overlap regions.

We obtained two epochs of late-time spectroscopy with the FOcal Reducer and low-dispersion Spectrograph 2 instrument (days~+326.0 and +623.8). Each observation was performed with the 300V grism and a 1\farcs0 wide slit. The integration time of each epoch was $12\times1200$~s. The data were reduced with the reduction pipeline \program{PypeIt} \citep{Prochaska2020a, Prochaska2020b}.

\paragraph{\textbf{10\,m Keck Telescope}~}~
We obtained one epoch with the DEep Imaging Multi-Object Spectrograph \citep[DEIMOS;][]{Faber2003a} on 7
October 2023 (day~+49.5) and another epoch with the 
Low Resolution Imaging Spectrometer \citep[LRIS;][]{Oke1995a} on 17 October
2023 (day~+58.4). The DEIMOS observation utilized the low-resolution grating 600ZD, an 0\farcs8-wide slit, and the GG455 order-blocking filter. The integration time was set to 900 s. The LRIS spectrum was acquired with the B400/3400 blue-side grism and the R400/8500 red-side grating, dichroic 560, and a 1\farcs0-wide slit. The integration time was 1200 s. All spectra were reduced in a standard fashion with the reduction pipeline \program{PypeIt} and \program{LPipe} \citep{Perley2019a}.

\paragraph{\textbf{8.4\,m Large Binocular Telescope}~}~
We also obtained one epoch of spectroscopy on 28 January 2025 (day~+483.8) with the Multi-Object Double Spectrographs MODS-1 and MODS-2 \citep{Pogge2010a} mounted on the LBT. The observation consisted of $9\times1200$~s exposures. ABBA offsets were used in the dual-grating mode (G400L and G670L), providing wavelength coverage 3200--9500~\AA\ and a slit mask with a width of $1\farcs2$. We processed the spectral data using the Spectroscopic Interactive Pipeline and Graphical Interface (SIPGI) tool \citep{sipgi2022}, and we applied corrections for bad pixels and cosmic rays, followed by bias subtraction and flat-field correction. The wavelength calibration was calculated from comparison-lamp frames, also to correct for optical distortion. Finally, the wavelength- and flux-calibrated, sky-subtracted 2D spectra were combined, and the 1D spectrum of each source was extracted. The spectra were flux-calibrated using a sensitivity function derived from spectrophotometric standard stars.

\paragraph{\textbf{2.0\,m Liverpool Telescope}~}~
We obtained a spectrum with the SPectrograph for the Rapid Acquisition of Transients \citep[SPRAT;][]{Piascik2014a} at the LT using a blue-optimized setup, which was reduced using a custom \textsc{python} pipeline. This pipeline is based on the packages \program{lacosmic} \citep{vanDokkum2001a}, \program{NumPy} \citep{vanderWalt2011a}, \program{SciPy} \citep{Scipy}, and \program{Astropy} \citep{Astropy2022a}. Aperture sizes and polynomial fits for sky subtraction and trace fitting were set manually to optimize the S/N. Table 1 from La Palma Technical Note\footnote{\url{https://www.ing.iac.es/Astronomy/observing/manuals/ps/tech_notes/tn031.pdf}} No. 31 was used to correct for differences in airmass between the science and standard-star frames.

\paragraph{\textbf{6.5\,m Magellan~}}~We obtained a high-resolution spectrum with the Magellan Echellette \citep[MagE;][]{Marshall2008a} spectrograph mounted on the Magellan Clay Telescope on 21 October 2023 (day~+61.7). MagE provides wavelength coverage 3200--10,000~\AA\ and, with the 0\farcs85 slit employed for these observations, delivers a spectral resolving power of $R \approx 4800$. The data were reduced using the \program{PypeIt} pipeline.

\paragraph{\textbf{6.5\,m MMT~}}~We obtained optical spectroscopy of \sn\ with the Binospec spectrograph \citep{Fabricant2019a} mounted on the 6.5\,m MMT Observatory. The observations were acquired on 10 and 15 January 2024 (day~+135.3 and 139.8) and 5 May 2024 (day~+296.2) using the 270 line mm$^{-1}$ grism with a central wavelength of 6500~\AA\ and a $1''$-wide slit, providing wavelength coverage 3850--9150~\AA\ at a resolving power of $R\approx1340$. The data were reduced using \program{PypeIt}.

\paragraph{\textbf{2.56\,m Nordic Optical Telescope}~}~
We acquired two epochs of low-resolution spectroscopy with ALFOSC on 29 August 2023 and 24 September 2023 (days~+13.6 and +37.2). The spectra were obtained with a 1\farcs0-wide slit and grism \#4. The data were reduced using \program{PypeIt}.

\paragraph{\textbf{Palomar 60-inch Telescope}~}~
The SED Machine at the Palomar 60-inch Telescope is a very low resolution ($R\approx 100$) integral-field unit covering the wavelength range 3650--10,000~\AA\ \citep{Blagorodnova2018a}. We acquired a series of spectra between 18 June and 16 October 2023 (days~$-51.4$ to $+57.4$). All data were reduced with \program{HyperGal} \citep{Lezmy2022a} that uses the output from \program{pysedm} \citep{Rigault2019a, Kim2022a}. In this paper, we present only the highest-quality spectra for which no other observations are available. All reduced spectra, irrespective of their quality, are available on \program{\textsc{WISeREP}}.

\paragraph{\textbf{Remarks}~}~
All spectra were calibrated against spectroscopic standard stars observed during the same night, and the flux scale was tied to our multiband photometry. Also, spectra were corrected for Milky Way extinction. A host-attenuation correction was not applied as it is negligible (Section~\ref{sec:redshift_host_attenuation}). The absolute flux-calibrated spectra are available on \textsc{WISeREP} \citep{Yaron2012a}.

\subsection{Spectropolarimetry}\label{obs:specpol}

We obtained one epoch of spectropolarimetry on 24 September 2023 (day~+37.3) with VLT/FORS2. 
The data were reduced using a custom pipeline \citep{Wichern2026a}, which extracts ordinary and extraordinary beam spectra optimally within regions of equal size ($4\times$ FWHM of the ordinary beam). Three cycles of spectra were obtained over the half-wave plate angles $0\fdg0$, $22\fdg5$, $45\fdg0$, and $67\fdg5$, and these were combined using inverse-variance weighting. 

The observations were taken with a $65\%$ full-moon illumination at an angle of $\sim87^\circ$ from the Moon. The high, scattered lunar background may have influenced the results. We investigated this by measuring the local background level in different regions alongside each beam. Estimating the background level farther away from each beam resulted in more significant polarization increasing toward the blue. When placing the background regions (with a width of 10 pixels, similar to the measurement aperture) right next to the aperture of each beam, the wavelength dependence of the polarization is minimal, and the Moon does not appear to have a significant effect in the presented polarimetry. Regardless, we exclude $<5000$~\AA, where the Moon's influence was strongest, from the analysis.

The spectropolarimetry was calibrated against a standard star. Polarisation bias correction was performed following \citet{Plaszczynski2014a}. In addition, the flux scale of the intensity spectrum was tied to the multiband photometry. A host-attenuation correction was not applied as it is negligible (Section~\ref{sec:redshift_host_attenuation}). The absolute flux-calibrated intensity spectrum is available on \textsc{WISeREP}.

\subsection{Radio Observations}\label{appendix:radio}

We observed the field of \sn at radio wavelengths with the Arcminute Microkelvin Imager Large Array \citep[AMI-LA;][]{Zwart2008a, Hickish2018a} at the Mullard Radio Astronomy Observatory (United Kingdom) on 15 November 2023 and 5 December 2023, corresponding to 84.4 and 102.6 rest-frame days after $r$-band peak light, respectively. AMI-LA is a radio interferometer consisting of eight 12.8\,m (diameter) antennas, providing 28 baselines ranging from 18 to 110~m and operating over a 5~GHz bandwidth centered at 15.5~GHz. 

The data were reduced, flagged, and calibrated using \textsc{reduce\_dc}, a customised AMI-LA data-reduction pipeline \citep{Perrott2013a}. Phase calibration was performed using short, interleaved observations. Imaging was carried out with the \textsc{Common Astronomy Software Applications} \citep[CASA;][]{McMullin2007a} task \textsc{CLEAN}, and the image RMS noise was measured using the CASA task \textsc{IMSTAT}.
No source is detected at the position of \sn down to a flux density of 48~$\mu$Jy (RMS) at a central frequency of 15.5~GHz.

\subsection{X-ray Observations}\label{appendix:xrays}

While monitoring \sn with UVOT, \swift also observed the field with its onboard X-ray telescope XRT between 0.3 and 10 keV in photon-counting mode \citep{Burrows2005a}. We analyzed these data with the online tools of the UK \swift team\footnote{\href{https://www.swift.ac.uk/user_objects/}{https://www.swift.ac.uk/user\_objects}} that use the software package \program{HEASoft} version 6.26.1 and methods described by \citet{Evans2007a, Evans2009a}. \sn was not detected at any epoch. As these two epochs were taken within 3 observer-frame days of each other, we combine them to go deeper.

The $3\sigma$ count-rate limit is $3\times10^{-3}~\rm s^{-1}$ (0.3--10~keV) on day~+30.3. To convert this limit into a flux, we assume a power-law spectrum with a photon index\footnote{The photon index is defined as the power-law index of the photon flux density ($N(E)\propto E^{-\Gamma}$).} of $\Gamma=2$ and a Galactic neutral hydrogen column density of $3.91\times10^{20}$~cm$^{-2}$ \citep{HI4PI2016a}. The count-rate limit translates to an unabsorbed flux of $<1.2\times10^{-13}~{\rm erg\,cm}^{-2}\,{\rm s}^{-1}$ in the range 0.3--10 keV. 

\section{Host Galaxies of 87A-like Supernovae}\label{appendix:87A-like}

\citet{Sit2023a} reported 13 long-rising Type~II SNe from the volume-limited Census of the Local Universe survey. These authors gave some host properties, but not the host photometry. Following the methodology of \citet{Schulze2021a}, we retrieved the science-ready images from the \textit{Galaxy Evolution Explorer} (\galex) general release 6/7 \citep{Martin2005a}, the
SDSS DR 9 \citep{Ahn2012a}), the Pan-STARRS DR1 \citep{Chambers2016a}, the 2MASS \citep{Skrutskie2006a}, and reprocessed {\it WISE} images from the unWISE archive \citep{Lang2014a}. The unWISE images include data from the {\it NEOWISE} Reactivation mission \citep{Mainzer2014a, Meisner2017a} through Year 6/7. The optical photometry was complemented by images from the DESI Legacy Imaging Surveys \citep[Legacy Surveys, LS;][]{Dey2018a} DR10, where data were missing. We measured the host brightness using \program{LAMBDAR} \citep[\program{Lambda Adaptive Multi-Band Deblending Algorithm in R};][]{Wright2016a} and utilized tools developed by \citet{Schulze2021a}. This sample was complemented with two luminous 87A-like SNe (OGLE-73, \citealt{Terreran2017a}; SN\,2020faa, \citealt{Yang2021a}). The photometry of the OGLE object was measured from Legacy Survey images, and the photometry of SN\,2020faa and iPTF14hls were taken from \citet{Yang2021a} and \citet{Schulze2021a}, which also employed the methods from \citet{Schulze2021a}. All measurements are summarized in Table~\ref{tab:host_87A_phot} in the Appendix.

We modeled the host galaxy SEDs with \program{Prospector} under the same assumptions as the host of \sn (Section~\ref{sec:host_galaxy}).  Table~\ref{tab:host_87A_prospector} in the Appendix summarizes the key properties of the SED modeling. 

\newpage
\startlongtable
%\begin{table*}
%\centering

\begin{deluxetable}{ccccc}
\tablecolumns{5}
\tablecaption{Photometry of SN\,2023iop. 
\label{tab:SN2023iop:photometry}}
\tablehead{
  \colhead{MJD} & \colhead{Telescope/} & \colhead{Filter} & \colhead{Brightness} & \colhead{Brightness} \\
  & \colhead{Instrument} & &\colhead{(observed, mag)} & \colhead{(corrected, mag)}
  }
\startdata
    $60075.4837^{+1.0016}_{-1.9984}$ & ZTF &$ r $& $19.52 \pm 0.10$ & $19.40 \pm 0.10$ \\
    $60079.4753^{+0.9960}_{-0.0040}$ & ZTF &$ r $& $19.33 \pm 0.05$ & $19.22 \pm 0.05$ \\
    $60081.4749^{+0.9950}_{-0.0050}$ & ZTF &$ r $& $19.30 \pm 0.05$ & $19.18 \pm 0.05$ \\
    $60082.9725^{+0.5009}_{-0.4991}$ & ZTF &$ r $& $19.31 \pm 0.04$ & $19.19 \pm 0.04$ \\
    $60083.4747^{+0.9990}_{-0.0010}$ & ZTF &$ r $& $19.22 \pm 0.06$ & $19.10 \pm 0.06$ \\
    $60093.6000$ & ATLAS &$ o $& $19.01 \pm 0.09$ & $18.90 \pm 0.09$ \\
    $60095.5900$ & ATLAS &$ o $& $19.00 \pm 0.07$ & $18.89 \pm 0.07$ \\
    $60101.6000$ & ATLAS &$ o $& $18.72 \pm 0.18$ & $18.61 \pm 0.18$ \\
    $60105.5900$ & ATLAS &$ o $& $18.68 \pm 0.07$ & $18.57 \pm 0.07$ \\
    $60107.5900$ & ATLAS &$ o $& $18.66 \pm 0.13$ & $18.55 \pm 0.13$ \\
    $60109.6000$ & ATLAS &$ c $& $19.12 \pm 0.12$ & $18.97 \pm 0.12$ \\
    $60110.4000$ & ATLAS &$ c $& $19.23 \pm 0.11$ & $19.08 \pm 0.11$ \\
    $60110.4288^{+1.0000}_{-0.0000}$ & ZTF &$ i $& $18.51 \pm 0.08$ & $18.42 \pm 0.08$ \\
    $60111.5900$ & ATLAS &$ c $& $19.22 \pm 0.07$ & $19.08 \pm 0.07$ \\
    $60113.3931$ & P60/SEDm+RC &$ r $& $18.97 \pm 0.08$ & $18.76 \pm 0.08$ \\
    $60113.4131^{+1.0000}_{-0.0000}$ & ZTF &$ i $& $18.46 \pm 0.07$ & $18.37 \pm 0.07$ \\
    $60113.5900$ & ATLAS &$ c $& $19.09 \pm 0.07$ & $18.95 \pm 0.07$ \\
    $60115.5900$ & ATLAS &$ c $& $18.95 \pm 0.07$ & $18.80 \pm 0.07$ \\
    $60116.1342$ & LT/IO:O &$ g $& $19.08 \pm 0.02$ & $18.96 \pm 0.02$ \\
    $60116.1353$ & LT/IO:O &$ r $& $18.80 \pm 0.01$ & $18.63 \pm 0.01$ \\
    $60116.1363$ & LT/IO:O &$ i $& $18.54 \pm 0.02$ & $18.49 \pm 0.02$ \\
    $60116.4295^{+1.0000}_{-0.0000}$ & ZTF &$ i $& $18.46 \pm 0.07$ & $18.38 \pm 0.07$ \\
    $60117.6000$ & ATLAS &$ c $& $19.00 \pm 0.08$ & $18.86 \pm 0.08$ \\
    $60120.4494^{+1.0000}_{-0.0000}$ & ZTF &$ i $& $18.36 \pm 0.06$ & $18.28 \pm 0.06$ \\
    $60123.4143^{+1.0000}_{-0.0000}$ & ZTF &$ g $& $19.25 \pm 0.06$ & $19.09 \pm 0.06$ \\
    $60123.4463^{+1.0000}_{-0.0000}$ & ZTF &$ r $& $18.69 \pm 0.04$ & $18.58 \pm 0.04$ \\
    $60123.5900$ & ATLAS &$ c $& $18.96 \pm 0.06$ & $18.81 \pm 0.06$ \\
    $60125.3843^{+1.0000}_{-0.0000}$ & ZTF &$ g $& $19.21 \pm 0.11$ & $19.05 \pm 0.11$ \\
    $60125.4390^{+1.0000}_{-0.0000}$ & ZTF &$ r $& $18.78 \pm 0.04$ & $18.67 \pm 0.04$ \\
    $60125.5800$ & ATLAS &$ o $& $18.52 \pm 0.04$ & $18.42 \pm 0.04$ \\
    $60129.5900$ & ATLAS &$ o $& $18.55 \pm 0.08$ & $18.44 \pm 0.08$ \\
    $60131.6100$ & ATLAS &$ o $& $18.52 \pm 0.07$ & $18.41 \pm 0.07$ \\
    $60133.4040^{+1.0000}_{-0.0000}$ & ZTF &$ r $& $18.58 \pm 0.05$ & $18.47 \pm 0.05$ \\
    $60133.4377^{+1.0000}_{-0.0000}$ & ZTF &$ i $& $18.24 \pm 0.05$ & $18.16 \pm 0.05$ \\
    $60133.4498^{+1.0000}_{-0.0000}$ & ZTF &$ g $& $18.93 \pm 0.07$ & $18.77 \pm 0.07$ \\
    $60133.5900$ & ATLAS &$ o $& $18.52 \pm 0.07$ & $18.41 \pm 0.07$ \\
    $60135.4611^{+1.0000}_{-0.0000}$ & ZTF &$ r $& $18.52 \pm 0.04$ & $18.41 \pm 0.04$ \\
    $60137.4472^{+1.0000}_{-0.0000}$ & ZTF &$ i $& $18.27 \pm 0.05$ & $18.19 \pm 0.05$ \\
    $60137.5800$ & ATLAS &$ o $& $18.48 \pm 0.04$ & $18.38 \pm 0.04$ \\
    $60138.4306^{+1.0000}_{-0.0000}$ & ZTF &$ g $& $19.01 \pm 0.05$ & $18.85 \pm 0.05$ \\
    $60139.5600$ & ATLAS &$ c $& $18.89 \pm 0.06$ & $18.75 \pm 0.06$ \\
    $60140.3000$ & ATLAS &$ o $& $18.35 \pm 0.05$ & $18.25 \pm 0.05$ \\
    $60140.4127^{+1.0000}_{-0.0000}$ & ZTF &$ g $& $19.01 \pm 0.05$ & $18.85 \pm 0.05$ \\
    $60141.4490^{+1.0000}_{-0.0000}$ & ZTF &$ i $& $18.29 \pm 0.05$ & $18.21 \pm 0.05$ \\
    $60141.5700$ & ATLAS &$ c $& $18.90 \pm 0.06$ & $18.75 \pm 0.06$ \\
    $60143.4189^{+1.0000}_{-0.0000}$ & ZTF &$ g $& $18.97 \pm 0.05$ & $18.81 \pm 0.05$ \\
    $60143.4488^{+1.0000}_{-0.0000}$ & ZTF &$ r $& $18.55 \pm 0.04$ & $18.43 \pm 0.04$ \\
    $60144.4632^{+1.0000}_{-0.0000}$ & ZTF &$ i $& $18.13 \pm 0.04$ & $18.05 \pm 0.04$ \\
    $60145.4377^{+0.9995}_{-0.0005}$ & ZTF &$ r $& $18.48 \pm 0.02$ & $18.37 \pm 0.02$ \\
    $60145.5600$ & ATLAS &$ c $& $18.82 \pm 0.05$ & $18.68 \pm 0.05$ \\
    $60146.4481^{+1.0000}_{-0.0000}$ & ZTF &$ r $& $18.48 \pm 0.03$ & $18.36 \pm 0.03$ \\
    $60147.4244^{+1.0000}_{-0.0000}$ & ZTF &$ i $& $18.20 \pm 0.04$ & $18.12 \pm 0.04$ \\
    $60147.5500$ & ATLAS &$ c $& $18.90 \pm 0.06$ & $18.76 \pm 0.06$ \\
    $60148.3861^{+1.0000}_{-0.0000}$ & ZTF &$ r $& $18.49 \pm 0.03$ & $18.37 \pm 0.03$ \\
    $60148.4352^{+1.0000}_{-0.0000}$ & ZTF &$ g $& $19.03 \pm 0.05$ & $18.86 \pm 0.05$ \\
    $60149.4344^{+0.9998}_{-0.0002}$ & ZTF &$ r $& $18.43 \pm 0.02$ & $18.32 \pm 0.02$ \\
    $60149.5400$ & ATLAS &$ c $& $18.73 \pm 0.05$ & $18.59 \pm 0.05$ \\
    $60150.4327^{+1.0000}_{-0.0000}$ & ZTF &$ g $& $18.91 \pm 0.04$ & $18.75 \pm 0.04$ \\
    $60150.4438^{+1.0000}_{-0.0000}$ & ZTF &$ r $& $18.53 \pm 0.04$ & $18.42 \pm 0.04$ \\
    $60150.4688^{+1.0000}_{-0.0000}$ & ZTF &$ i $& $18.19 \pm 0.04$ & $18.10 \pm 0.04$ \\
    $60151.4569$ & P60/SEDm+RC &$ r $& $18.59 \pm 0.05$ & $18.42 \pm 0.05$ \\
    $60151.5400$ & ATLAS &$ c $& $18.73 \pm 0.05$ & $18.59 \pm 0.05$ \\
    $60152.4706^{+1.0000}_{-0.0000}$ & ZTF &$ g $& $19.02 \pm 0.05$ & $18.86 \pm 0.05$ \\
    $60153.5400$ & ATLAS &$ c $& $18.88 \pm 0.06$ & $18.73 \pm 0.06$ \\
    $60155.4018^{+1.0000}_{-0.0000}$ & ZTF &$ i $& $18.13 \pm 0.04$ & $18.04 \pm 0.04$ \\
    $60155.4356^{+1.0000}_{-0.0000}$ & ZTF &$ g $& $18.94 \pm 0.04$ & $18.78 \pm 0.04$ \\
    $60155.4733^{+1.0000}_{-0.0000}$ & ZTF &$ r $& $18.39 \pm 0.03$ & $18.28 \pm 0.03$ \\
    $60155.5300$ & ATLAS &$ o $& $18.36 \pm 0.05$ & $18.25 \pm 0.05$ \\
    $60157.5100$ & ATLAS &$ o $& $18.24 \pm 0.07$ & $18.13 \pm 0.07$ \\
    $60159.3936^{+1.0000}_{-0.0000}$ & ZTF &$ r $& $18.40 \pm 0.06$ & $18.29 \pm 0.06$ \\
    $60159.4343^{+1.0000}_{-0.0000}$ & ZTF &$ i $& $18.19 \pm 0.06$ & $18.10 \pm 0.06$ \\
    $60162.4004^{+1.0000}_{-0.0000}$ & ZTF &$ g $& $18.82 \pm 0.06$ & $18.66 \pm 0.06$ \\
    $60162.4257^{+1.0000}_{-0.0000}$ & ZTF &$ i $& $18.16 \pm 0.05$ & $18.08 \pm 0.05$ \\
    $60162.4448^{+1.0000}_{-0.0000}$ & ZTF &$ r $& $18.37 \pm 0.05$ & $18.26 \pm 0.05$ \\
    $60164.4200$ & ATLAS &$ o $& $18.42 \pm 0.11$ & $18.31 \pm 0.11$ \\
    $60168.4009^{+1.0000}_{-0.0000}$ & ZTF &$ g $& $19.02 \pm 0.04$ & $18.86 \pm 0.04$ \\
    $60168.4100$ & ATLAS &$ c $& $18.82 \pm 0.07$ & $18.68 \pm 0.07$ \\
    $60168.4548^{+1.0000}_{-0.0000}$ & ZTF &$ r $& $18.49 \pm 0.03$ & $18.38 \pm 0.03$ \\
    $60169.0000$ & ATLAS &$ o $& $18.49 \pm 0.10$ & $18.38 \pm 0.10$ \\
    $60169.0100$ & ATLAS &$ o $& $18.40 \pm 0.08$ & $18.29 \pm 0.08$ \\
    $60170.3619^{+1.0000}_{-0.0000}$ & ZTF &$ g $& $19.08 \pm 0.04$ & $18.92 \pm 0.04$ \\
    $60170.4011^{+1.0000}_{-0.0000}$ & ZTF &$ r $& $18.45 \pm 0.03$ & $18.33 \pm 0.03$ \\
    $60171.3963^{+1.0000}_{-0.0000}$ & ZTF &$ g $& $18.98 \pm 0.04$ & $18.82 \pm 0.04$ \\
    $60171.4900$ & ATLAS &$ c $& $18.71 \pm 0.07$ & $18.57 \pm 0.07$ \\
    $60173.2314$ & P60/SEDm+RC &$ r $& $18.49 \pm 0.05$ & $18.32 \pm 0.05$ \\
    $60173.4214^{+1.0000}_{-0.0000}$ & ZTF &$ i $& $18.22 \pm 0.04$ & $18.14 \pm 0.04$ \\
    $60173.4900$ & ATLAS &$ c $& $18.75 \pm 0.06$ & $18.61 \pm 0.06$ \\
    $60175.4800$ & ATLAS &$ c $& $18.73 \pm 0.06$ & $18.59 \pm 0.06$ \\
    $60177.4600$ & ATLAS &$ c $& $18.75 \pm 0.04$ & $18.60 \pm 0.04$ \\
    $60179.4600$ & ATLAS &$ c $& $18.77 \pm 0.05$ & $18.63 \pm 0.05$ \\
    $60182.2495$ & P60/SEDm+RC &$ r $& $18.58 \pm 0.05$ & $18.40 \pm 0.05$ \\
    $60182.2764$ & P60/SEDm+RC &$ g $& $<16.70$ & $<16.51$ \\
    $60182.2791$ & P60/SEDm+RC &$ r $& $18.55 \pm 0.02$ & $18.37 \pm 0.02$ \\
    $60182.2818$ & P60/SEDm+RC &$ i $& $18.24 \pm 0.02$ & $18.20 \pm 0.02$ \\
    $60182.3679^{+1.0000}_{-0.0000}$ & ZTF &$ i $& $18.08 \pm 0.04$ & $18.00 \pm 0.04$ \\
    $60182.4047^{+1.0000}_{-0.0000}$ & ZTF &$ g $& $19.10 \pm 0.05$ & $18.94 \pm 0.05$ \\
    $60183.3490^{+0.1007}_{-0.8993}$ & ZTF &$ r $& $18.44 \pm 0.02$ & $18.32 \pm 0.02$ \\
    $60183.6200$ & ATLAS &$ c $& $18.88 \pm 0.16$ & $18.73 \pm 0.16$ \\
    $60184.3330^{+0.9995}_{-0.0005}$ & ZTF &$ r $& $18.48 \pm 0.02$ & $18.37 \pm 0.02$ \\
    $60184.4224^{+1.0000}_{-0.0000}$ & ZTF &$ g $& $19.04 \pm 0.05$ & $18.88 \pm 0.05$ \\
    $60185.4024^{+1.0000}_{-0.0000}$ & ZTF &$ i $& $18.15 \pm 0.05$ & $18.06 \pm 0.05$ \\
    $60186.3500$ & ATLAS &$ o $& $18.57 \pm 0.14$ & $18.47 \pm 0.14$ \\
    $60186.3983^{+1.0000}_{-0.0000}$ & ZTF &$ r $& $18.55 \pm 0.06$ & $18.44 \pm 0.06$ \\
    $60186.4462^{+1.0000}_{-0.0000}$ & ZTF &$ g $& $19.08 \pm 0.13$ & $18.92 \pm 0.13$ \\
    $60187.4400$ & ATLAS &$ o $& $18.43 \pm 0.16$ & $18.33 \pm 0.16$ \\
    $60191.5300$ & ATLAS &$ o $& $18.38 \pm 0.06$ & $18.28 \pm 0.06$ \\
    $60192.3700^{+0.9659}_{-0.0341}$ & ZTF &$ g $& $19.10 \pm 0.04$ & $18.94 \pm 0.04$ \\
    $60192.4116^{+1.0000}_{-0.0000}$ & ZTF &$ i $& $18.14 \pm 0.04$ & $18.06 \pm 0.04$ \\
    $60192.4240^{+1.0000}_{-0.0000}$ & ZTF &$ r $& $18.56 \pm 0.05$ & $18.44 \pm 0.05$ \\
    $60194.3204^{+1.0000}_{-0.0000}$ & ZTF &$ r $& $18.53 \pm 0.04$ & $18.42 \pm 0.04$ \\
    $60194.3500$ & ATLAS &$ o $& $18.37 \pm 0.08$ & $18.27 \pm 0.08$ \\
    $60194.4093^{+1.0000}_{-0.0000}$ & ZTF &$ g $& $19.07 \pm 0.05$ & $18.91 \pm 0.05$ \\
    $60195.3002^{+1.0000}_{-0.0000}$ & ZTF &$ i $& $18.05 \pm 0.04$ & $17.96 \pm 0.04$ \\
    $60195.5100$ & ATLAS &$ o $& $18.30 \pm 0.03$ & $18.19 \pm 0.03$ \\
    $60196.3408^{+1.0000}_{-0.0000}$ & ZTF &$ r $& $18.46 \pm 0.03$ & $18.34 \pm 0.03$ \\
    $60196.3811$ & P60/SEDm+RC &$ r $& $18.58 \pm 0.04$ & $18.41 \pm 0.04$ \\
    $60196.4104^{+0.9081}_{-0.0919}$ & ZTF &$ g $& $19.09 \pm 0.02$ & $18.93 \pm 0.02$ \\
    $60197.4500$ & ATLAS &$ c $& $18.91 \pm 0.06$ & $18.77 \pm 0.06$ \\
    $60199.5000$ & ATLAS &$ c $& $18.98 \pm 0.08$ & $18.84 \pm 0.08$ \\
    $60200.2200$ & ATLAS &$ o $& $18.36 \pm 0.05$ & $18.25 \pm 0.05$ \\
    $60200.3840^{+1.0000}_{-0.0000}$ & ZTF &$ g $& $19.22 \pm 0.04$ & $19.06 \pm 0.04$ \\
    $60200.4108^{+1.0000}_{-0.0000}$ & ZTF &$ r $& $18.50 \pm 0.03$ & $18.39 \pm 0.03$ \\
    $60201.2432^{+1.0000}_{-0.0000}$ & ZTF &$ i $& $18.13 \pm 0.04$ & $18.05 \pm 0.04$ \\
    $60201.4300$ & ATLAS &$ c $& $18.94 \pm 0.06$ & $18.79 \pm 0.06$ \\
    $60202.3244^{+1.0000}_{-0.0000}$ & ZTF &$ r $& $18.50 \pm 0.03$ & $18.39 \pm 0.03$ \\
    $60202.4112^{+1.0000}_{-0.0000}$ & ZTF &$ g $& $19.25 \pm 0.04$ & $19.09 \pm 0.04$ \\
    $60203.1549$ & P60/SEDm+RC &$ r $& $18.62 \pm 0.06$ & $18.44 \pm 0.06$ \\
    $60203.1818$ & P60/SEDm+RC &$ u $& $<20.20$ & $<19.98$ \\
    $60203.1859$ & P60/SEDm+RC &$ g $& $19.32 \pm 0.03$ & $19.11 \pm 0.03$ \\
    $60203.1886$ & P60/SEDm+RC &$ r $& $18.62 \pm 0.01$ & $18.44 \pm 0.01$ \\
    $60203.1913$ & P60/SEDm+RC &$ i $& $18.28 \pm 0.02$ & $18.21 \pm 0.02$ \\
    $60203.3272$ & P60/SEDm+RC &$ u $& $<20.70$ & $<20.48$ \\
    $60203.4439$ & \swift/UVOT &$ w1 $& $<21.25$ & $<20.96$ \\
    $60203.4447$ & \swift/UVOT &$ u $& $<20.17$ & $<19.95$ \\
    $60203.4453$ & \swift/UVOT & b & $<19.34$ & $<19.15$ \\
    $60203.4472$ & \swift/UVOT &$ w2 $& $<21.97$ & $<21.56$ \\
    $60203.4512$ & \swift/UVOT &$ m2 $& $<21.94$ & $<21.52$ \\
    $60203.4800$ & ATLAS &$ c $& $18.92 \pm 0.14$ & $18.78 \pm 0.14$ \\
    $60204.3387^{+1.0000}_{-0.0000}$ & ZTF &$ g $& $19.18 \pm 0.04$ & $19.02 \pm 0.04$ \\
    $60204.3717^{+1.0000}_{-0.0000}$ & ZTF &$ i $& $18.04 \pm 0.04$ & $17.95 \pm 0.04$ \\
    $60204.3805^{+1.0000}_{-0.0000}$ & ZTF &$ r $& $18.54 \pm 0.03$ & $18.43 \pm 0.03$ \\
    $60204.8984$ & \swift/UVOT & v & $<18.39$ & $<18.26$ \\
    $60205.4400$ & ATLAS &$ c $& $19.01 \pm 0.06$ & $18.87 \pm 0.06$ \\
    $60206.4082^{+1.0000}_{-0.0000}$ & ZTF &$ g $& $19.30 \pm 0.05$ & $19.14 \pm 0.05$ \\
    $60206.4352^{+1.0000}_{-0.0000}$ & ZTF &$ r $& $18.58 \pm 0.03$ & $18.47 \pm 0.03$ \\
    $60207.2768^{+1.0000}_{-0.0000}$ & ZTF &$ i $& $18.17 \pm 0.04$ & $18.08 \pm 0.04$ \\
    $60207.3924$ & P60/SEDm+RC &$ r $& $18.73 \pm 0.06$ & $18.54 \pm 0.06$ \\
    $60207.4194$ & P60/SEDm+RC &$ g $& $19.36 \pm 0.06$ & $19.17 \pm 0.06$ \\
    $60207.4220$ & P60/SEDm+RC &$ r $& $18.65 \pm 0.02$ & $18.46 \pm 0.02$ \\
    $60207.4247$ & P60/SEDm+RC &$ i $& $18.17 \pm 0.02$ & $18.11 \pm 0.02$ \\
    $60207.4800$ & ATLAS &$ c $& $18.94 \pm 0.05$ & $18.79 \pm 0.05$ \\
    $60209.2840^{+1.0000}_{-0.0000}$ & ZTF &$ r $& $18.62 \pm 0.05$ & $18.51 \pm 0.05$ \\
    $60209.3829^{+0.9998}_{-0.0002}$ & ZTF &$ g $& $19.26 \pm 0.05$ & $19.10 \pm 0.05$ \\
    $60209.4700$ & ATLAS &$ c $& $18.98 \pm 0.08$ & $18.84 \pm 0.08$ \\
    $60211.3677^{+1.0000}_{-0.0000}$ & ZTF &$ i $& $18.16 \pm 0.04$ & $18.07 \pm 0.04$ \\
    $60211.4000$ & ATLAS &$ o $& $18.41 \pm 0.05$ & $18.30 \pm 0.05$ \\
    $60212.2500$ & ATLAS &$ o $& $18.45 \pm 0.08$ & $18.34 \pm 0.08$ \\
    $60212.3924^{+1.0000}_{-0.0000}$ & ZTF &$ r $& $18.58 \pm 0.03$ & $18.47 \pm 0.03$ \\
    $60212.4148^{+1.0000}_{-0.0000}$ & ZTF &$ g $& $19.35 \pm 0.04$ & $19.19 \pm 0.04$ \\
    $60213.4600$ & ATLAS &$ o $& $18.49 \pm 0.06$ & $18.38 \pm 0.06$ \\
    $60214.0758$ & NOT/NOTCam &$ J $& $18.68 \pm 0.03$ & $18.64 \pm 0.03$ \\
    $60214.0895$ & NOT/NOTCam &$ H $& $19.00 \pm 0.04$ & $18.98 \pm 0.04$ \\
    $60214.2348$ & P60/SEDm+RC &$ r $& $18.78 \pm 0.26$ & $18.57 \pm 0.26$ \\
    $60214.2617$ & P60/SEDm+RC &$ g $& $<16.00$ & $<15.88$ \\
    $60214.2644$ & P60/SEDm+RC &$ r $& $18.73 \pm 0.03$ & $18.51 \pm 0.03$ \\
    $60214.2671$ & P60/SEDm+RC &$ i $& $18.21 \pm 0.02$ & $18.15 \pm 0.02$ \\
    $60214.2744^{+1.0000}_{-0.0000}$ & ZTF &$ r $& $18.61 \pm 0.06$ & $18.49 \pm 0.06$ \\
    $60214.3541^{+1.0000}_{-0.0000}$ & ZTF &$ i $& $18.12 \pm 0.06$ & $18.04 \pm 0.06$ \\
    $60214.3792^{+1.0000}_{-0.0000}$ & ZTF &$ g $& $19.33 \pm 0.13$ & $19.17 \pm 0.13$ \\
    $60216.1747^{+1.0000}_{-0.0000}$ & ZTF &$ r $& $18.58 \pm 0.09$ & $18.47 \pm 0.09$ \\
    $60216.2340^{+1.0000}_{-0.0000}$ & ZTF &$ g $& $19.35 \pm 0.18$ & $19.19 \pm 0.18$ \\
    $60219.4800$ & ATLAS &$ o $& $18.62 \pm 0.14$ & $18.52 \pm 0.14$ \\
    $60220.2701^{+1.0000}_{-0.0000}$ & ZTF &$ i $& $18.13 \pm 0.06$ & $18.05 \pm 0.06$ \\
    $60220.3170^{+1.0000}_{-0.0000}$ & ZTF &$ r $& $18.65 \pm 0.06$ & $18.53 \pm 0.06$ \\
    $60220.3888^{+1.0000}_{-0.0000}$ & ZTF &$ g $& $19.42 \pm 0.13$ & $19.26 \pm 0.13$ \\
    $60221.4500$ & ATLAS &$ o $& $18.49 \pm 0.05$ & $18.38 \pm 0.05$ \\
    $60222.2154^{+1.0000}_{-0.0000}$ & ZTF &$ g $& $19.45 \pm 0.08$ & $19.29 \pm 0.08$ \\
    $60222.2489^{+1.0000}_{-0.0000}$ & ZTF &$ r $& $18.76 \pm 0.06$ & $18.65 \pm 0.06$ \\
    $60222.9548$ & NOT/NOTCam &$ H $& $18.95 \pm 0.05$ & $18.93 \pm 0.05$ \\
    $60222.9721$ & NOT/NOTCam &$ J $& $18.69 \pm 0.04$ & $18.65 \pm 0.04$ \\
    $60223.2680^{+1.0000}_{-0.0000}$ & ZTF &$ i $& $18.23 \pm 0.06$ & $18.15 \pm 0.06$ \\
    $60223.4300$ & ATLAS &$ o $& $18.64 \pm 0.05$ & $18.54 \pm 0.05$ \\
    $60224.2100$ & ATLAS &$ o $& $18.54 \pm 0.07$ & $18.43 \pm 0.07$ \\
    $60224.2435^{+1.0000}_{-0.0000}$ & ZTF &$ r $& $18.64 \pm 0.03$ & $18.53 \pm 0.03$ \\
    $60224.2571^{+1.0000}_{-0.0000}$ & ZTF &$ g $& $19.42 \pm 0.06$ & $19.26 \pm 0.06$ \\
    $60224.4016$ & P60/SEDm+RC &$ r $& $18.85 \pm 0.12$ & $18.65 \pm 0.12$ \\
    $60224.4082$ & P60/SEDm+RC &$ r $& $18.86 \pm 0.02$ & $18.70 \pm 0.02$ \\
    $60224.4113$ & P60/SEDm+RC &$ i $& $18.35 \pm 0.03$ & $18.28 \pm 0.03$ \\
    $60225.4200$ & ATLAS &$ c $& $19.38 \pm 0.09$ & $19.24 \pm 0.09$ \\
    $60226.3044$ & P60/SEDm+RC &$ r $& $18.71 \pm 0.06$ & $18.52 \pm 0.06$ \\
    $60226.3076$ & P60/SEDm+RC &$ r $& $18.93 \pm 0.02$ & $18.70 \pm 0.02$ \\
    $60226.3105$ & P60/SEDm+RC &$ g $& $19.67 \pm 0.03$ & $19.53 \pm 0.03$ \\
    $60226.3135$ & P60/SEDm+RC &$ i $& $18.30 \pm 0.02$ & $18.30 \pm 0.02$ \\
    $60226.3200^{+1.0000}_{-0.0000}$ & ZTF &$ i $& $18.24 \pm 0.05$ & $18.16 \pm 0.05$ \\
    $60227.2498^{+1.0000}_{-0.0000}$ & ZTF &$ g $& $19.59 \pm 0.08$ & $19.43 \pm 0.08$ \\
    $60227.2571^{+1.0000}_{-0.0000}$ & ZTF &$ r $& $18.79 \pm 0.05$ & $18.68 \pm 0.05$ \\
    $60227.4300$ & ATLAS &$ c $& $19.47 \pm 0.10$ & $19.33 \pm 0.10$ \\
    $60228.2566$ & P60/SEDm+RC &$ r $& $18.88 \pm 0.11$ & $18.58 \pm 0.11$ \\
    $60228.2627$ & P60/SEDm+RC &$ r $& $18.92 \pm 0.02$ & $18.66 \pm 0.02$ \\
    $60228.2657$ & P60/SEDm+RC &$ i $& $18.30 \pm 0.03$ & $18.27 \pm 0.03$ \\
    $60229.4100$ & ATLAS &$ c $& $19.36 \pm 0.08$ & $19.22 \pm 0.08$ \\
    $60230.1600$ & ATLAS &$ o $& $18.79 \pm 0.06$ & $18.68 \pm 0.06$ \\
    $60230.2920^{+1.0000}_{-0.0000}$ & ZTF &$ r $& $18.90 \pm 0.05$ & $18.78 \pm 0.05$ \\
    $60230.2980^{+1.0000}_{-0.0000}$ & ZTF &$ i $& $18.44 \pm 0.08$ & $18.36 \pm 0.08$ \\
    $60230.3482^{+1.0000}_{-0.0000}$ & ZTF &$ g $& $19.70 \pm 0.08$ & $19.54 \pm 0.08$ \\
    $60231.4100$ & ATLAS &$ c $& $19.37 \pm 0.08$ & $19.23 \pm 0.08$ \\
    $60232.2362^{+1.0000}_{-0.0000}$ & ZTF &$ g $& $19.70 \pm 0.06$ & $19.54 \pm 0.06$ \\
    $60232.2561^{+1.0000}_{-0.0000}$ & ZTF &$ r $& $18.85 \pm 0.05$ & $18.74 \pm 0.05$ \\
    $60233.2881$ & P60/SEDm+RC &$ r $& $19.08 \pm 0.07$ & $18.84 \pm 0.07$ \\
    $60233.2916$ & P60/SEDm+RC &$ g $& $19.94 \pm 0.04$ & $19.74 \pm 0.04$ \\
    $60233.2949$ & P60/SEDm+RC &$ r $& $19.07 \pm 0.02$ & $18.83 \pm 0.02$ \\
    $60233.2982$ & P60/SEDm+RC &$ i $& $18.48 \pm 0.95$ & $18.48 \pm 0.95$ \\
    $60233.3040^{+1.0000}_{-0.0000}$ & ZTF &$ i $& $18.31 \pm 0.05$ & $18.23 \pm 0.05$ \\
    $60233.4000$ & ATLAS &$ c $& $19.54 \pm 0.09$ & $19.40 \pm 0.09$ \\
    $60234.2128^{+1.0000}_{-0.0000}$ & ZTF &$ g $& $19.97 \pm 0.08$ & $19.81 \pm 0.08$ \\
    $60234.2474^{+1.0000}_{-0.0000}$ & ZTF &$ r $& $18.96 \pm 0.04$ & $18.85 \pm 0.04$ \\
    $60235.4000$ & ATLAS &$ c $& $19.49 \pm 0.08$ & $19.35 \pm 0.08$ \\
    $60236.2413^{+1.0000}_{-0.0000}$ & ZTF &$ g $& $20.08 \pm 0.11$ & $19.92 \pm 0.11$ \\
    $60236.3043^{+1.0000}_{-0.0000}$ & ZTF &$ i $& $18.54 \pm 0.06$ & $18.46 \pm 0.06$ \\
    $60236.3220^{+1.0000}_{-0.0000}$ & ZTF &$ r $& $19.05 \pm 0.05$ & $18.94 \pm 0.05$ \\
    $60237.3900$ & ATLAS &$ c $& $19.56 \pm 0.08$ & $19.42 \pm 0.08$ \\
    $60238.1749^{+1.0000}_{-0.0000}$ & ZTF &$ r $& $19.01 \pm 0.05$ & $18.89 \pm 0.05$ \\
    $60238.2308^{+1.0000}_{-0.0000}$ & ZTF &$ g $& $19.91 \pm 0.08$ & $19.75 \pm 0.08$ \\
    $60239.3032^{+1.0000}_{-0.0000}$ & ZTF &$ i $& $18.58 \pm 0.05$ & $18.49 \pm 0.05$ \\
    $60239.4000$ & ATLAS &$ o $& $18.95 \pm 0.07$ & $18.85 \pm 0.07$ \\
    $60241.3900$ & ATLAS &$ o $& $19.03 \pm 0.14$ & $18.92 \pm 0.14$ \\
    $60242.1976^{+1.0000}_{-0.0000}$ & ZTF &$ r $& $19.16 \pm 0.10$ & $19.05 \pm 0.10$ \\
    $60242.2752^{+1.0000}_{-0.0000}$ & ZTF &$ i $& $18.46 \pm 0.06$ & $18.38 \pm 0.06$ \\
    $60242.3678^{+1.0000}_{-0.0000}$ & ZTF &$ g $& $19.75 \pm 0.16$ & $19.59 \pm 0.16$ \\
    $60243.8801$ & LT/IO:O &$ i $& $19.31 \pm 0.31$ & $19.25 \pm 0.31$ \\
    $60243.8817$ & LT/IO:O & z & $19.21 \pm 0.28$ & $19.14 \pm 0.28$ \\
    $60245.1893^{+1.0000}_{-0.0000}$ & ZTF &$ r $& $19.42 \pm 0.19$ & $19.30 \pm 0.19$ \\
    $60245.3232^{+1.0000}_{-0.0000}$ & ZTF &$ i $& $18.91 \pm 0.16$ & $18.83 \pm 0.16$ \\
    $60246.1103$ & NOT/NOTCam &$ J $& $18.98 \pm 0.08$ & $18.94 \pm 0.08$ \\
    $60246.1239$ & NOT/NOTCam &$ H $& $19.31 \pm 0.11$ & $19.29 \pm 0.11$ \\
    $60247.2280$ & P60/SEDm+RC &$ g $& $<18.70$ & $<18.48$ \\
    $60247.2321$ & P60/SEDm+RC &$ r $& $<18.70$ & $<18.48$ \\
    $60247.4200$ & ATLAS &$ o $& $19.05 \pm 0.13$ & $18.95 \pm 0.13$ \\
    $60247.7317^{+2.5601}_{-2.4399}$ & ZTF &$ g $& $20.39 \pm 0.16$ & $20.23 \pm 0.16$ \\
    $60249.1605$ & P60/SEDm+RC &$ g $& $<19.40$ & $<19.18$ \\
    $60249.1645$ & P60/SEDm+RC &$ r $& $19.54 \pm 0.05$ & $19.30 \pm 0.05$ \\
    $60249.2709^{+1.0000}_{-0.0000}$ & ZTF &$ i $& $18.84 \pm 0.12$ & $18.76 \pm 0.12$ \\
    $60249.4500$ & ATLAS &$ o $& $19.30 \pm 0.11$ & $19.20 \pm 0.11$ \\
    $60249.8616$ & LT/IO:O &$ g $& $20.48 \pm 0.09$ & $20.40 \pm 0.09$ \\
    $60249.8639$ & LT/IO:O &$ r $& $19.61 \pm 0.07$ & $19.34 \pm 0.07$ \\
    $60249.8654$ & LT/IO:O &$ i $& $18.94 \pm 0.06$ & $18.90 \pm 0.06$ \\
    $60249.8669$ & LT/IO:O & z & $19.12 \pm 0.10$ & $19.06 \pm 0.10$ \\
    $60250.2540^{+1.0000}_{-0.0000}$ & ZTF &$ r $& $19.43 \pm 0.08$ & $19.32 \pm 0.08$ \\
    $60251.3900$ & ATLAS &$ o $& $19.50 \pm 0.12$ & $19.39 \pm 0.12$ \\
    $60252.2097^{+1.0000}_{-0.0000}$ & ZTF &$ g $& $20.74 \pm 0.18$ & $20.58 \pm 0.18$ \\
    $60252.2350^{+1.0000}_{-0.0000}$ & ZTF &$ i $& $19.02 \pm 0.10$ & $18.94 \pm 0.10$ \\
    $60252.2912^{+1.0000}_{-0.0000}$ & ZTF &$ r $& $19.49 \pm 0.07$ & $19.38 \pm 0.07$ \\
    $60253.3600$ & ATLAS &$ c $& $20.23 \pm 0.19$ & $20.09 \pm 0.19$ \\
    $60254.1200$ & ATLAS &$ o $& $19.12 \pm 0.16$ & $19.01 \pm 0.16$ \\
    $60254.1444^{+1.0000}_{-0.0000}$ & ZTF &$ r $& $19.53 \pm 0.08$ & $19.41 \pm 0.08$ \\
    $60254.1861^{+1.0000}_{-0.0000}$ & ZTF &$ g $& $20.74 \pm 0.15$ & $20.58 \pm 0.15$ \\
    $60255.2294^{+1.0000}_{-0.0000}$ & ZTF &$ i $& $19.06 \pm 0.11$ & $18.98 \pm 0.11$ \\
    $60255.8261$ & LT/IO:O &$ g $& $20.61 \pm 0.18$ & $20.53 \pm 0.18$ \\
    $60255.8283$ & LT/IO:O &$ r $& $19.77 \pm 0.15$ & $19.50 \pm 0.15$ \\
    $60255.8298$ & LT/IO:O &$ i $& $19.17 \pm 0.15$ & $19.12 \pm 0.15$ \\
    $60256.8445$ & LT/IO:O & z & $19.31 \pm 0.12$ & $19.24 \pm 0.12$ \\
    $60257.1715^{+1.0000}_{-0.0000}$ & ZTF &$ r $& $19.78 \pm 0.13$ & $19.67 \pm 0.13$ \\
    $60257.3600$ & ATLAS &$ c $& $20.39 \pm 0.19$ & $20.25 \pm 0.19$ \\
    $60258.2449^{+1.0518}_{-0.9482}$ & ZTF &$ g $& $20.59 \pm 0.15$ & $20.43 \pm 0.15$ \\
    $60258.2646^{+1.0000}_{-0.0000}$ & ZTF &$ i $& $19.26 \pm 0.12$ & $19.18 \pm 0.12$ \\
    $60259.1332^{+1.0000}_{-0.0000}$ & ZTF &$ r $& $19.65 \pm 0.09$ & $19.53 \pm 0.09$ \\
    $60259.3600$ & ATLAS &$ c $& $20.30 \pm 0.19$ & $20.16 \pm 0.19$ \\
    $60260.1300$ & ATLAS &$ o $& $19.85 \pm 0.18$ & $19.75 \pm 0.18$ \\
    $60260.1733^{+1.0000}_{-0.0000}$ & ZTF &$ i $& $19.10 \pm 0.10$ & $19.02 \pm 0.10$ \\
    $60260.8519$ & LT/IO:O &$ g $& $20.77 \pm 0.11$ & $20.69 \pm 0.11$ \\
    $60260.8541$ & LT/IO:O &$ r $& $19.90 \pm 0.06$ & $19.63 \pm 0.06$ \\
    $60260.8556$ & LT/IO:O &$ i $& $19.28 \pm 0.07$ & $19.24 \pm 0.07$ \\
    $60260.8572$ & LT/IO:O & z & $19.66 \pm 0.12$ & $19.60 \pm 0.12$ \\
    $60261.1559^{+1.0000}_{-0.0000}$ & ZTF &$ r $& $19.67 \pm 0.09$ & $19.55 \pm 0.09$ \\
    $60263.1521^{+1.0000}_{-0.0000}$ & ZTF &$ i $& $19.15 \pm 0.11$ & $19.07 \pm 0.11$ \\
    $60263.3817^{+21.8258}_{-2.1742}$ & ZTF &$ g $& $21.07 \pm 0.18$ & $20.91 \pm 0.18$ \\
    $60265.9312$ & LT/IO:O &$ g $& $20.99 \pm 0.08$ & $20.91 \pm 0.08$ \\
    $60265.9350$ & LT/IO:O &$ r $& $19.97 \pm 0.06$ & $19.69 \pm 0.06$ \\
    $60265.9368$ & LT/IO:O &$ i $& $19.31 \pm 0.06$ & $19.28 \pm 0.06$ \\
    $60265.9385$ & LT/IO:O & z & $19.59 \pm 0.11$ & $19.53 \pm 0.11$ \\
    $60266.1100$ & ATLAS &$ o $& $19.64 \pm 0.16$ & $19.53 \pm 0.16$ \\
    $60267.3300$ & ATLAS &$ o $& $19.64 \pm 0.14$ & $19.53 \pm 0.14$ \\
    $60269.4600$ & ATLAS &$ o $& $19.09 \pm 0.20$ & $18.98 \pm 0.20$ \\
    $60269.9515$ & LT/IO:O &$ g $& $20.87 \pm 0.69$ & $20.79 \pm 0.69$ \\
    $60269.9554$ & LT/IO:O &$ r $& $19.77 \pm 0.19$ & $19.48 \pm 0.19$ \\
    $60269.9571$ & LT/IO:O &$ i $& $19.11 \pm 0.12$ & $19.08 \pm 0.12$ \\
    $60269.9588$ & LT/IO:O & z & $19.34 \pm 0.15$ & $19.27 \pm 0.15$ \\
    $60277.7997$ & LT/IO:O &$ r $& $20.20 \pm 0.23$ & $19.90 \pm 0.23$ \\
    $60277.8018$ & LT/IO:O &$ i $& $19.39 \pm 0.08$ & $19.36 \pm 0.08$ \\
    $60277.8039$ & LT/IO:O & z & $19.51 \pm 0.13$ & $19.45 \pm 0.13$ \\
    $60277.9075$ & LT/IO:O &$ r $& $20.35 \pm 0.14$ & $20.05 \pm 0.14$ \\
    $60277.9099$ & LT/IO:O &$ i $& $19.44 \pm 0.07$ & $19.41 \pm 0.07$ \\
    $60281.0783$ & P60/SEDm+RC &$ g $& $<21.60$ & $<21.38$ \\
    $60281.0810$ & P60/SEDm+RC &$ r $& $20.22 \pm 0.06$ & $19.94 \pm 0.06$ \\
    $60281.0837$ & P60/SEDm+RC &$ i $& $19.43 \pm 0.05$ & $19.39 \pm 0.05$ \\
    $60281.2685$ & MMT/MMIRS &$ J $& $19.52 \pm 0.01$ & $19.48 \pm 0.01$ \\
    $60284.1500^{+1.0000}_{-0.0000}$ & ZTF &$ r $& $20.06 \pm 0.10$ & $19.94 \pm 0.10$ \\
    $60284.2031^{+1.0000}_{-0.0000}$ & ZTF &$ i $& $19.64 \pm 0.14$ & $19.56 \pm 0.14$ \\
    $60284.2900$ & ATLAS &$ o $& $19.85 \pm 0.14$ & $19.75 \pm 0.14$ \\
    $60285.8618$ & LT/IO:O &$ g $& $21.44 \pm 0.16$ & $21.37 \pm 0.16$ \\
    $60285.8659$ & LT/IO:O &$ r $& $20.38 \pm 0.08$ & $20.07 \pm 0.08$ \\
    $60285.8680$ & LT/IO:O &$ i $& $19.64 \pm 0.07$ & $19.62 \pm 0.07$ \\
    $60285.8699$ & LT/IO:O & z & $19.89 \pm 0.17$ & $19.83 \pm 0.17$ \\
    $60288.2119^{+0.9650}_{-1.0350}$ & ZTF &$ r $& $20.68 \pm 0.18$ & $20.57 \pm 0.18$ \\
    $60288.2800$ & ATLAS &$ o $& $19.77 \pm 0.12$ & $19.67 \pm 0.12$ \\
    $60289.1796^{+10.9642}_{-1.0358}$ & ZTF &$ g $& $21.93 \pm 0.41$ & $21.77 \pm 0.41$ \\
    $60290.1770^{+1.0000}_{-0.0000}$ & ZTF &$ r $& $20.56 \pm 0.18$ & $20.44 \pm 0.18$ \\
    $60292.2700$ & ATLAS &$ o $& $20.05 \pm 0.16$ & $19.94 \pm 0.16$ \\
    $60293.8700$ & LT/IO:O &$ g $& $21.93 \pm 0.18$ & $21.85 \pm 0.18$ \\
    $60293.8741$ & LT/IO:O &$ r $& $20.71 \pm 0.09$ & $20.40 \pm 0.09$ \\
    $60293.8761$ & LT/IO:O &$ i $& $19.85 \pm 0.07$ & $19.83 \pm 0.07$ \\
    $60293.8781$ & LT/IO:O & z & $20.22 \pm 0.20$ & $20.16 \pm 0.20$ \\
    $60295.1452^{+5.0000}_{-0.0000}$ & ZTF &$ i $& $19.37 \pm 0.22$ & $19.29 \pm 0.22$ \\
    $60305.8104$ & LT/IO:O &$ i $& $20.08 \pm 0.11$ & $20.06 \pm 0.11$ \\
    $60305.8130$ & LT/IO:O & z & $20.39 \pm 0.20$ & $20.32 \pm 0.20$ \\
    $60305.8458$ & NOT/NOTCam &$ J $& $20.43 \pm 0.08$ & $20.40 \pm 0.08$ \\
    $60305.8592$ & NOT/NOTCam &$ H $& $21.31 \pm 0.24$ & $21.28 \pm 0.24$ \\
    $60307.8453$ & LT/IO:O &$ g $& $22.76 \pm 0.38$ & $22.68 \pm 0.38$ \\
    $60307.8494$ & LT/IO:O &$ r $& $21.04 \pm 0.14$ & $20.73 \pm 0.14$ \\
    $60307.8515$ & LT/IO:O &$ i $& $20.61 \pm 0.12$ & $20.60 \pm 0.12$ \\
    $60307.8534$ & LT/IO:O & z & $20.82 \pm 0.34$ & $20.76 \pm 0.34$ \\
    $60315.8414$ & LT/IO:O &$ r $& $<17.52$ & $<17.21$ \\
    $60315.8438$ & LT/IO:O &$ i $& $<18.35$ & $<18.33$ \\
    $60315.8462$ & LT/IO:O & z & $<17.28$ & $<17.21$ \\
    $60316.8897$ & LT/IO:O &$ r $& $<20.79$ & $<20.48$ \\
    $60316.8920$ & LT/IO:O &$ i $& $20.64 \pm 0.22$ & $20.63 \pm 0.22$ \\
    $60316.8944$ & LT/IO:O & z & $20.73 \pm 0.42$ & $20.66 \pm 0.42$ \\
    $60332.1552^{+19.9801}_{-21.0199}$ & ZTF &$ r $& $21.39 \pm 0.24$ & $21.28 \pm 0.24$ \\
    $60353.8325$ & NOT/NOTCam &$ J $& $20.88 \pm 0.20$ & $20.84 \pm 0.20$ \\
    $60353.8458$ & NOT/NOTCam &$ H $& $21.23 \pm 0.49$ & $21.21 \pm 0.49$ \\
    $60356.1197^{+16.9981}_{-3.0019}$ & ZTF &$ r $& $21.66 \pm 0.23$ & $21.54 \pm 0.23$ \\
    $60441.4771^{+63.0145}_{-8.9855}$ & ZTF &$ r $& $21.97 \pm 0.21$ & $21.86 \pm 0.21$ \\
    $60472.4627^{+7.9987}_{-16.0013}$ & ZTF &$ g $& $22.46 \pm 0.24$ & $22.29 \pm 0.24$ \\
    $60488.1945$ & NOT/NOTCam &$ J $& $21.65 \pm 0.12$ & $21.61 \pm 0.12$ \\
    $60488.2083$ & NOT/NOTCam &$ H $& $22.06 \pm 0.24$ & $22.04 \pm 0.24$ \\
    $60494.3570$ & VLT/FORS2 &$ g $& $22.69 \pm 0.25$ & $22.37 \pm 0.25$ \\
    $60494.3660$ & VLT/FORS2 &$ r $& $21.74 \pm 0.20$ & $21.80 \pm 0.20$ \\
    $60494.3740$ & VLT/FORS2 &$ i $& $21.86 \pm 0.22$ & $21.74 \pm 0.22$ \\
    $60494.3860$ & VLT/FORS2 & z & $22.20 \pm 0.21$ & $22.13 \pm 0.21$ \\
    $60494.4551^{+12.9892}_{-13.0108}$ & ZTF &$ g $& $22.48 \pm 0.22$ & $22.31 \pm 0.22$ \\
    $60507.4680^{+19.9861}_{-50.0139}$ & ZTF &$ i $& $21.61 \pm 0.23$ & $21.52 \pm 0.23$ \\
    $60519.9404^{+9.4707}_{-10.5293}$ & ZTF &$ g $& $22.29 \pm 0.18$ & $22.13 \pm 0.18$ \\
    60520.4199 & Magellan/IMACS	 &$ r $&$ 22.25 \pm 0.07 $&$ 21.93 \pm 0.07 $\\
    60520.4255 & Magellan/IMACS	 &$ g $&$ 22.83 \pm 0.09 $&$ 22.61 \pm 0.09 $\\
    $60523.4279^{+14.9667}_{-16.0333}$ & ZTF &$ r $& $22.21 \pm 0.22$ & $22.09 \pm 0.22$ \\
    $60528.3040$ & VLT/FORS2 &$ g $& $22.82 \pm 0.24$ & $22.51 \pm 0.24$ \\
    $60528.3130$ & VLT/FORS2 &$ r $& $21.82 \pm 0.21$ & $21.89 \pm 0.21$ \\
    $60528.3210$ & VLT/FORS2 &$ i $& $21.65 \pm 0.19$ & $21.53 \pm 0.19$ \\
    $60528.3330$ & VLT/FORS2 & z & $22.25 \pm 0.07$ & $22.18 \pm 0.07$ \\
    $60549.1780$ & VLT/FORS2 &$ g $& $23.45 \pm 0.07$ & $23.14 \pm 0.07$ \\
    $60549.1860$ & VLT/FORS2 &$ r $& $22.08 \pm 0.18$ & $22.35 \pm 0.18$ \\
    $60549.1950$ & VLT/FORS2 &$ i $& $21.55 \pm 0.05$ & $21.88 \pm 0.05$ \\
    $60549.2070$ & VLT/FORS2 & z & $22.75 \pm 0.32$ & $22.69 \pm 0.32$ \\
    $60553.3446^{+24.0344}_{-22.9656}$ & ZTF &$ g $& $23.19 \pm 0.24$ & $23.03 \pm 0.24$ \\
    $60568.4567$ & MMT/MMIRS &$ J $& $22.81 \pm 0.07$ & $22.77 \pm 0.07$ \\
    $60585.2330$ & VLT/FORS2 &$ g $& $24.24 \pm 0.17$ & $23.92 \pm 0.17$ \\
    $60585.2410$ & VLT/FORS2 &$ r $& $23.37 \pm 0.19$ & $23.43 \pm 0.19$ \\
    $60585.2500$ & VLT/FORS2 &$ i $& $22.76 \pm 0.25$ & $22.64 \pm 0.25$ \\
    $60585.2620$ & VLT/FORS2 & z & $<23.34$ & $<23.28$ \\
    60620.0765 & Magellan/IMACS	 &$ r $&$ <23.64 $&$ <23.30 $\\
    $60850.3510$ & VLT/FORS2 &$ g $& $<24.71$ & $<24.40$ \\
    $60850.3600$ & VLT/FORS2 &$ r $& $<24.39$ & $<24.46$ \\
    $60850.3690$ & VLT/FORS2 &$ i $& $<23.66$ & $<23.54$ \\
    $60850.3800$ & VLT/FORS2 & z & $<23.42$ & $<23.36$ \\
\enddata
\tablecomments{Magnitudes are not corrected for Milky Way extinction and are reported in the AB system. The third column lists the brightness before applying the $s$ correction and correcting for Milky Way extinction, while the fourth column lists the brightness after applying both corrections. All uncertainties are reported at $1\sigma$ confidence. Upper limits are reported at $3\sigma$ confidence. 
}
\end{deluxetable}
\clearpage
\begin{table}
\centering
\caption{\textit{WISE} photometry of SN\,2023iop.}
\label{tab:SN2023iop:photometry:wise}
\begin{tabular}{ccc}
\hline\hline
MJD & Filter & $F_\nu$ (mJy) \\
\hline
55380.8853 &$ W1 $& $0.004 \pm 0.005$ \\
55380.8853 &$ W2 $& $0.007 \pm 0.014$ \\
55558.7050 &$ W1 $& $0.002 \pm 0.006$ \\
55558.7050 &$ W2 $& $0.015 \pm 0.013$ \\
56656.3342 &$ W1 $& $-0.014 \pm 0.007$ \\
56656.3342 &$ W2 $& $0.015 \pm 0.014$ \\
56844.4202 &$ W1 $& $-0.001 \pm 0.007$ \\
56844.4202 &$ W2 $& $-0.012 \pm 0.014$ \\
57021.8451 &$ W1 $& $-0.022 \pm 0.009$ \\
57021.8451 &$ W2 $& $-0.012 \pm 0.018$ \\
57208.3320 &$ W1 $& $0.002 \pm 0.009$ \\
57208.3321 &$ W2 $& $0.000 \pm 0.019$ \\
57382.5125 &$ W1 $& $-0.017 \pm 0.009$ \\
57382.5125 &$ W2 $& $0.009 \pm 0.017$ \\
57573.2511 &$ W1 $& $-0.002 \pm 0.009$ \\
57573.2511 &$ W2 $& $-0.010 \pm 0.019$ \\
57741.7115 &$ W1 $& $0.005 \pm 0.008$ \\
57741.7115 &$ W2 $& $0.031 \pm 0.017$ \\
57937.3997 &$ W1 $& $0.004 \pm 0.008$ \\
57937.3997 &$ W2 $& $0.027 \pm 0.020$ \\
58102.3487 &$ W1 $& $-0.012 \pm 0.009$ \\
58102.3487 &$ W2 $& $-0.000 \pm 0.017$ \\
58303.0987 &$ W1 $& $0.001 \pm 0.009$ \\
58303.1640 &$ W2 $& $0.018 \pm 0.020$ \\
58465.5517 &$ W1 $& $-0.006 \pm 0.009$ \\
58465.5517 &$ W2 $& $0.010 \pm 0.019$ \\
58668.7789 &$ W1 $& $-0.002 \pm 0.009$ \\
58668.7789 &$ W2 $& $-0.025 \pm 0.023$ \\
58830.1960 &$ W1 $& $-0.007 \pm 0.009$ \\
58830.1960 &$ W2 $& $-0.027 \pm 0.019$ \\
59032.8332 &$ W1 $& $-0.011 \pm 0.010$ \\
59032.8332 &$ W2 $& $0.012 \pm 0.025$ \\
59194.3273 &$ W1 $& $-0.006 \pm 0.009$ \\
59194.3273 &$ W2 $& $-0.002 \pm 0.019$ \\
59398.5701 &$ W1 $& $-0.033 \pm 0.009$ \\
59398.5701 &$ W2 $& $0.002 \pm 0.022$ \\
59559.8952 &$ W1 $& $0.001 \pm 0.007$ \\
59559.8952 &$ W2 $& $-0.006 \pm 0.018$ \\
59764.2210 &$ W1 $& $-0.015 \pm 0.010$ \\
59764.2210 &$ W2 $& $0.012 \pm 0.024$ \\
59925.6254 &$ W1 $& $-0.003 \pm 0.009$ \\
59925.6254 &$ W2 $& $0.021 \pm 0.019$ \\
60128.3509 &$ W1 $& $0.021 \pm 0.011$ \\
60128.3509 &$ W2 $& $0.022 \pm 0.022$ \\
60293.7897 &$ W1 $& $0.016 \pm 0.009$ \\
60293.7897 &$ W2 $& $0.041 \pm 0.019$ \\
60495.4396 &$ W1 $& $-0.013 \pm 0.014$ \\
60495.4396 &$ W2 $& $-0.050 \pm 0.043$ \\
\hline
\end{tabular}
\end{table}
\clearpage
\begin{table*}
\caption{Spectroscopic observations of SN2023iop}\label{tab:spectroscopy}
\hspace{-2.125cm}
\begin{tabular}{lccccccc}
\hline\hline
MJD & Phase & Telescope/ & Disperser & Slit & Wavelength  & Spectral  & Exp. \\
 & (days) & Instrument & & width (arcsec) & range (\AA) & resolution $R$ & time (s) \\
\hline
60113.394 & -51.4 & P60/SEDm            & \nodata & IFU & 4650 -- 9200 & 100 & 2250 \\
60151.457 & -16.8 & P60/SEDm            & \nodata & IFU & 4650 -- 9200 & 100 & 2250 \\
60173.232 &  +2.9 & P60/SEDm            & \nodata & IFU & 4650 -- 9200 & 100 & 2250 \\
60182.250 & +11.1 & P60/SEDm            & \nodata & IFU & 4650 -- 9200 & 100 & 2250 \\
60183.111 & +11.9 & LT/SPRAT            & blue    & 1.8 & 4020 -- 7960 & 350 & 1500 \\
60185.004 & +13.6 & NOT/ALFOSC          & Gr\#4   & 1.0 & 3600 -- 9600 & 360 &2400  \\
60196.382 & +23.9 & P60/SEDm            & \nodata & IFU & 4650 -- 9200 & 100 & 2250 \\
60203.155 & +30.1 & P60/SEDm            & \nodata & IFU & 4650 -- 9200 & 100 & 2250 \\
60207.393 & +33.9 & P60/SEDm            & \nodata & IFU & 4650 -- 9200 & 100 & 2250 \\
60211.034 & +37.2 & NOT/ALFOSC          & Gr\#4   & 1.0 & 3600 -- 9600 & 360 & 2400 \\
60211.094 & +37.3 & VLT/FORS2$^\dagger$ & 300V    & 1.0 & 4200 -- 9000 & 440 & 8640 \\
60214.235 & +40.1 & P60/SEDm            & \nodata & IFU & 4650 -- 9200 & 100 & 2250 \\
60224.402 & +49.4 & P60/SEDm            & \nodata & IFU & 4650 -- 9200 & 100 & 220  \\
60224.539 & +49.5 & Keck/DEIMOS         & 600ZD   & 1.0 & 5430 -- 10130 & 2000 & 900  \\
60226.305 & +51.1 & P60/SEDm            & \nodata & IFU & 4650 -- 9200 & 100 & 200  \\
60228.257 & +52.9 & P60/SEDm            & \nodata & IFU & 4650 -- 9200 & 100 & 200  \\
60233.289 & +57.4 & P60/SEDm            & \nodata & IFU & 4650 -- 9200 & 100 & 230  \\
60234.298 & +58.4 & Keck/LRIS           & 400/3400, 400/8500 & 1.0 & 3076 -- 9350 & 600/1200 & 1200 \\
60238.042 & +61.7 & Magellan/MagE       & \nodata & 0.85 & 3200 -- 10000 & 4800 & 5400 \\
60249.019 & +71.7 & VLT/X-shooter       & \nodata & 1.0,0.9,0.9 & 3000 -- 24800 & 5400/8900/5600 & 3600 \\
60280.049 & +99.9 & VLT/X-shooter       & \nodata & 1.0,0.9,0.9 & 3000 -- 24800 & 5400/8900/5600 & 3600 \\
60319.128 &+135.3 & MMT/Binospec        & G270 & 1.0 & 3850 -- 9150 & 1340 & 3600 \\
60324.075 &+139.8 & MMT/Binospec        & G270 & 1.0 & 3850 -- 9150 & 1340 & 3600 \\
60496.390 &+296.2 & MMT/Binospec        & G270 & 1.0 & 3850 -- 9150 & 1340 & 7200 \\
60529.264 &+326.0 & VLT/FORS2           & 300V & 1.0         & 3800 -- 9600   & 440            & 14400 \\
60555.505 &+349.9 & Keck/LRIS           & 600/4000, 400/8500 & 1.0, 1.0 & 3400 -- 10275 & 1900/1200 & 3050 \\
60589.337 &+380.6 & Keck/LRIS           & 400/3400, 400/8500 & 1.0, 1.0 & 3076 -- 9350 & 600/1200 & 7275 \\
60678.217 &+461.2 & Keck/LRIS           & 400/3400, 400/8500 & 1.0, 1.0 & 3076 -- 9350 & 600/1200 & 3600 \\
60703.077 &+483.8 & LBT/MODS            & G400L, G670L  & 1.2       & 3200 -- 9500 & 1400, 2300 & 10800 \\
60857.342 &+623.8 & VLT/FORS2           & 300V & 1.0    & 3800 -- 9600   & 440            & 14400 \\
\hline
\end{tabular}
\tablecomments{The modified Julian dates quote the beginning of each spectroscopic observation.
The phase is reported in the rest frame with respect to the time of $r$-band maximum light at MJD = 60169.63. For the multi-arm instruments LRIS and X-shooter, we report the exposure time and spectral resolution of each detector. The wavelength ranges and spectral resolution values are taken from instrument manuals and reported in the observer frame. The spectral resolutions refer to the unbanned data. The $^\dagger$ indicates the epoch of the VLT/FORS2 spectropolarimetry.
}
\end{table*}
\clearpage
\begin{table}
\caption{\sn's host galaxy photometry}\label{tab:host_phot}
%\centering
\hspace{-1.25cm}
\begin{tabular}{llcc}
\hline\hline
Telescope & Instrument  & Filter & Brightness   \\
  & &        & (mag)        \\
\hline
CFHT & MegaPrime   & $u$&$23.39 \pm 0.17$\\
VLT & FORS2   & $g\_{\rm HIGH   }$&$22.21 \pm 0.03$\\
VLT & FORS2   & $R\_{\rm SPECIAL}$&$21.70 \pm 0.03$\\
VLT & FORS2   & $I\_{\rm BESSELL}$&$21.58 \pm 0.05$\\
VLT & FORS2   & $z\_{\rm SPECIAL}$&$21.73 \pm 0.07$\\
LBT & LUCI    & $J               $&$21.47 \pm 0.10$\\
LBT & LUCI    & $K_{\rm s}              $&$21.75 \pm 0.13$\\
\hline
\end{tabular}
\tablecomments{All measurements are in the AB system and not corrected for extinction. Their uncertainties are reported at the $1\sigma$ confidence level.}
\end{table}

\clearpage
\startlongtable
\begin{deluxetable}{ccc}
\tablecolumns{3}
\tablecaption{Photometric measurements of SN\,1987A-like host galaxies. \label{tab:host_87A_phot}}
\tablehead{
  \colhead{Survey} & \colhead{Filter} & \colhead{Brightness (mag)}
}
\startdata
\multicolumn{3}{c}{\textbf{OGLE-73} ($z=0.1225$)}\\
\tableline
LS &$ g $& $22.42 \pm 0.03$ \\
LS &$ r $& $21.90 \pm 0.03$ \\
LS &$ i $& $21.63 \pm 0.04$ \\
LS &$ z $& $21.62 \pm 0.05$ \\
\tableline
\multicolumn{3}{c}{\textbf{SN\,2018cub} ($z=0.044$)}\\
\tableline
SDSS &$ u $& $16.58 \pm 0.02$ \\
SDSS &$ g $& $15.72 \pm 0.01$ \\
SDSS &$ r $& $15.42 \pm 0.02$ \\
SDSS &$ i $& $15.17 \pm 0.02$ \\
SDSS &$ z $& $14.97 \pm 0.03$ \\
Pan-STARRS &$ g $& $15.74 \pm 0.01$ \\
Pan-STARRS &$ r $& $15.39 \pm 0.03$ \\
Pan-STARRS &$ i $& $15.24 \pm 0.05$ \\
Pan-STARRS &$ z $& $15.16 \pm 0.06$ \\
Pan-STARRS &$ y $& $15.10 \pm 0.07$ \\
2MASS &$ J $& $15.06 \pm 0.05$ \\
2MASS &$ H $& $14.76 \pm 0.06$ \\
2MASS &$ K $& $15.02 \pm 0.07$ \\
\wise &$ W1 $& $15.37 \pm 0.02$ \\
\wise &$ W2 $& $15.89 \pm 0.03$ \\
\tableline
\multicolumn{3}{c}{\textbf{SN\,2018lrq} ($z=0.025$)}\\
\tableline
\galex &$ FUV $& $19.10 \pm 0.13$ \\
\galex &$ NUV $& $18.81 \pm 0.04$ \\
SDSS &$ u $& $16.95 \pm 0.06$ \\
SDSS &$ g $& $15.59 \pm 0.02$ \\
SDSS &$ r $& $14.88 \pm 0.02$ \\
SDSS &$ i $& $14.47 \pm 0.02$ \\
SDSS &$ z $& $14.14 \pm 0.02$ \\
Pan-STARRS &$ g $& $15.60 \pm 0.03$ \\
Pan-STARRS &$ r $& $14.89 \pm 0.02$ \\
Pan-STARRS &$ i $& $14.53 \pm 0.02$ \\
Pan-STARRS &$ z $& $14.27 \pm 0.02$ \\
Pan-STARRS &$ y $& $14.02 \pm 0.05$ \\
2MASS &$ H $& $13.35 \pm 0.02$ \\
\tableline
\multicolumn{3}{c}{\textbf{SN\,2018ego} ($z=0.037$)}\\
\tableline
\galex &$ FUV $& $18.49 \pm 0.11$ \\
\galex &$ NUV $& $18.05 \pm 0.04$ \\
SDSS &$ u $& $17.13 \pm 0.06$ \\
SDSS &$ g $& $15.99 \pm 0.04$ \\
SDSS &$ r $& $15.50 \pm 0.05$ \\
SDSS &$ i $& $15.22 \pm 0.03$ \\
SDSS &$ z $& $14.97 \pm 0.06$ \\
\wise &$ W1 $& $15.38 \pm 0.02$ \\
\wise &$ W2 $& $15.91 \pm 0.03$ \\
\tableline
\multicolumn{3}{c}{\textbf{SN\,2018lsg} ($z=0.025$)}\\
\tableline
Pan-STARRS &$ g $& $14.85 \pm 0.02$ \\
Pan-STARRS &$ r $& $14.31 \pm 0.06$ \\
Pan-STARRS &$ i $& $14.11 \pm 0.04$ \\
Pan-STARRS &$ z $& $13.89 \pm 0.02$ \\
Pan-STARRS &$ y $& $13.68 \pm 0.04$ \\
2MASS &$ H $& $13.37 \pm 0.01$ \\
2MASS &$ J $& $13.58 \pm 0.01$ \\
\wise &$ W1 $& $14.07 \pm 0.01$ \\
\wise &$ W2 $& $14.40 \pm 0.01$ \\
\tableline
\multicolumn{3}{c}{\textbf{SN\,2018imj} ($z=0.013$)}\\
\tableline
Pan-STARRS &$ g $& $13.56 \pm 0.09$ \\
Pan-STARRS &$ r $& $12.67 \pm 0.10$ \\
Pan-STARRS &$ i $& $12.23 \pm 0.08$ \\
Pan-STARRS &$ z $& $11.94 \pm 0.07$ \\
Pan-STARRS &$ y $& $11.67 \pm 0.09$ \\
2MASS &$ H $& $11.34 \pm 0.01$ \\
2MASS &$ J $& $11.28 \pm 0.01$ \\
\wise &$ W1 $& $11.90 \pm 0.01$ \\
\wise &$ W2 $& $12.48 \pm 0.01$ \\
\tableline
\multicolumn{3}{c}{\textbf{SN\,2018hna} ($z=0.002$)}\\
\tableline
SDSS &$ u $& $14.94 \pm 0.03$ \\
SDSS &$ g $& $14.16 \pm 0.03$ \\
SDSS &$ r $& $13.91 \pm 0.03$ \\
SDSS &$ i $& $13.70 \pm 0.03$ \\
SDSS &$ z $& $13.70 \pm 0.09$ \\
\tableline
\multicolumn{3}{c}{\textbf{SN\,2019bsw} ($z=0.029$)}\\
\tableline
Pan-STARRS &$ g $& $16.05 \pm 0.02$ \\
Pan-STARRS &$ r $& $15.63 \pm 0.02$ \\
Pan-STARRS &$ i $& $15.45 \pm 0.02$ \\
Pan-STARRS &$ z $& $15.33 \pm 0.02$ \\
Pan-STARRS &$ y $& $15.14 \pm 0.05$ \\
\wise &$ W1 $& $15.84 \pm 0.02$ \\
\wise &$ W2 $& $16.49 \pm 0.07$ \\
\tableline
\multicolumn{3}{c}{\textbf{SN\,2019zfc} ($z=0.031$)}\\
\tableline
\galex &$ FUV $& $21.52 \pm 0.18$ \\
\galex &$ NUV $& $21.49 \pm 0.13$ \\
SDSS &$ u $& $20.51 \pm 0.41$ \\
SDSS &$ g $& $19.39 \pm 0.08$ \\
SDSS &$ r $& $19.28 \pm 0.12$ \\
SDSS &$ i $& $18.98 \pm 0.13$ \\
SDSS &$ z $& $19.00 \pm 0.43$ \\
Pan-STARRS &$ g $& $19.43 \pm 0.16$ \\
Pan-STARRS &$ r $& $19.16 \pm 0.06$ \\
Pan-STARRS &$ i $& $18.93 \pm 0.06$ \\
Pan-STARRS &$ z $& $19.00 \pm 0.16$ \\
Pan-STARRS &$ y $& $18.85 \pm 0.47$ \\
\tableline
\multicolumn{3}{c}{\textbf{SN\,2020oem} ($z=0.042$)}\\
\tableline
\galex &$ FUV $& $19.34 \pm 0.04$ \\
\galex &$ NUV $& $19.14 \pm 0.03$ \\
SDSS &$ u $& $18.27 \pm 0.06$ \\
SDSS &$ z $& $16.21 \pm 0.07$ \\
Pan-STARRS &$ g $& $17.12 \pm 0.04$ \\
Pan-STARRS &$ r $& $16.73 \pm 0.02$ \\
Pan-STARRS &$ i $& $16.50 \pm 0.02$ \\
Pan-STARRS &$ z $& $16.41 \pm 0.04$ \\
Pan-STARRS &$ y $& $16.25 \pm 0.05$ \\
LS &$ g $& $17.00 \pm 0.00$ \\
LS &$ r $& $16.60 \pm 0.01$ \\
LS &$ z $& $16.25 \pm 0.01$ \\
\tableline
\multicolumn{3}{c}{\textbf{SN\,2020abah} ($z=0.03$)}\\
\tableline
\galex &$ FUV $& $18.19 \pm 0.09$ \\
\galex &$ NUV $& $17.49 \pm 0.04$ \\
SDSS &$ u $& $16.49 \pm 0.02$ \\
SDSS &$ g $& $15.23 \pm 0.01$ \\
SDSS &$ r $& $14.63 \pm 0.01$ \\
SDSS &$ i $& $14.30 \pm 0.01$ \\
SDSS &$ z $& $14.06 \pm 0.02$ \\
Pan-STARRS &$ g $& $15.19 \pm 0.01$ \\
Pan-STARRS &$ r $& $14.60 \pm 0.01$ \\
Pan-STARRS &$ i $& $14.33 \pm 0.00$ \\
Pan-STARRS &$ z $& $14.10 \pm 0.01$ \\
Pan-STARRS &$ y $& $13.90 \pm 0.02$ \\
\wise &$ W1 $& $14.74 \pm 0.04$ \\
\wise &$ W2 $& $15.37 \pm 0.05$ \\
\tableline
\multicolumn{3}{c}{\textbf{SN\,2021mju} ($z=0.028$)}\\
\tableline
SDSS &$ u $& $17.87 \pm 0.07$ \\
SDSS &$ g $& $16.57 \pm 0.02$ \\
SDSS &$ r $& $15.87 \pm 0.01$ \\
Pan-STARRS &$ g $& $16.44 \pm 0.03$ \\
Pan-STARRS &$ r $& $15.84 \pm 0.02$ \\
Pan-STARRS &$ i $& $15.49 \pm 0.02$ \\
Pan-STARRS &$ z $& $15.26 \pm 0.02$ \\
Pan-STARRS &$ y $& $15.06 \pm 0.03$ \\
\wise &$ W1 $& $15.58 \pm 0.04$ \\
\wise &$ W2 $& $16.17 \pm 0.03$ \\
\tableline
\multicolumn{3}{c}{\textbf{SN\,2021skm} ($z=0.031$)}\\
\tableline
\galex &$ FUV $& $19.92 \pm 0.20$ \\
\galex &$ NUV $& $19.31 \pm 0.07$ \\
SDSS &$ u $& $18.21 \pm 0.05$ \\
SDSS &$ g $& $17.15 \pm 0.01$ \\
SDSS &$ r $& $16.61 \pm 0.02$ \\
SDSS &$ i $& $16.31 \pm 0.02$ \\
SDSS &$ z $& $16.01 \pm 0.04$ \\
Pan-STARRS &$ g $& $17.13 \pm 0.02$ \\
Pan-STARRS &$ r $& $16.55 \pm 0.03$ \\
Pan-STARRS &$ i $& $16.33 \pm 0.01$ \\
Pan-STARRS &$ z $& $16.07 \pm 0.02$ \\
Pan-STARRS &$ y $& $15.96 \pm 0.05$ \\
2MASS &$ J $& $15.06 \pm 0.05$ \\
2MASS &$ H $& $14.76 \pm 0.06$ \\
\wise &$ W1 $& $16.46 \pm 0.03$ \\
\wise &$ W2 $& $16.99 \pm 0.03$ \\
\tableline
\multicolumn{3}{c}{\textbf{SN\,2021wun} ($z=0.023$)}\\
\tableline
\galex &$ FUV $& $19.86 \pm 0.15$ \\
\galex &$ NUV $& $19.17 \pm 0.06$ \\
SDSS &$ u $& $18.59 \pm 0.06$ \\
SDSS &$ g $& $17.67 \pm 0.02$ \\
SDSS &$ r $& $17.28 \pm 0.02$ \\
SDSS &$ i $& $17.06 \pm 0.03$ \\
SDSS &$ z $& $16.73 \pm 0.09$ \\
Pan-STARRS &$ g $& $17.72 \pm 0.06$ \\
Pan-STARRS &$ r $& $17.26 \pm 0.03$ \\
Pan-STARRS &$ i $& $17.13 \pm 0.02$ \\
Pan-STARRS &$ z $& $17.06 \pm 0.14$ \\
Pan-STARRS &$ y $& $16.88 \pm 0.13$ \\
\wise &$ W1 $& $17.71 \pm 0.05$ \\
\wise &$ W2 $& $18.50 \pm 0.09$ \\
\enddata
\tablecomments{All measurements are in the AB system and not corrected for extinction. Uncertainties are reported at the $1\sigma$ confidence level.}
\end{deluxetable}
\clearpage
\begin{table*}
\caption{Properties of the SN\,1987A-like host galaxies measured with \program{Prospector}.}\label{tab:host_87A_prospector}
\hspace{-1.5cm}
%\scriptsize
\begin{tabular}{lcccccccc}
\hline\hline
IAU Name  & $\chi^2/\rm{n.o.f.}$  &$ m_r $&$ M_r   $&$\log~\rm SFR  $&$ \log M   $& Age &$ E(B-V)$ \\
&&(mag) & (mag) & ($M_\odot\,\rm yr^{-1}$) & ($M_\odot$) & (Gyr) & (mag)\\
\hline
OGLE2014-SN-73 &$ 2.1/4 $&$21.80^{+0.07}_{-0.07}$&$-17.25^{+0.05}_{-0.06}$&$-0.89^{+0.35}_{-0.36}$ & $8.39^{+0.18}_{-0.22}$ & $2.9^{+5.2}_{-2.1}$ & $0.15^{+0.09}_{-0.08}$\\
SN\,2018cub   &$ 19.3/15 $&$15.41^{+0.04}_{-0.02}$&$-21.25^{+0.04}_{-0.02}$&$0.58^{+0.06}_{-0.06}$ & $10.17^{+0.06}_{-0.11}$ & $9.6^{+2.9}_{-4.8}$ & $0.13^{+0.02}_{-0.02}$\\
SN\,2018lrq   &$ 14.2/13 $&$14.89^{+0.04}_{-0.03}$&$-20.42^{+0.03}_{-0.05}$&$0.20^{+0.76}_{-0.39}$ & $10.20^{+0.13}_{-0.07}$ & $2.6^{+3.0}_{-1.1}$ & $0.33^{+0.24}_{-0.12}$\\
SN\,2018ego   &$ 2.5/9 $&$15.37^{+0.06}_{-0.04}$&$-20.82^{+0.06}_{-0.04}$&$0.12^{+0.10}_{-0.11}$ & $10.14^{+0.14}_{-0.20}$ & $5.4^{+4.8}_{-3.3}$ & $0.15^{+0.03}_{-0.03}$\\
SN\, 2018lsg &$ 5.4/9 $&$14.24^{+0.04}_{-0.05}$&$-21.06^{+0.04}_{-0.05}$&$0.40^{+0.16}_{-0.25}$ & $10.24^{+0.13}_{-0.13}$ & $5.5^{+5.3}_{-3.5}$ & $0.17^{+0.06}_{-0.06}$\\
SN\,2018imj   &$ 14.1/9 $&$12.18^{+0.06}_{-0.05}$&$-21.65^{+0.06}_{-0.05}$&$-3.17^{+2.75}_{-8.27}$ & $10.86^{+0.15}_{-0.21}$ & $5.5^{+4.3}_{-2.8}$ & $0.25^{+0.04}_{-0.05}$\\
SN\,2018hna   &$ 5.0/5 $&$13.82^{+0.04}_{-0.05}$&$-15.92^{+0.04}_{-0.05}$&$-1.54^{+0.31}_{-0.20}$ & $7.89^{+0.13}_{-0.17}$ & $5.2^{+5.1}_{-3.6}$ & $0.09^{+0.10}_{-0.06}$\\
SN\,2019bsw  &$ 1.3/7 $&$15.55^{+0.04}_{-0.04}$&$-20.05^{+0.04}_{-0.04}$&$-0.51^{+0.30}_{-1.04}$ & $9.72^{+0.15}_{-0.17}$ & $3.3^{+5.4}_{-1.8}$ & $0.07^{+0.05}_{-0.04}$\\
SN\,2019zfc   &$ 5.3/12 $&$18.85^{+0.05}_{-0.05}$&$-17.00^{+0.05}_{-0.07}$&$-0.87^{+0.31}_{-0.24}$ & $8.10^{+0.19}_{-0.24}$ & $1.9^{+3.4}_{-1.4}$ & $0.14^{+0.08}_{-0.06}$\\
SN\,2020oem   &$ 7.9/12 $&$16.68^{+0.05}_{-0.06}$&$-19.86^{+0.04}_{-0.04}$&$0.13^{+0.23}_{-0.40}$ & $9.59^{+0.13}_{-0.12}$ & $4.0^{+4.3}_{-2.3}$ & $0.19^{+0.06}_{-0.08}$\\
SN\,2020abah  &$ 20.9/14 $&$14.55^{+0.03}_{-0.02}$&$-21.14^{+0.03}_{-0.02}$&$-0.20^{+0.12}_{-0.12}$ & $10.50^{+0.06}_{-0.13}$ & $10.2^{+2.5}_{-4.2}$ & $0.09^{+0.04}_{-0.04}$\\
SN\,2021mju   &$ 5.9/10 $&$15.68^{+0.03}_{-0.03}$&$-19.86^{+0.03}_{-0.03}$&$-0.78^{+0.26}_{-1.48}$ & $9.96^{+0.12}_{-0.18}$ & $6.5^{+4.5}_{-3.7}$ & $0.11^{+0.09}_{-0.07}$\\
SN\,2021skm   &$ 16.2/16 $&$16.42^{+0.04}_{-0.03}$&$-19.36^{+0.03}_{-0.04}$&$-0.52^{+0.11}_{-0.13}$ & $9.64^{+0.10}_{-0.15}$ & $7.6^{+4.2}_{-3.8}$ & $0.11^{+0.04}_{-0.06}$\\
SN\,2021wun   &$ 25.0/14 $&$17.16^{+0.04}_{-0.02}$&$-17.93^{+0.04}_{-0.02}$&$-1.26^{+0.10}_{-0.09}$ & $8.89^{+0.12}_{-0.18}$ & $6.5^{+4.3}_{-3.6}$ & $0.04^{+0.03}_{-0.02}$\\
\hline
\end{tabular}
\tablecomments{All uncertainties are reported at the $1\sigma$ confidence level. The $\chi^2$ value was measured from the best fit, and ``n.o.f.'' stands for the numbers of filters used in the fit. The apparent magnitudes are model-predicted apparent magnitudes. }
\end{table*}
\clearpage
\begin{table}
\centering
\caption{Bolometric light curve of SN\,2023iop and the evolution of the blackbody properties.}
\label{tab:lc:bolometric}
\begin{tabular}{cccc}
\hline\hline
Phase & $L_{\rm bol}$ & $\log R_{\rm BB}$ & $T_{\rm BB}$ \\
(d) & ($10^{43}$\,erg\,s$^{-1}$) & (cm) & (K) \\
\hline
$-48.0$ & $2.96 \pm 0.20$ & $15.58 \pm 0.01$ & $7032 \pm 85$ \\
$-41.4$ & $2.76 \pm 0.78$ & $15.63 \pm 0.04$ & $6593 \pm 318$ \\
$-32.7$ & $3.40 \pm 0.90$ & $15.68 \pm 0.04$ & $6530 \pm 302$ \\
$-28.3$ & $3.50 \pm 0.69$ & $15.71 \pm 0.03$ & $6403 \pm 203$ \\
$-23.9$ & $3.45 \pm 0.53$ & $15.74 \pm 0.02$ & $6185 \pm 165$ \\
$-19.1$ & $3.56 \pm 0.42$ & $15.71 \pm 0.02$ & $6445 \pm 130$ \\
$-14.5$ & $3.68 \pm 0.51$ & $15.75 \pm 0.02$ & $6185 \pm 143$ \\
$-6.9$ & $3.98 \pm 0.94$ & $15.70 \pm 0.04$ & $6665 \pm 279$ \\
$-0.5$ & $3.44 \pm 0.57$ & $15.73 \pm 0.03$ & $6242 \pm 168$ \\
$+2.4$ & $3.61 \pm 0.63$ & $15.72 \pm 0.03$ & $6347 \pm 184$ \\
$+12.0$ & $3.44 \pm 0.34$ & $15.76 \pm 0.02$ & $6069 \pm 104$ \\
$+21.4$ & $3.55 \pm 0.40$ & $15.82 \pm 0.02$ & $5729 \pm 104$ \\
$+25.4$ & $3.35 \pm 0.44$ & $15.76 \pm 0.02$ & $5993 \pm 124$ \\
$+30.1$ & $3.24 \pm 0.23$ & $15.81 \pm 0.01$ & $5649 \pm 66$ \\
$+34.1$ & $3.29 \pm 0.28$ & $15.87 \pm 0.01$ & $5316 \pm 74$ \\
$+39.5$ & $3.26 \pm 0.37$ & $15.88 \pm 0.02$ & $5230 \pm 95$ \\
$+44.2$ & $3.39 \pm 0.96$ & $15.94 \pm 0.05$ & $4973 \pm 219$ \\
$+48.7$ & $2.78 \pm 0.37$ & $15.84 \pm 0.02$ & $5303 \pm 113$ \\
$+52.5$ & $2.82 \pm 0.22$ & $15.90 \pm 0.01$ & $4944 \pm 60$ \\
$+57.4$ & $2.64 \pm 0.33$ & $15.93 \pm 0.02$ & $4742 \pm 83$ \\
$+61.2$ & $2.37 \pm 0.44$ & $15.90 \pm 0.03$ & $4773 \pm 132$ \\
$+66.8$ & $2.19 \pm 0.83$ & $15.83 \pm 0.06$ & $5044 \pm 309$ \\
$+72.1$ & $1.55 \pm 0.32$ & $15.80 \pm 0.04$ & $4816 \pm 150$ \\
$+75.7$ & $1.69 \pm 0.55$ & $15.91 \pm 0.06$ & $4341 \pm 199$ \\
$+79.4$ & $1.24 \pm 0.36$ & $15.72 \pm 0.05$ & $4963 \pm 223$ \\
$+83.1$ & $1.15 \pm 0.28$ & $15.73 \pm 0.04$ & $4828 \pm 179$ \\
$+88.9$ & $1.10 \pm 0.22$ & $15.76 \pm 0.04$ & $4612 \pm 139$ \\
$+97.9$ & $0.99 \pm 0.60$ & $15.73 \pm 0.10$ & $4671 \pm 466$ \\
$+102.2$ & $1.10 \pm 0.48$ & $15.84 \pm 0.08$ & $4229 \pm 273$ \\
$+106.4$ & $0.82 \pm 0.28$ & $15.75 \pm 0.06$ & $4335 \pm 211$ \\
$+112.6$ & $0.75 \pm 0.28$ & $15.79 \pm 0.07$ & $4088 \pm 219$ \\
$+124.5$ & $0.44 \pm 0.25$ & $15.60 \pm 0.10$ & $4441 \pm 395$ \\
$+294.4$ & $0.14 \pm 0.09$ & \nodata & \nodata \\
$+325.2$ & $0.15 \pm 0.09$ & \nodata & \nodata \\
$+344.1$ & $0.10 \pm 0.02$ & \nodata & \nodata \\
$+376.8$ & $0.04 \pm 0.03$ & \nodata & \nodata \\
\hline
\end{tabular}
\tablecomments{The phase is reported in the rest frame with respect to the time of $r$-band maximum light at MJD = 60169.63.}
\end{table}
\clearpage
\begin{table*}
\caption{Evolution of SN features (\ion{Fe}{2}\,$\lambda$5018 and H$\alpha$).}
\label{tab:line_params}
\hspace{-2cm}
\begin{tabular}{ccccccccc}
\hline\hline
 & \multicolumn{2}{c}{Velocity} & \multicolumn{3}{c}{FWHM} & \multicolumn{3}{c}{EW} \\
\cline{2-3}\cline{4-6}\cline{7-9}
Phase & Fe~{\sc ii} & H$\alpha$ abs. & Fe~{\sc ii} & H$\alpha$ abs. & H$\alpha$ em. & Fe~{\sc ii} & H$\alpha$ abs. & H$\alpha$ em. \\
(day) & (km\,s$^{-1}$) & (km\,s$^{-1}$) & (km\,s$^{-1}$) & (km\,s$^{-1}$) & (km\,s$^{-1}$) & (\AA) & (\AA) & (\AA) \\
\hline
+13.6  & $5597 \pm 94$ & $7925 \pm 75$ & $2485 \pm 265$ & $4755 \pm 213$ & $9484 \pm 1251$ & $9.7 \pm 5.2$ & $53.8 \pm 17.4$ & $267.6 \pm 51.4$ \\
+37.2  & $5341 \pm 94$ & $8065 \pm 75$ & $2700 \pm 266$ & $4874 \pm 213$ & $8897 \pm 1249$ & $7.7 \pm 1.5$ & $61.1 \pm 10.6$ & $292.4 \pm 41.2$ \\
+49.5  & $5400 \pm 17$ & $8060 \pm 13$ & $3529 \pm 49$ & $4677 \pm 38$ & $8283 \pm 57$ & $9.6 \pm 0.2$ & $58.0 \pm 0.3$ & $341.6 \pm 1.1$ \\
+58.4  & $4041 \pm 60$ & $7954 \pm 25$ & $2733 \pm 169$ & $3826 \pm 69$ & $7537 \pm 60$ & $9.1 \pm 3.5$ & $52.9 \pm 18.4$ & $599.3 \pm 2.8$ \\
+61.7  & $4843 \pm 56$ & $7565 \pm 55$ & $2898 \pm 157$ & $4469 \pm 155$ & $8891 \pm 105$ & $8.0 \pm 3.1$ & $63.6 \pm 14.3$ & $355.3 \pm 3.2$ \\
+71.7  & $4894 \pm 136$ & $8187 \pm 104$ & $3414 \pm 383$ & $4497 \pm 293$ & $8292 \pm 76$ & $6.8 \pm 0.1$ & $59.9 \pm 0.2$ & $457.2 \pm 2.2$ \\
+99.9  & $4087 \pm 136$ & $8192 \pm 104$ & $3136 \pm 383$ & $4164 \pm 293$ & $8274 \pm 63$ & $6.7 \pm 0.1$ & $55.3 \pm 0.2$ & $635.2 \pm 2.3$ \\
+135.3 & $3360 \pm 141$ & $7740 \pm 110$ & $1836 \pm 400$ & $3058 \pm 310$ & $7808 \pm 69$ & $1.3 \pm 0.0$ & $30.1 \pm 0.0$ & $901.9 \pm 15.8$ \\
+139.8 & $2841 \pm 141$ & $7967 \pm 110$ & $1487 \pm 400$ & $2605 \pm 310$ & $7560 \pm 182$ & $1.9 \pm 0.0$ & $21.6 \pm 0.0$ & $885.1 \pm 22.3$ \\
+296.2 & $3183 \pm 35$ & \ldots & $542 \pm 100$ & \ldots & $6617 \pm 124$ & $12.8 \pm 1.5$ & \ldots & $949.2 \pm 26.4$ \\
+326.0 & $3333 \pm 91$ & \ldots & $2555 \pm 258$ & \ldots & $6694 \pm 574$ & $14.5 \pm 4.6$ & \ldots & $1032.0 \pm 73.8$ \\
+349.9 & \ldots & \ldots & \ldots & \ldots & $5750 \pm 427$ & \ldots & \ldots & $466.0 \pm 32.2$ \\
+380.6$^{a}$ & \ldots & \ldots & \ldots & \ldots & $3648 \pm 376$ & \ldots & \ldots & $241.4 \pm 13.6$ \\
+380.6$^{b}$ & \ldots & \ldots & \ldots & \ldots & $4333 \pm 345$ & \ldots & \ldots & $470.1 \pm 20.9$ \\
\hline
\end{tabular}
\tablecomments{The abbreviations ``abs.'' and ``em.'' stand for the absorption and emission components of a P-Cygni profile, respectively. 
The line properties at day +380.6 are sensitive to how the data reduction was performed. We report the values of a reduction with \textsc{LPipe} ($^a$) and \textsc{PypeIt} ($^b$). The phase is reported in the rest frame with respect to the time of $r$-band maximum light at MJD = 60169.63.
}
\end{table*}

\end{document}